\documentclass[dvips,preprint,12pt]{elsarticle}
\usepackage{graphicx, algorithmic, algorithm2e,  psfrag, amssymb, amsthm, lineno, amsfonts, amsmath, pstricks-add, bm, float, comment, amsthm, color, hyperref}
\RestyleAlgo{boxruled}
\usepackage[caption=false]{subfig}
\hypersetup{colorlinks,breaklinks,
           linkcolor=blue,urlcolor=blue,
           anchorcolor=blue,citecolor=blue}
\newtheorem{thm}{Theorem}
\newtheorem{lem}{Lemma}
\newtheorem{cor}{Corollary}

\newtheorem{remark}{Remark}
\journal{Journal of Computational Physics}
\begin{document}
\begin{frontmatter}
\title{Basis Adaptive Sample Efficient Polynomial Chaos (BASE-PC)}
\author{Jerrad Hampton}
\author{Alireza Doostan\corref{cor1}}
\ead{alireza.doostan@colorado.edu}
\cortext[cor1]{Corresponding Author: Alireza Doostan}

\address{Aerospace Engineering Sciences Department, University of Colorado, Boulder, CO 80309, USA}

\begin{abstract}
For a large class of orthogonal basis functions, there has been a recent identification {\color{black}of expansion} methods for computing accurate, stable approximations of a {\color{black}quantity of interest}.  This paper presents, within the context of {\color{black}uncertainty quantification}, a practical implementation using basis adaptation, and coherence motivated sampling, which under assumptions has satisfying guarantees. This implementation is referred to as Basis Adaptive Sample Efficient Polynomial Chaos (BASE-PC). A key component of this is the use of anisotropic polynomial order which admits evolving global bases for approximation in {\color{black}an} efficient manner, leading to consistently stable approximation for a practical class of smooth functionals. This fully adaptive, non-intrusive method, requires no \textit{a priori} information of the solution, and has satisfying theoretical guarantees of recovery. A key contribution to stability is the use of a presented correction sampling for coherence-optimal sampling in order to improve stability and accuracy {\color{black}within the adaptive basis scheme}. Theoretically, the method may dramatically reduce the impact of dimensionality in function approximation, and numerically the method is demonstrated to perform well on problems with dimension up to $1000$.
\end{abstract}
\begin{keyword}
Polynomial Chaos \sep Orthogonal Polynomials \sep Uncertainty Quantification \sep Compressive Sensing \sep Basis Adaptation \sep Importance Sampling
\end{keyword}
\end{frontmatter}

\section{\texorpdfstring{Introduction}{Introduction}}
\label{sec:intro}

A reliable approach to analyzing complex engineering systems requires understanding how various Quantities of Interest (QoI) depend upon system inputs that are often uncertain; where a poor understanding will lead to poor executive decisions. Uncertainty Quantification (UQ)~\cite{Ghanem91a,LeMaitre10,Xiu10a} is a field that aims at addressing these issues in a practical and rigorous manner, giving a meaningful characterization of uncertainties from the available information and admitting efficient propagation of these uncertainties for a quantitative validation of model predictions.

Probability is a natural framework for modeling uncertainty, wherein we assume uncertain inputs are represented by a $d$-dimensional random vector $\bm{\Xi}:=(\Xi_1,\cdots,\Xi_d)$ with some joint probability density function $f(\bm{\xi})$ supported on $\Omega$, where we further assume that the coordinates of $\bm{\Xi}$ are independent. In this manner, the scalar QoI to be approximated, here denoted by $u(\bm{\Xi})$, is modeled as a fixed but unknown function of the input. In this work we approximate $u(\bm{\Xi})$, assumed to have finite variance, by a spectral expansion in multivariate basis functions, each of which is denoted by $\psi_k(\bm{\Xi})$, and are naturally chosen to be orthogonal with respect to the distribution of $\bm{\Xi}$~\cite{Xiu02,Soize05}. We focus here on the case that $\psi_k$ are polynomials, a method referred to as a Polynomial Chaos (PC) expansion~\cite{Ghanem91a,Xiu02},
\begin{align}
\label{Eq:PCEDef}
u(\bm{\Xi}) &= \mathop{\sum}\limits_{k=0}^\infty c_k \psi_{k}(\bm{\Xi}).
\end{align}
We note that the independence assumption for the coordinates of $\bm{\Xi}$ may be removed if care is taken in prescribing orthogonal basis functions $\psi_k$, although we do not consider any such examples here. 

For computation, we allow an arbitrary number of input dimensions $d$ but assume $u$ can be accurately approximated in some relatively small set of basis functions. Let $\bm{k} = (k_1,\cdots,k_d)$ be a vector such that $k_i\in{\mathbb{N}\cup\{0\}}$ represents the order of the polynomial $\psi_{k_i}(\Xi_i)$, which is orthonormal with respect to the distribution of $\Xi_i$. For instance, when $\Xi_i$ follows a uniform or Gaussian distribution, $\psi_{k_i}(\Xi_i)$ are normalized Legendre or Hermite polynomials, respectively~\cite{Xiu02,Soize05}. For a $d$-dimensional vector $\bm{k}$, the $d$-dimensional polynomial $\psi_{\bm{k}}(\bm{\Xi})$ is then constructed by the tensorization of $\psi_{k_i}(\Xi_i)$, where $k_i$ is the $i$th coordinate of $\bm{k}$. Specifically,
\begin{align*}
\psi_{\bm{k}}(\bm{\Xi})=\mathop{\prod}\limits_{i=1}^d\psi_{k_i}(\Xi_i).
\end{align*}
In this work we select basis functions in a manner that iteratively adjusts parameters that define a basis. Specifically, we consider a definition of anisotropic total order~\cite{AnisotropicTotalOrderCite} using one parameter, $p_i$, per dimension. We combine these into a vector, $\bm{p}:= (p_1,\cdots,p_d)$, so that an order-$\bm{p}$ basis is defined by a related set of $\bm{k} = (k_1,\cdots,k_d)$, specifically
\begin{align}
\label{eqn:general_order_def}
\mathcal{B}_{\bm{p}}:= \left\{\psi_{\bm{k}}\bigg| \mathop{\sum}\limits_{i=1}^d \frac{k_i}{p_i} \le 1\right\}.
\end{align}
This basis definition has a number of parameters that scales with dimension, and which we will repeatedly modify to improve the quality of our polynomial approximation. We note that if all $p_i = p$, then the order-$\bm{p}$ basis is identical to a total order basis of order $p$. We also note that this basis can have an additional hyperbolicity parameter associated with it as considered in~\cite{namedBASPC}, although we do not consider any such parameter here. Heuristically, we expect most $p_i$ to be low and only a few to be relatively high, allowing a basis that faithfully approximates the QoI with relatively few basis functions compared to a total order basis with an order that is able to achieve the same accuracy in the reconstruction. Often, the subscript on $\mathcal{B}$ is omitted; replaced with a scalar index related to iterative adjustment; or replaced with a bound on approximation error achieved in that basis; and this should not be confusing in context. For the remainder of this text, we refer to an order-$\bm{p}$ basis as an anisotropic order basis.

We use $|\mathcal{B}|$ to denote the total number of basis functions in a set $\mathcal{B}$, indexed in an arbitrary manner for $k=\{1,\cdots,|\mathcal{B}|\}$, while the vector $\bm{k}$ specifically identifies the basis function by determining the order in each dimension. This facilitates a polynomial surrogate approximation to $u$ for any basis set $\mathcal{B}$, given by
\begin{align}
\label{Eq:PCETrunc}
u(\bm{\Xi}) &\approx \mathop{\sum}\limits_{k=1}^{|\mathcal{B}|} c_k \psi_{k}(\bm{\Xi}).
\end{align}
The error introduced by this truncation is referred to as {\it truncation error}, and converges to zero -- in the mean squares sense as basis functions are added -- when 
\begin{align}
\label{eqn:exact_projection_coefficients}
 c_k = \mathbb{E}(u(\bm\Xi)\psi_{k}(\bm\Xi)).
\end{align}
 Here, $\mathbb{E}$ denotes the mathematical expectation operator. Without any \textit{a priori} information as to what $\mathcal{B}$ should be, we seek to identify $\mathcal{B}$ based solely on solution characteristics as revealed by computed coefficients, $\{c_k\}$.

Identifying an optimal $\mathcal{B}$ first involves identifying a scalar quantity to optimize. {\color{black} In the present work, this quantity is related to a cross-validated error computed via $\ell_1$-minimization using non-intrusive methodology \cite{Doostan10b,Doostan11a}}. Specifically, for a fixed basis, to identify the PC coefficients $\bm{c}=(c_1,\cdots,c_{|\mathcal{B}|})^T$ in (\ref{Eq:PCETrunc}) we consider a sampling-based method. This method does not require changes to deterministic solvers for $u$ as we generate realizations of $\bm{\Xi}$ to identify $u(\bm{\Xi})$, or perform a related importance sampling as in~\cite{Hampton14, Hampton15}. We denote the $i$th such realizations as $\bm{\xi}^{(i)}$ and $u(\bm{\xi}^{(i)})$, respectively. We let $N$ denote the number of samples of the QoI which we utilize, and define,
\begin{align}
\label{eqn:psi_u}
\bm{u}&:=(u(\bm{\xi}^{(1)}),\cdots,u(\bm{\xi}^{(N)}))^T;\\
\label{eqn:psi_specific}
\bm{\Psi}(i,j)&:=\psi_{j}(\bm{\xi}^{(i)}),
\end{align}
where we refer to $\bm\Psi$ as the {\it measurement matrix} {\color{black}associated with $\mathcal{B}$}. These definitions imply the matrix equality $\bm{\Psi}\bm{c}=\bm{u}$, or more generally that this equality holds approximately. We also introduce a diagonal positive-definite matrix $\bm{W}$ such that {\color{black}$\bm{W}(i,i)=w(\bm{\xi}^{(i)})$, a function of $\bm{\xi}^{(i)}$, is determined by our sampling strategy in the manner of basis-dependent importance sampling; see ~\cite{Hampton14, Hampton15} and Section \ref{subsec:sample_id}}. This weighting and the corresponding importance sampling are described in Section~\ref{subsec:sample_id}. Here we {\color{black} employ compressive sampling, specifically} {\color{black} the} {\color{black} Basis Pursuit Denoising~\cite{Chen98,Chen01,Donoho06b,Bruckstein09} interpretation of $\ell_1$-minimization, to compute $\hat{\bm{c}}$, our identified coefficients, }
\begin{align}
\label{eqn:ell1}
\hat{\bm{c}} := \mathop{\mbox{argmin}}_{\bm{c}}\|\bm{c}\|_1 \mbox{ subject to } \|\bm{W}(\bm{u}-\bm{\Psi}\bm{c})\|_2 \le \delta\|\bm{Wu}\|_2,
\end{align}
where $\delta$ is set via cross-validation {\color{black}\cite{Doostan11a}}. The optimization in (\ref{eqn:ell1}) may be solved efficiently via interior point methods, where we utilize here an implementation of SPGL1~\cite{SPGL1} that is slightly modified for repeated utilization, as our method depends on repeatedly computing these coefficients for various bases. We refer to $\bm{W\Psi}$, as the \textit{design matrix}, denoted by $\bm{D}$, i.e. 
\begin{align}
\label{eqn:design_def}
\bm{D} &:= \bm{W\Psi}.
\end{align}
We note that we use (\ref{eqn:ell1}) here to compute coefficients, and that this is motivated from the robustness of compressive sensing wherein the number of samples is small compared to the number of basis functions. That is solutions to (\ref{eqn:ell1}) are robust to including unnecessary basis functions, defined as basis functions whose inclusion does not significantly reduce the error of the reconstructed surrogate. This is important, as though we seek to limit the number of unnecessary basis functions, computing solutions via (\ref{eqn:ell1}) insures that having unnecessary basis functions has a relatively small effect on the number of samples needed to compute an accurate surrogate. We note that our method to identify this basis generally reduces the number of basis functions considerably, potentially to the point where the number of basis functions is exceeded by the number of samples. In this sense, the method presented here is not clearly interpreted in terms of compressive sensing, although the theoretical guarantees with regards to sparsity concerning solutions computed via $\ell_1$-minimizations still apply.

Recalling (\ref{Eq:PCETrunc}), we denote our surrogate approximation to $u$ in terms of these computed coefficients, $\{\hat{c}_k\}$, by
\begin{align}
\label{eqn:uhat_def}
\hat{u}(\bm{\Xi}) := \mathop{\sum}\limits_{k=1}^{|\mathcal{B}|} \hat{c}_k \psi_{k}(\bm{\Xi}).
\end{align}
Here the surrogate reconstruction of $u$, denoted $\hat{u}$, is computed iteratively via solution to (\ref{eqn:ell1}) repeated over different potential reconstruction bases and available samples. We measure our error by relative root-mean-square error (RRMSE), defined by
\begin{align}
\label{eqn:RRMSE_def}
\mbox{RRMSE}(\hat{u}) := \frac{\sqrt{\mathbb{E}(\hat{u}(\bm{\Xi})-u(\bm{\Xi}))^2}}{\sqrt{\mathbb{E}(u^2(\bm{\Xi}))}}.
\end{align}
Our identification of a basis is done so as to minimize a validated estimate of $\mbox{RRMSE}(\hat{u})$, i.e. we select a basis that with its corresponding computed coefficients returns the lowest estimate of $\mbox{RRMSE}(\hat{u})$ from the set of considered bases. This estimate is computed from repeated solution of (\ref{eqn:ell1}) for different subsamples of our total pool of available samples.  The class of potential anisotropic order bases depends on the computed coefficients,  $\{\hat{c}_k\}$, as well as the dimension and order of the associated basis functions. Heuristically, $p_k$ is increased in dimensions with basis functions having high order in that dimension and large magnitude solution coefficients. Conversely, $p_k$ is decreased in dimensions where basis functions having high order in that dimension are associated with solution coefficients having low magnitude.

This is a heuristic similar to that utilized in~\cite{dai2008subspace}, and typically favors dimensions with more local variance as in the approach of~\cite{wan05}. From this coefficient magnitude information, the basis is adapted from a basis denoted $\mathcal{B}_0$ to one denoted $\mathcal{B}_1$. If specified, this basis adaptation also includes an increase to the {\color{black}dimension of} the PC basis. During this step, several potential $\mathcal{B}_1$ are generated and tested. From this set of potential bases the basis giving the lowest cross-validated approximation error is kept. With this error minimizing basis, new samples are identified that assure a low coherence for the aggregate samples with respect to this basis as in~\cite{Hampton14, Hampton15}. With these additional samples, the basis may then be updated again, and the process of basis adaptation and sample identification may be repeated in an iterative manner.

{\color{black} Recalling $c_k$ from (\ref{eqn:exact_projection_coefficients}), we assume the error model} 
\begin{align}
\label{eqn:pre_error}
u(\bm{\Xi})&=\mathop{\sum}\limits_{k=1}^{|\mathcal{B}|} c_k \psi_{k}(\bm{\Xi}) + \epsilon(\bm{\Xi}),\\
&\approx\mathop{\sum}\limits_{k=1}^{|\mathcal{B}|} \hat{c}_k \psi_{k}(\bm{\Xi}) + \epsilon(\bm{\Xi}),\nonumber\\
&=\hat{u}(\bm{\Xi}) + \epsilon(\bm{\Xi}), \nonumber
\end{align}
noting that the robustness of solutions with regards to model or measurement errors has been investigated~\cite{CDL13,CandesPlan, Hampton14, Hampton15}. Generally, we seek to guarantee that $\mbox{RRMSE}(\hat{u})$ is close to $\sqrt{\mathbb{E}(\epsilon^2(\bm{\Xi}))}/\sqrt{\mathbb{E}(u^2(\bm{\Xi}))}$. As the number of basis functions used for our approximation increases, the error arising from performing regression with an incomplete set of basis functions is shown for examples to converge to zero more rapidly than for comparable non-adaptive bases, both in terms of the number of samples needed to compute the approximation, and with regards to the number of basis functions used in the approximation. 

In summary, to achieve any specified approximation error, the design and measurement matrices for the basis adaptive approach require significantly fewer entries than the corresponding non-adaptive approach. These methods are referred to collectively as Basis Adaptive Sample Efficient Polynomial Chaos (BASE-PC), and are presented in detail in Section~\ref{sec:sampling}. While Compressive Sensing~\cite{Candes06a,Donoho06b,Elad10a,Eldar12a} can handle a relatively large set of basis functions using the sparsity promoted in solutions to (\ref{eqn:ell1}), and do so within the context of UQ~\cite{Doostan10b,Doostan11a,Blatman11,Mathelin12a,Yan12,Yang13,Karagiannis14,Peng14,schiavazzi14,Sargsyan14}, the number of basis functions is still responsible for algorithmic bottlenecks, and a reduction of $|\mathcal{B}|$ through the shaping of the operative basis can produce significant gains in accuracy~\cite{wan05, Blatman11, Jakeman15, namedBASPC, dai2008subspace}.

Though not considered here, as in~\cite{Adcock15, RauhutWard15}, an independent column weighting, $\bm{V}$, may be used to reduce the contribution of higher order polynomials and give a more stable approximation for high order models, particularly if interpolation is desired in place of the regression considered here. Further, the results of~\cite{Adcock15} may assist with identifying appropriate ratios of samples to basis functions for stable, alias-free approximations in such cases. We also note that the inclusion of derivative information as in~\cite{Peng16} falls within the coherence and coherence-optimal sampling {\color{black}framework, although} we do not consider any examples that utilize derivative information here. Noting that a truncation to a finite-dimensional problem is necessary for computation, $d$ may be infinite within similar contexts as in~\cite{AdcockHansen15}, although we assume in this work that some truncation to a finite dimension $d$ is identified before computation is performed. The infinite dimensional results and framework of~\cite{AdcockHansen15} also directly corresponds to our use of $\ell_1$-minimization on subsets of the infinite set of basis functions which exists in the context of polynomial approximation, even when $d$ is finite.

\subsection{\texorpdfstring{Contributions of This Work}{Contributions of This Work}}
\label{subsec:Contribution}
This work combines and advances several results from recent developments in PC into a single practical implementation designed to promote stability and convergence with theoretical guarantees. As an extension of {\color{black}previous related work, the main} contributions of this study are as follows. 

The sampling distributions in~\cite{Hampton14, Hampton15} are given expanded utility to the practical case where the reconstruction basis may change. This is done by identifying a novel correction sampling that retains all previously generated samples, while giving aggregate sample pools from an appropriate distribution that guarantees a stability in the approximations, i.e. that allows an adaptation of the sampling distribution to similarly adapting bases. This proposed use of correction sampling within importance sampling is novel to the authors' knowledge.

This method also provides an approach to adaptive PC that builds upon and differs conceptually from adaptations in the stochastic space~\cite{LeMaitre03a, wan05, wan2006multi}, and utilizes a different approach to basis adaptivity when compared to other proposed methods for adapting the basis~\cite{dai2008subspace, Blatman11, Jakeman15, NiAdaptive, namedBASPC}. Key to this adaptation is the use of anisotropic total order, which is described by $d$ parameters, allowing for an efficient approach to adaptation, while being robust with regards to the functions it is capable of approximating. Specifically, it uses a global basis that is a specific version of those considered in~\cite{namedBASPC}, while using different methods for sampling and  basis identification. This basis avoids more specific adaptations as in~\cite{dai2008subspace, Blatman11}, which can lead to bases whose descriptions are more complex. Our adaptation of the basis also combines a heuristic for coefficient magnitude similar to that in~\cite{dai2008subspace}, and a minimization of estimated $\mbox{RRMSE}(\hat{u})$ similar to that in~\cite{namedBASPC}, that is also novel to the authors' knowledge. We note that the {\color{black}BASE-PC method here} should not be confused with the independently developed BASPC of~\cite{namedBASPC}, which has a similar acronym and purpose, as well as similarity in several computations. It also differs from the approach of~\cite{alemazkoor16} which focuses on identifying which dimensions are to be included into the approximation. {\color{black} A key difference between the approach here and other approaches is that the approach here is able to exploit sparsity, but does not explicitly depend upon it, and is capable of recovering both sparse and non-sparse solutions. It is suspected that many of the above methods too have this property, although this work demonstrates said property explicitly.}

We also provide significant theoretical justification for the BASE-PC method, which can be {\color{black} possibly extended to other adaptive} approaches. Under some justifiable assumptions we provide theoretical guarantees for both the basis and sample adaptive approaches used here. {\color{black} This analysis also expands to the case of non-sparse recovery which is a critical property for the basis adaptation approach, and fills a gap in analysis within the current basis adaptation literature.} Further, we identify a set of functionals that under some assumptions are recovered by the BASE-PC method with a number of samples that does not depend on $d$, the dimension of the random inputs. In this case, the number of elements in the approximating basis also does not depend on $d$. This result is of interest with regards to the so-called curse-of-dimensionality associated with computations regarding high dimensional problems. 

{\color{black}The organization of this paper has Section~\ref{sec:sampling} describing the implementation of BASE-PC in detail with an algorithmic description of components critical for driving the basis and sample adaptations; Section~\ref{sec:examples} presenting numerical examples; and Section~\ref{sec:theory} providing theoretical justifications for the repeated iteration of the BASE-PC method.}
\section{\texorpdfstring{BASE-PC Implementation Details}{BASE-PC Implementation Details}}
\label{sec:sampling}
Here we present a detailed account of the BASE-PC iteration and its constituent functions presented in pseudocode, including default parameters. The implementation described here is that used in the examples of Section~\ref{sec:examples}. These computations are divided into three categories corresponding to three subsections: Those computations associated with the evaluation and identification of the basis are presented in Section~\ref{subsec:basis_id}; those computations used for identification of new sample points are presented in Section~\ref{subsec:sample_id}; and those computations which identify the surrogate approximation for a given basis and sample set are presented briefly in Section~\ref{subsec:coef_id}. All of these components are utilized in a main iteration as described in Section~\ref{subsec:main_iteration}.
\subsection{\texorpdfstring{Basis Evaluation and Update}{Basis Evaluation and Update}}
\label{subsec:basis_id}
For each input dimension, the identification of the one-dimensional orthonormal polynomials are given by the appropriate three-term recursion in a computationally efficient manner. We refer to this basic one-dimensional identification of a particular order by \textit{basis\_eval\_1d}({\tt type}, $p$, $\xi$), where {\tt type} determines the appropriate polynomial family; $p$ refers to the maximal order polynomial to be computed in that dimension; and $\xi$ refers to the point at which evaluation is occurring. 

The identification of the multi-dimensional orthonormal polynomials is referred to as \textit{basis\_eval}($\mathcal{B}$,$\bm{\xi}$), where $\mathcal{B}$ represents a description of the basis at which evaluation is occurring, including relevant order information, and $\bm{\xi}$ is the point at which the basis should be evaluated. This function identifies each one-dimensional evaluation via \textit{basis\_eval\_1d}, before multiplying them appropriately to identify the evaluation of each basis function at the input.

It is necessary for bases of arbitrary anisotropic order to be constructed, and we refer to this function as \textit{basis\_id}($\bm{p}$), where $\bm{p}$ is as in (\ref{eqn:general_order_def}), identifying the requested anisotropic basis. For brevity, a specific algorithm is not presented here, though the construction is explained in some detail relative to the construction of a total order basis.

First, the identification of the basis is done by sorting $\bm{p}$ by dimension in a descending manner, so that $p_{(1)}$ corresponds to the maximal coordinate of $\bm{p}$. A loop is initialized so as to identify the total order basis of $p_{(1)}$, and each such basis function is tested to see if it meets the prescribed anisotropic order criteria. This determines whether or not the basis function is a member of the prescribed anisotropic total order basis, and it is added if it is a member. Due to the sorting of orders, basis functions may be efficiently discarded, in that one failed test guarantees the failure of potentially many other basis functions, so that the number of tests is kept small. In this way, when $p_{(1)}$ is large but many other orders are small, relatively few basis functions {\color{black}need to be tested} when compared to the potentially large size of the total order basis having potentially large order and dimension. Hence, the identification of the basis is computationally tractable even when the requested anisotropic order basis has a high order in some dimensions, and a large total number of dimensions of relatively low order. We note that in such a case iterating over the full total order basis  associated with order $p_{(1)}$ would be infeasible due to the combinatorially large number of basis functions of a total order basis when both dimension and order are large. We note that this sorting of dimension based on the order of the anisotropic order basis is not kept for the remainder of what occurs, being used only for the construction of the basis.

For a given basis and set of {\color{black}input samples} $\{\bm{\xi}^{(k)}\}_{k=1}^N$, we can form the measurement matrix $\bm{\Psi}$ that evaluates each basis function at each input, as in (\ref{eqn:psi_specific}). With an additional weight matrix $\bm{W}$ that is diagonal and positive-definite, we can form $\bm{D}=\bm{W}\bm{\Psi}$.

For a given basis, when the surrogate {\color{black}coefficients, $\bm{c}$,} have been identified, we may remove $m$ basis functions coinciding with small {\color{black}entries} of $\bm{c}$. This allows us to shape and adapt the basis as per our heuristic of removing basis functions that have correspondingly small coefficient. We refer to this as basis contraction. We do this using a method called \textit{basis\_contract}($\mathcal{B}$,$\bm{c}$,$m$), and presented in Algorithm~\ref{alg:contract}. The parameter {\color{black}$m$ is} looped over during the basis adaptation procedure. We note that in the case that multiple minimizing $|c_i|$ exist, we choose the one with smallest index $i$.
\begin{algorithm}[htb]
\begin{algorithmic}
\STATE Set $\mathcal{R}=\emptyset$ \% Will contain basis functions to remove.
\FOR{$k \le m$}
\STATE Set $k =\mathop{\arg\min}\limits_{i\in\mathcal{B}\setminus\mathcal{R}}|c_i|$. \% Minimize over elements $\mathcal{B}$ not in $\mathcal{R}$.
\STATE{Set $\mathcal{R} = \mathcal{R}\cup\{k\}$.} \% Add basis function to be removed.
\ENDFOR 
\STATE Return $\mathcal{B}\setminus\mathcal{R}$ \% Contracted basis is elements of $\mathcal{B}$ not in $\mathcal{R}$.
\end{algorithmic}
\caption{\textit{basis\_contract}($\mathcal{B}$,$\bm{c}$,$m$): Returns contraction of input basis, using information from a computed solution.}
\label{alg:contract}
\end{algorithm}

Adjoint to contraction of the basis is expansion of the basis, through a function referred to as \textit{basis\_expand}($\mathcal{B}$), and presented in Algorithm~\ref{alg:expand}. We note that \textit{basis\_expand} expands general bases that do not coincide with anisotropic order bases, specifically bases that have had a number of basis elements removed via \textit{basis\_contract}. The parameter $\gamma$ in \textit{basis\_expand} controls the relative expansion of the basis, with higher values leading to larger bases. For the examples in Section~\ref{sec:examples} $\gamma=1.5$ is larger for the low dimensional problem of Sections~\ref{subsec:franke} and $\gamma=1.3$ is used for the low-dimensional problem in~\ref{subsec:surface_adsorption}. Similarly, $\gamma=1.01$ is smaller for the  problems of Sections~\ref{subsec:cavity} and~\ref{subsec:sine_exponential_1000d}. The larger $\gamma$ helps accelerate adaptation when the dimensions are smaller and the orders are expected to be relatively larger, while in the higher dimensional case it becomes more important to restrain the number of basis functions as the typical order of basis in any given dimension is low. Generally, small values of $\gamma$ will work well, at the potential cost of needing more basis adaptation iterations. 

We also include a certain number of new dimensions at order $1$, denoted {\tt dim\_add}, which is set to $20$ for the examples in Section~\ref{sec:examples}. The modification for {\tt dim\_add} is most important for the example in Section~\ref{subsec:sine_exponential_1000d}. The other examples have $20$ or fewer dimensions, and this constraint simply enforces that the minimal order in each dimension for those problems is 1, i.e. there is at least a linear term in each dimension.
\begin{algorithm}[htb]
\begin{algorithmic}
\STATE Set $\bm{p} = \bm{0}$. \% Will hold order information.
\FOR{$\bm{k}$ such that $\psi_{\bm{k}}\in\mathcal{B}$} 
\STATE Set $\bm{p} = \mathop{\max}(\bm{p},\bm{k}).$ \% Maximum is taken coordinate-wise. 
\ENDFOR
\STATE Add up to {\tt dim\_add} dimensions to $\bm{p}$ at order $1$.
\STATE $\mathcal{B} = $\textit{basis\_id}($\lceil\gamma\bm{p}\rceil$). \% Ceiling function is taken coordinate-wise.
\end{algorithmic}
\caption{\textit{basis\_expand}($\mathcal{B}$): Returns expansion of input basis.}
\label{alg:expand}
\end{algorithm}

In the examples, \textit{basis\_contract} and \textit{basis\_expand} are used in tandem, repeatedly expanding further contracted bases. These {\color{black}contracted bases} are further contracted by removing additional basis functions, leading to different expanded bases, and choosing the basis from a number of these by selecting which one produces a minimal validated error in surrogate approximation. As basis stability and obtaining the lowest available errors are a priority, it is reasonable to admit more solution solves. Hence, a basis can be selected at each iteration from a set of candidate bases that minimizes an estimate of the RRMSE, a process which we refer to as basis validation. The algorithm to do this validation is summarized in Algorithm~\ref{alg:validate}, and is referred to as \textit{basis\_validate}($\mathcal{B}_0$,$\bm{c}_0$). For the computations here, {\tt max\_strikes} $ = 6$, where this parameter is instrumental for identifying the size of candidate bases we have to select from, where we stop identifying candidate bases with confidence that expansion of further contracted bases is unlikely to produce a basis with a lower estimate of RRMSE. Further, the basis adaptation procedure of expanding a contracted basis may be performed efficiently by noting that \textit{basis\_contract} need only remove one new element of an already sorted coefficient vector $\bm{c}$ and new coefficients, and error estimates need only be computed when \textit{basis\_expand} produces a new basis. Here, a strike is an event where a validated error does not fall below the minimum achieved validated error. For computational efficiency the algorithm terminates if too many strikes are accumulated, resetting the strike counter if a new minimum is achieved.
\begin{algorithm}[htb]
\begin{algorithmic}
\STATE Let $n = |\mathcal{B}_0|$. \% The number of basis elements in $\mathcal{B}_0$.
\STATE Set $m=0$, {\tt strikes} $ = 0$, and {\tt min\_error} $= \infty$.
\WHILE{$m\le n\ \&\ $ {\tt strikes} $<$ {\tt max\_strikes}}
\STATE Set $\mathcal{B}_m = $\textit{basis\_expand}(\textit{basis\_contract}($\mathcal{B}_0$,$\bm{c}$,$m$)).
\IF{$\mathcal{B}_m\ne\mathcal{B}_{m-1}$}
\STATE Evaluate all samples and QoI for $\mathcal{B}_m$ to get $\bm{D}_m$ and $\bm{W}_m\bm{u}$.\\
\STATE Compute surrogate coefficients $\bm{c}_m$ and estimate of RRMSE $\epsilon_m$. \\
\% Surrogate computation details are presented in Section~\ref{subsec:coef_id}.
\IF{$\epsilon_m <$ {\tt min\_error}}\STATE{{\tt min\_error} $= \epsilon_m\ \&\ $ {\tt strikes} $ = 0$.}
\ELSE\STATE{{\tt strikes} $=$ {\tt strikes} $ + 1$.}
\ENDIF
\ENDIF
\STATE $m = m+1$.
\ENDWHILE
\STATE Return basis achieving minimal validated error.
\end{algorithmic}
\caption{\textit{basis\_validate}($\mathcal{B}_0$,$\bm{c}_0$): Returns validated basis from set of potential bases.}
\label{alg:validate}
\end{algorithm}

For cases of moderate dimensionality, the prescribed methods are sufficient. However when nearly linear scaling in dimension is required, it is useful to provide an upper bound on the orders prescribed for each dimension, a method referred to as \textit{basis\_upper\_bound} and presented in Algorithm~\ref{alg:upper_bound}. This algorithm is only used for the example in Section~\ref{subsec:sine_exponential_1000d}, but is important there as without it, the number of basis functions during the basis expansion phase would quickly grow too large for tractable computation. We also note that this algorithm can be used by first ordering $\bm{p}$ in descending order, although we do not do so here, as the dimensionality in Section~\ref{subsec:sine_exponential_1000d} is already loosely sorted in a descending order of importance.
\begin{algorithm}[htb]
\begin{algorithmic}
\STATE Let $i_k$ index the last coordinate of $\bm{p}$ having order $k$.
\STATE Initialize $v$
\STATE Let $k^{\star}$ be max $k$ such that $i_k$ is defined.
\FOR{$k\le k^{\star}$}
\STATE Set $v_{k} = i_k+ {\tt dim\_add}$.
\ENDFOR
\STATE Initialize $b$ \% Is the vector that bounds the order in each coordinate.
\FOR{$k\le k^{\star}$}
\STATE Set $b(1:v_{k}) = k$. \% Set first $v_k$ entries of $b$ to $k$.
\ENDFOR
\STATE Set $b(1:{\tt dim\_add}) = b(1:{\tt dim\_add})+1$. \% Increase order for first dimensions.
\end{algorithmic}
\caption{\textit{basis\_upper\_bound}: Returns coordinate-wise upper bound on $\bm{p}$.}
\label{alg:upper_bound}
\end{algorithm}
This use of an upper bound on order at each iteration can prevent quadratic scaling in dimension from including 2nd order terms for a large number of dimensions. A linear or even constant approximation may be sufficient for most dimensions, and only a few dimensions need basis functions of higher order. Moreover these bounds may be systematically adjusted at each iteration, without a priori assumptions about an ideal basis for approximation. We note that another alternative to reduce the expansion of basis functions is to initialize $m$ in Algorithm~\ref{alg:validate} to some integer greater than $0$, although we do not consider doing so here. Adjusting this parameter would also reduce the size of expanded bases, and potentially reduce the number of bases for which estimates of the RRMSE need be computed.

After a solution has been updated in the new basis, we increase the number of samples used to compute coefficients. Motivated by a desire for a coherence-optimal sampling in our new basis, additional samples may be generated using the new basis as well as the basis used for sample generation in the previous iteration.  This process is particularly useful in certain cases where high order approximations are needed in one or more dimensions, leveraging the benefits of coherence-optimal sampling~\cite{Hampton14, Hampton15}, and not requiring \textit{a priori} knowledge about which dimensions require higher orders. Sometimes it is reasonable and practical to simply draw all samples from the same distribution, such as from the orthogonality distribution, and this provides a useful comparison for the examples in Section~\ref{sec:examples}.
\subsection{\texorpdfstring{Sample Generation}{Sample Generation}}
\label{subsec:sample_id}
In this work, when not sampling from an orthogonality distribution, sampling is done via Markov Chain Monte Carlo (MCMC) so as to minimize the coherence defined in~\cite{Hampton15}. We note that this distribution depends on the $\ell_2$-norm of the proposed vector of evaluated basis functions, as well as the orthogonality distribution. For each sample, denoted $\bm{\Xi}^{(k)}$, a weight $w^{(k)}$ is associated, so that in aggregate the design matrix $\bm{D}$ satisfies 
\begin{align}
\label{eqn:design_isotropy}
\mathbb{E}(\bm{D}^T\bm{D})=N\bm{I}.
\end{align}
For orthogonality distributions with infinite support, like the normal distribution, it is convnient to relax this requirement to holding only in an approximate sense~\cite{CandesPlan, Hampton14}.

Our implementation for drawing $N$ samples from a distribution $g$ is referred to as \textit{mcmc\_sample}($g$,$N$). We note that this implementation of MCMC does not utilize adaptive proposal distributions, perpetually drawing proposals from the orthogonality distribution, though this is not ideal for e.g. high-order Hermite polynomials and the normal distribution~\cite{Hampton14}. Our method tunes the sampling with a burn-in parameter. Several burn-in samples are repeated until a running average of the normalization constant for the distribution is stabilized, as this helps to insure a quality sample, and then these burn-in samples are discarded and not utilized as draws from the desired distribution. 

To improve the quality of sampling we also seek to limit the number of so-called collisions between samples, where a collision is defined to be when one MCMC sample is identical to the previous MCMC sample, which can arise when a large number of potential samples are rejected in sequence. To prevent this we draw more intermediate samples before accepting the next sample. An upper bound on this collision rate is enforced, specifically $\exp(-8)\approx 0.00033$, and duplicate samples are not kept. This number, having no particular significance, may be reduced if more accurate samples are needed. This imparts a negligible bias in the MCMC sample as the collision rate may be kept quite low without much computational burden. We note that our parameters produce a quality sampling from most coherence-optimal distributions, while being computationally quick. However, a more careful sampling that utilizes more resources may produce better results. It is also possible to use these generated samples as candidates for more specific experimental designs~\cite{dykstra71, shin16}. This design motivated approach is beyond the scope of this paper, and is a focus of future work.

The weight function $w(\bm{\xi})$ attached to every potential sample is related to the orthogonality distribution $f(\bm{\xi})$ and sampling distribution $g(\bm{\xi})$, as in~\cite{Hampton15}. For the initial sample, 
\begin{align}
\label{eqn:uncorrected_sample_dist}
g(\bm{\xi}) &= c_g\|\bm{\psi}(\bm{\xi})\|^2_2f(\bm{\xi}),
\end{align}
where $\bm{\psi}(\bm{\xi})$ is the row vector of realized basis functions evaluated at $\bm{\xi}$, $f(\bm{\xi})$ is the prescribed distribution for the uncertain inputs, {\color{black}and $c_g= |\mathcal{B}|^{-1}$ is the corresponding normalizing constant~\cite{Hampton15}.  As}
\begin{align*}
\mathbb{E}(\bm{D}^T\bm{D})_{i,j} = \int_{\Omega}w^2(\bm{\xi})\psi_i(\bm{\xi})\psi_j(\bm{\xi})g(\bm{\xi})d\bm{\xi}, 
\end{align*}
it follows that (\ref{eqn:design_isotropy}) is satisfied when $w(\bm{\xi}) = \sqrt{|\mathcal{B}|}\|\bm{\psi}(\bm{\xi})\|_2^{-1}$.

\subsubsection{\texorpdfstring{Correction Sampling}{Correction Sampling}}
\label{subsubsec:sample_id}
At each BASE-PC iteration, we consider two bases. The previous basis, denoted $\mathcal{B}_k$, and the current basis, denoted $\mathcal{B}_{k+1}$. Each basis has an associated coherence-optimal distribution from Section~\ref{subsec:sample_id}, which we denote $g_k$ and $g_{k+1}$, respectively. Our correction sampling assumes all previous samples were drawn from $g_k$, and wishes to draw additional samples maintaining (\ref{eqn:design_isotropy}), while having the aggregation of all samples be drawn in a way that resembles independent draws from $g_{k+1}$. This is done by implicitly defining the correction distribution $g_{k}^c$ by the identity
\begin{align}
\label{eqn:sample_split_dist}
(1-\alpha_k)g_k(\bm{\xi}) + \alpha_k g_k^c(\bm{\xi}) = g_{k+1}(\bm{\xi}).
\end{align}
Here $\alpha_k$ must be chosen large enough such that $g_k^c(\bm{\xi})\ge 0$ for all $\bm{\xi}$ in the relevant domain. Additionally, considering $g_{k+1}$ as a mixture of $g_k$ and $g_k^c$, $\alpha_k$ is connected to the sample sizes from the previous basis, denoted $N_k$; the new complete number of samples treated as if drawn from $g_{k+1}$, denoted $N_{k+1}$; and the number of correction samples used to do this, denoted,
\begin{align*}
N_k^c := N_{k+1}-N_k. 
\end{align*}
Interpreting (\ref{eqn:sample_split_dist}) in terms of this sampling idea,
\begin{align*}
\alpha_k = \frac{N_k^c}{N_{k+1}}.
\end{align*}
These requirements are combined as outlined in Algorithm~\ref{alg:corr_sample}, which generates $N_k^c$ transition samples, where $N_k^c$ is identified within an acceptable range of values. This algorithm requires a few parameters. There is a parameter for maximum sampling ratio, denoted {\tt max\_sample\_ratio} that enforces a maximum on how many correction samples are allowed as a ratio of the current sample size. In our examples, {\tt max\_sample\_ratio} $=1$, that is the sample size may at most double at each sampling. The primary benefit of setting {\tt max\_sample\_ratio} is to not require an impractical number of correction samples. Also, there is a minimum sampling ratio {\tt min\_sample\_ratio}, that bounds the minimum number of samples in the correction sample, relative to the current sample size, which is set a priori and varies for our examples between $0.1$ to $0.3$ depending on the computational budget. The main benefit of setting {\tt min\_sample\_ratio} is to reduce the number of iterations that would occur if a low number of samples were generated on each iteration.

Algorithm~\ref{alg:corr_sample} also specifies {\tt weight\_correction}, a variable that is used in the case that $\alpha$ in (\ref{eqn:sample_split_dist}) must be chosen larger than what {\tt max\_sample\_ratio} admits. Here, {\tt weight\_correction} artificially inflates $\alpha_k$ from (\ref{eqn:sample_split_dist}) by giving samples from the correction distribution higher weight, producing an effect similar to having more samples from that distribution. The factor, {\tt weight\_correction}, is multiplied to all rows of $\bm{D}$ corresponding to new samples generated by {\tt sample\_expand}, i.e. associated with the correction sampling. {\color{black}Its} primary role is to insure that (\ref{eqn:design_isotropy}) holds after the correction sampling. This multiplication is done for the next solution computation only, and for all subsequent samples the generated random variables are all assumed to have been drawn independently from the prescribed $g_k$. 

There are two reasons for this. First, the correction sampling assumes all previous samples are drawn from $g_k$, and maintaining previous weights contradicts this assumption. Further, for any given iteration, the violation of (\ref{eqn:design_isotropy}) that comes from misrepresenting previous weights vanishes as the overall sample size increases. Second, it is preferable that the aggregate sample not maintain lasting effects from previous samples. Having a few previous samples from a correction sampling that had attached to it a very large weight would potentially lead to the function being fit unnecessarily well at those points, at the detriment of other points in the domain. Stated another way, the RRMSE in the surrogate would be increased by inappropriately fitting some areas of the domain due to the persistence of weights.
\begin{algorithm}[htb]
\begin{algorithmic}
\STATE Set $\alpha_k = $ {\tt min\_sample\_ratio}.
\WHILE{$\alpha_k$-validated sample not generated}
\STATE Set $N_k^c = \lceil\alpha N_k\rceil$.
\STATE Set $\alpha_k = N_k^c/(N_k+N_k^c)$. \% Ceiling function changes $\alpha$ slightly.
\STATE Define $g_k^c$ via (\ref{eqn:sample_split_dist}).
\STATE Set $(\tilde{\alpha}_k,$ sample$)$ = \textit{mcmc\_sample}($g_k^c$,$N_k^c$).\% $\alpha_k$ may be increased.
\STATE \% The need to increase $\alpha_k$ is revealed during sampling.
\IF{$\tilde{\alpha}_k > \alpha_k$}
\STATE Set $\alpha_k = \tilde{\alpha}_k$. \% Increase $\alpha_k$ if needed.
\ELSE
\STATE Break while loop \% Here $\alpha_k$ has validated on the sample.
\ENDIF
\ENDWHILE
\IF{$\alpha_k >$ {\tt max\_sample\_ratio}}
\STATE Remove samples from $N_k^c$ so that $N_k^c/N_k < {\tt max\_sample\_ratio}$.
\STATE Set $\alpha^{\prime}_k = N_k^c/(N_k+N_k^c)$. \% Note that $\alpha^{\prime}_k < {\tt max\_sample\_ratio}$.
\STATE Set {\tt weight\_correction} $= \alpha_k^{-1}\alpha^{\prime}_k$. \% This is larger than $1$.
\STATE Set {\tt true\_sample\_ratio} $= \alpha^{\prime}_k$.
\ELSE
\STATE Set {\tt weight\_correction} $= 1$. \% No weight correction necessary.
\STATE Set {\tt true\_sample\_ratio} $= \alpha_k$.
\ENDIF
\end{algorithmic}
\caption{\textit{sample\_expand}($\mathcal{B}_k$, $\mathcal{B}_{k+1}$): Returns sample with correction to be used for next solution computation.}
\label{alg:corr_sample}
\end{algorithm}
\subsection{\texorpdfstring{Surrogate/Coefficient Identification}{Surrogate/Coefficient Identification}}
\label{subsec:coef_id}
With a basis and sample identified, a surrogate solution is identified by computing coefficients for each basis function. These coefficients are computed by solving (\ref{eqn:ell1}) with a cross-validated $\delta$~\cite{Doostan11a} to minimize a validated estimate of RRMSE, using a certain number of folds and a certain number of validation samples in each fold. Here the range of $\delta$ is given based on the previous validated error or an initial value. Specifically, the set of potential $\delta$ is $0$ and a set of $20$ tolerances that are spaced, evenly in a logarithmic scale, around the largest of the previous minimizing tolerance or validated error. Further, $24$ randomly generated partitions of the data are used to compute an estimate of the RRMSE and a corresponding $\delta$ for each partition. For each such partition, 80\% of samples are used for computation of the solution, while 20\% are used for validation. The number of partitions and percentage of validation samples are generally larger than needed for a relatively accurate estimation of {\color{black}error. We} note that the method of error estimation used here is closely related to the leave-one-out error estimate of~\cite{namedBASPC}. It may be useful in certain situations to consider other validation techniques, although this is sufficient for the examples here.

\subsection{\texorpdfstring{Main Iteration}{Main Iteration}}
\label{subsec:main_iteration}
The main iterative process then is to sequentially identify new bases for the surrogate approximation and new samples that are compatible with this sequence of bases so that the aggregate sample at each iteration mimics a coherence-optimal sample for the appropriate basis at that iteration. To clarify the presentation, we define \textit{initialize} as a function that produces some initial basis; a number of samples that are coherence-optimal for that basis; surrogate coefficients for that basis; and an estimate of RRMSE. For our examples, we initialize to a total-order basis with some small number of samples drawn from the coherence-optimal distribution, unless all samples are being drawn from the orthogonality distribution. The surrogate coefficients and RRMSE estimate are then computed in that basis for those samples as by the method described in Section~\ref{subsec:coef_id}. The BASE-PC algorithm is then described in Algorithm~\ref{alg:base_pc}, and referred to as \textit{base-pc\_loop}. Here {\tt max\_iterations} may be set based on convergence criteria. For our examples, the loop {\color{black}is run} until computational time grows large, although it is also reasonable to stop based on the RRMSE estimates as generated by \textit{basis\_validate}.

\begin{algorithm}[htb]
\begin{algorithmic}
\STATE Set $(\mathcal{B}_0,\bm{c}_0, \mathcal{S}) =$ \textit{initialize}(). \% We let $\mathcal{S}$ denote identified samples.
\FOR {$k=1:${\tt max\_iterations}}
\STATE $(\mathcal{B}_k,\bm{c}_k) = $\textit{basis\_validate}($\mathcal{B}_{k-1}$,$\bm{c}_{k-1}$).
\STATE $\mathcal{S} = $\textit{sample\_expand}($\mathcal{B}_k$, $\mathcal{B}_{k-1}$).
\ENDFOR
\end{algorithmic}
\caption{\textit{base-pc\_loop}: The main iteration for BASE-PC.}
\label{alg:base_pc}
\end{algorithm}
\section{Numerical Examples}
\label{sec:examples}
To investigate the numerical efficacy of the BASE-PC iteration, we investigate four problems. The first in Section~\ref{subsec:franke} is a low-dimensional smooth problem that is traditionally targeted for interpolation and regression problems. The second in Section~\ref{subsec:cavity} is a moderate dimensional problem with some characteristic coefficient decay often seen in engineering problems. The third in Section~\ref{subsec:sine_exponential_1000d} is a $1000$ dimensional manufactured problem that shows the BASE-PC method can be effective at dimensions not usually associated with PC accuracy. The final example in Section~\ref{subsec:surface_adsorption} is a low dimensional surface adsorption model that is not well suited to polynomial approximation, having many properties that may preclude it from use with PC, but demonstrating BASE-PC's improvement when polynomial approximation is of suspect accuracy, as occurs in many practical problems.

In all examples here the total order bases use only samples drawn from the orthogonality distribution as opposed to any coherence-optimal sampling. For the BASE-PC methods, sample adaptivity (SA) refers to use of the correction sampling distribution with coherence-optimal sampling as in~\cite{Hampton15}, while no sample adaptation (No SA) refers to using samples from the orthogonality distribution. In both cases, basis adaptation is performed in the same manner. In all cases validated RRMSE represents the estimated RRMSE as identified by BASE-PC, while RRMSE is a reference estimate of RRMSE computed using a large number of independently generated samples.
\subsection{Case I: Franke function}
\label{subsec:franke}
\begin{figure}[ht]
\centering
\includegraphics[scale = 0.65]{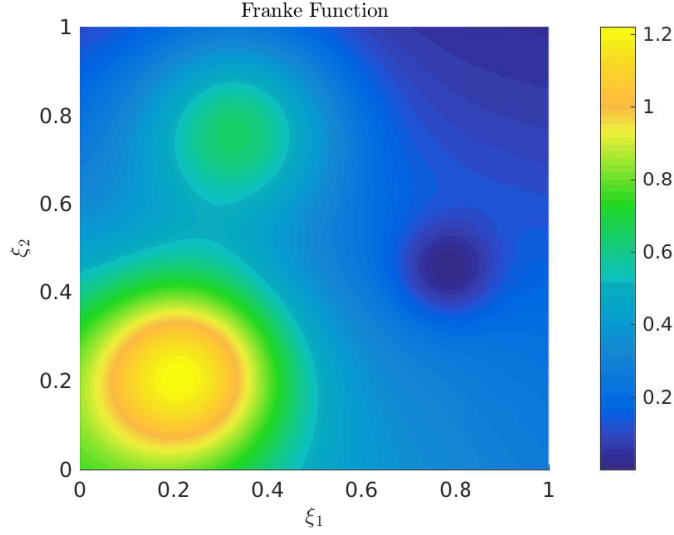}
\caption{The Franke function.}
\label{fig:franke_function}
\end{figure}

\begin{figure}[ht]
\centering
\subfloat[RRMSE vs QoI evaluations]{\includegraphics[scale = 0.5]{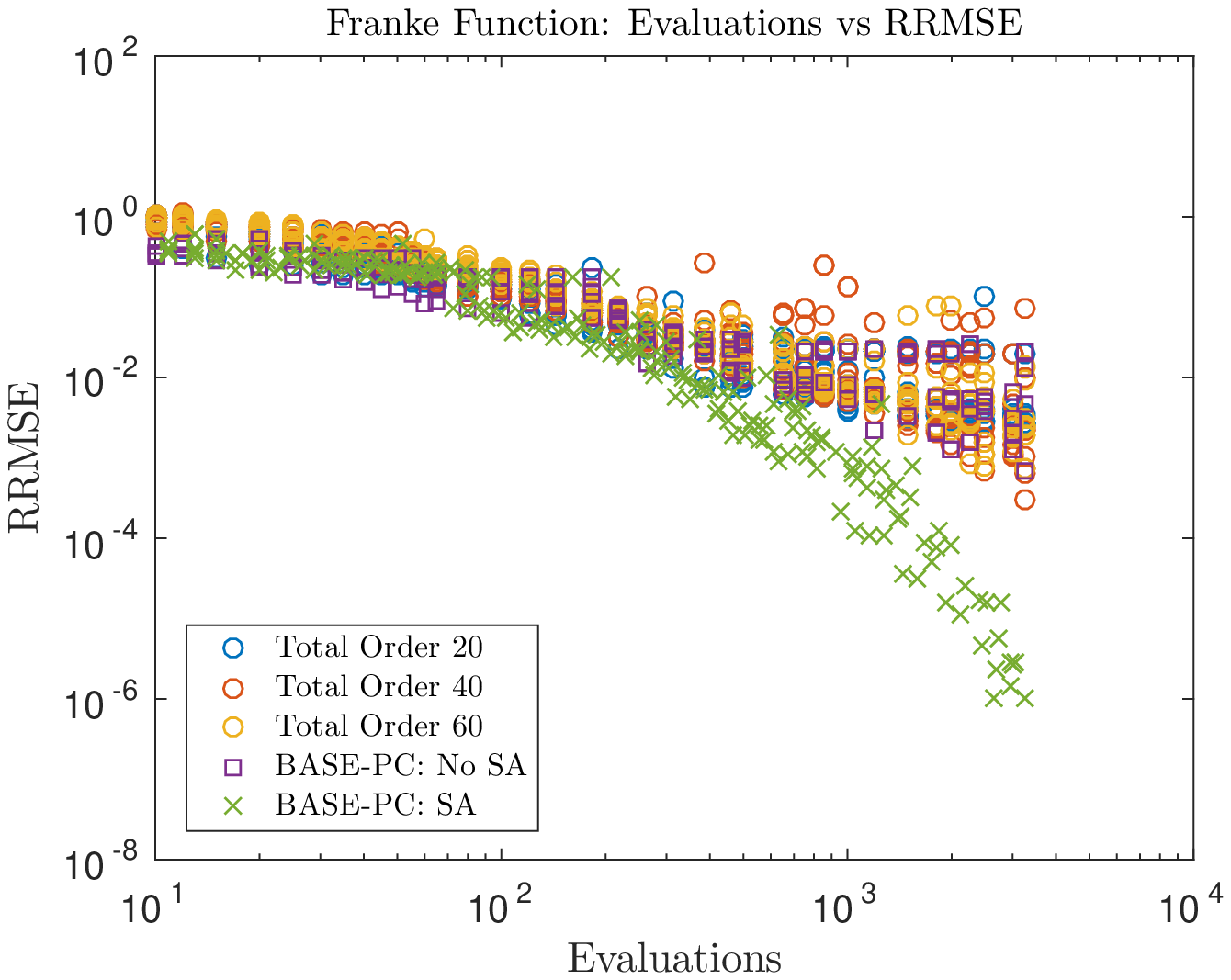}\label{fig:franke_nevals}}
\subfloat[RRMSE vs number of basis elements]{\includegraphics[scale = 0.5]{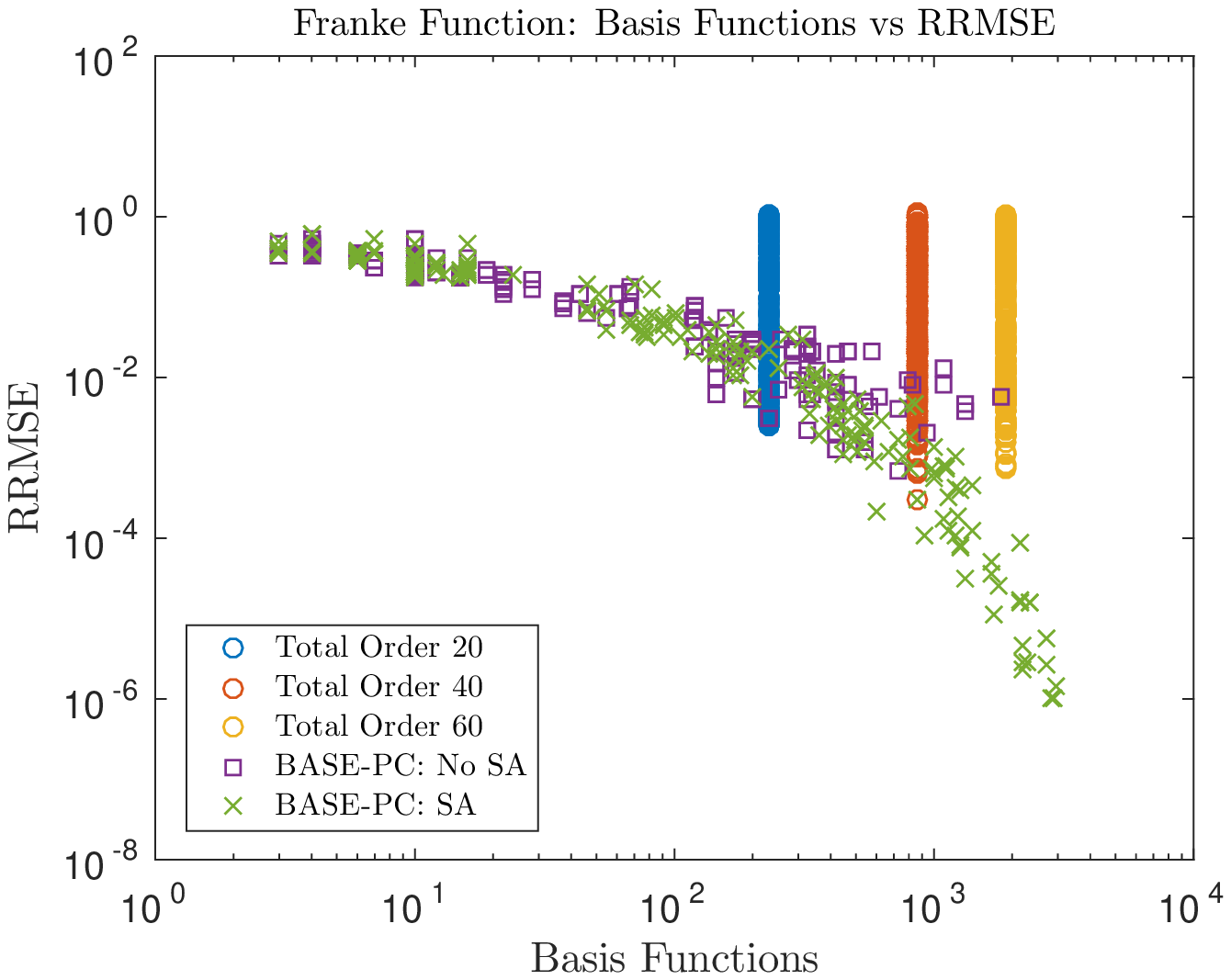}\label{fig:franke_nelems}}

\subfloat[QoI evaluations vs number of basis elements]{\includegraphics[scale = 0.5]{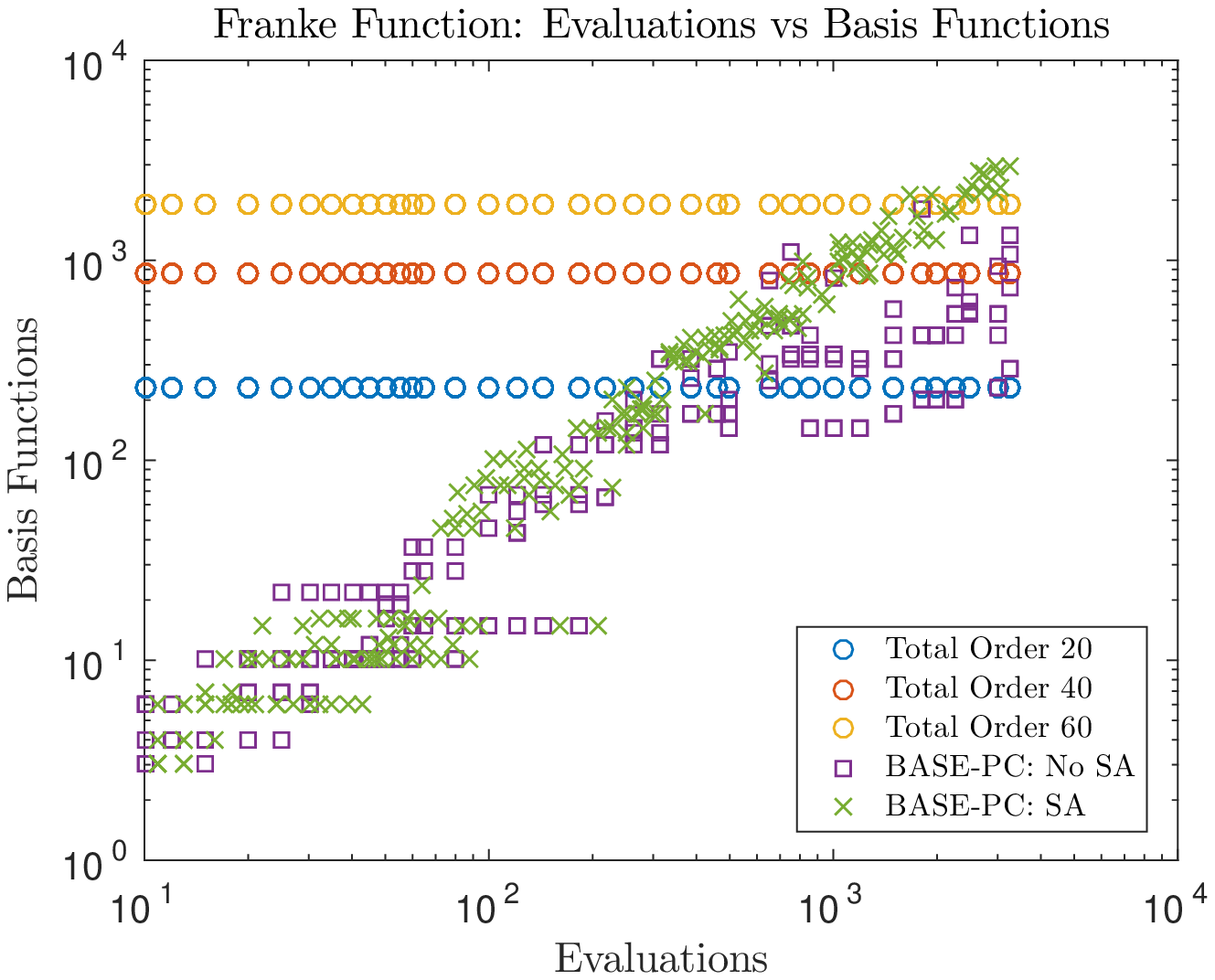}\label{fig:franke_nevals_nelems}}
\subfloat[RRMSE vs Estimated RRMSE]{\includegraphics[scale = 0.5]{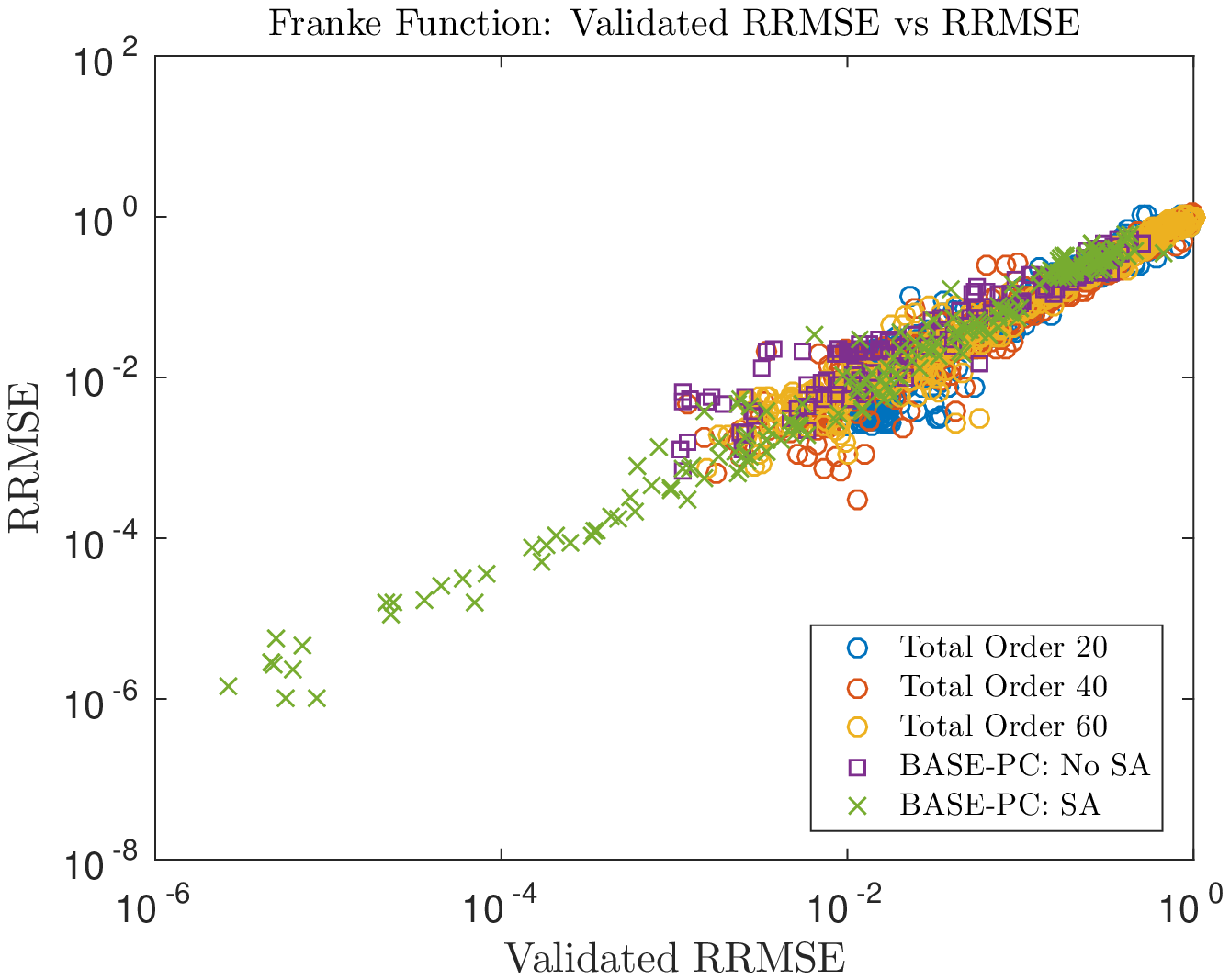}\label{fig:franke_corrleation}}
\caption{Comparisons of different methods for the Franke function.}
\label{fig:franke_basis}
\end{figure}

One function that is often used in regression or interpolation analysis is the Franke function~\cite{franke79}, which is a two dimensional function defined on $[0,1]\times[0,1]$ by
\begin{align}
 \label{eqn:franke_def}
 u(\bm{\Xi}) &:= \frac{3}{4}\exp\left(-\frac{(9\Xi_1-2)^2}{4}-\frac{(9\Xi_2-2)^2}{4}\right) + \frac{3}{4}\exp\left(-\frac{(9\Xi_1+1)^2}{49}-\frac{9\Xi_2+1}{10}\right)\\
&+ \frac{1}{2}\exp\left(-\frac{(9\Xi_1-7)^2}{4}-\frac{(9\Xi_2-3)^2}{4}\right) - \frac{1}{5}\exp\left(-(9\Xi_1-4)^2-(9\Xi_2-7)^2\right),\nonumber
\end{align}
and depicted in Figure~\ref{fig:franke_function}. The results of running the BASE-PC iterations are shown in Figure~\ref{fig:franke_basis}, demonstrating improvement for adapted bases over the use of total order bases in that each of the total order bases are only as accurate as the adaptive bases for a range of sample sizes. Specifically, when comparing the number of QoI evaluations to the RRMSE, we see that the sample adaptive BASE-PC iterations reliably outperform other methods, and that the BASE-PC iterations with and without sample adaptation use significantly fewer basis functions for the same level of accuracy when compared to the total order bases, and require no \textit{a priori} information about what order of basis to utilize. 

These plots show a gradual increase of the average number of basis functions included, as the number of QoI evaluations increases with the BASE-PC approach, a common theme in all the examples. This example also shows the benefit of sample adaptation, as higher sample sizes allow more exceptional accuracy when sample adaptation is performed. We note that in this case the order of adapted approximation remains comparable in both dimensions, so that the basis adaptivity behaves similarly to identifying a particular total order approximation.
\subsection{Case II: Stochastic heat driven cavity flow}
\label{subsec:cavity}
A practical case for consideration comes from temperature driven fluid flow in a cavity~\cite{LeMaitre02b, LeMaitre10, LeQuere91, Peng14}, where the QoI is a component of the velocity field at a fixed point and time. The left vertical wall has a uniform temperature $\tilde{T}_h$, referred to as the hot surface, while the right vertical wall has a variable temperature $\tilde{T}_c$, and is referred to as the cold surface; both walls are adiabatic. The reference temperature is defined as $\Delta\tilde{T}_{ref}:=\tilde{T}_h-\tilde{T}_c$. Let $\hat{\bm{y}}$ denote the unit normal vector in the vertical dimension. The non-dimensional governing equations are given by
\begin{equation}
\begin{aligned}
&\frac{\partial \bm{u}}{\partial t} + \bm{u}\cdot\nabla\bm{u}=-\nabla p + \frac{\text{Pr}}{\sqrt{\text{Ra}}}\nabla^2\bm{u}+\text{Pr}T\bm{\hat{y}},\\
& \nabla\cdot\bm{u}=0,\\ \label{eqn:cavity}
&\frac{\partial T}{\partial t}+ \nabla\cdot(\bm{u}T)=\frac{1}{\sqrt{\text{Ra}}}\nabla^2T,
\end{aligned}
\end{equation}
where $\bm{u}$ is velocity vector, $p$ is pressure, $T$ is normalized temperature and $t$ is time. Non-dimensional Prandtl and Rayleigh numbers are defined, respectively, as $\text{Pr}:=\tilde{\mu}c_p/\tilde{\kappa}$ and $\text{Ra}:=\tilde{\rho}\tilde{g}\beta\Delta\tilde{T}_{ref}\tilde{L}^3/(\tilde{\mu}\tilde{\kappa})$ where tilde denotes dimensional quantities: $\tilde{\rho}$ is density, $\tilde{L}$ is reference length, $\tilde{g}$ is gravitational acceleration, $\tilde{\mu}$, is molecular viscosity and $\tilde{\kappa}$ are is thermal conductivity. The coefficient of thermal expansion is $\beta=0.5$. In this example the Prandtl and Rayleigh numbers are given by $\text{Pr}=0.71$ and $\text{Ra}=10^6$.
\subsubsection{Stochastic Boundary Conditions}
\label{subsubsec:Sto_BC}
\begin{figure}[H]
\subfloat[Schematic figure for the problem]{\label{subfig:cavity}
\includegraphics[width=0.55\textwidth]{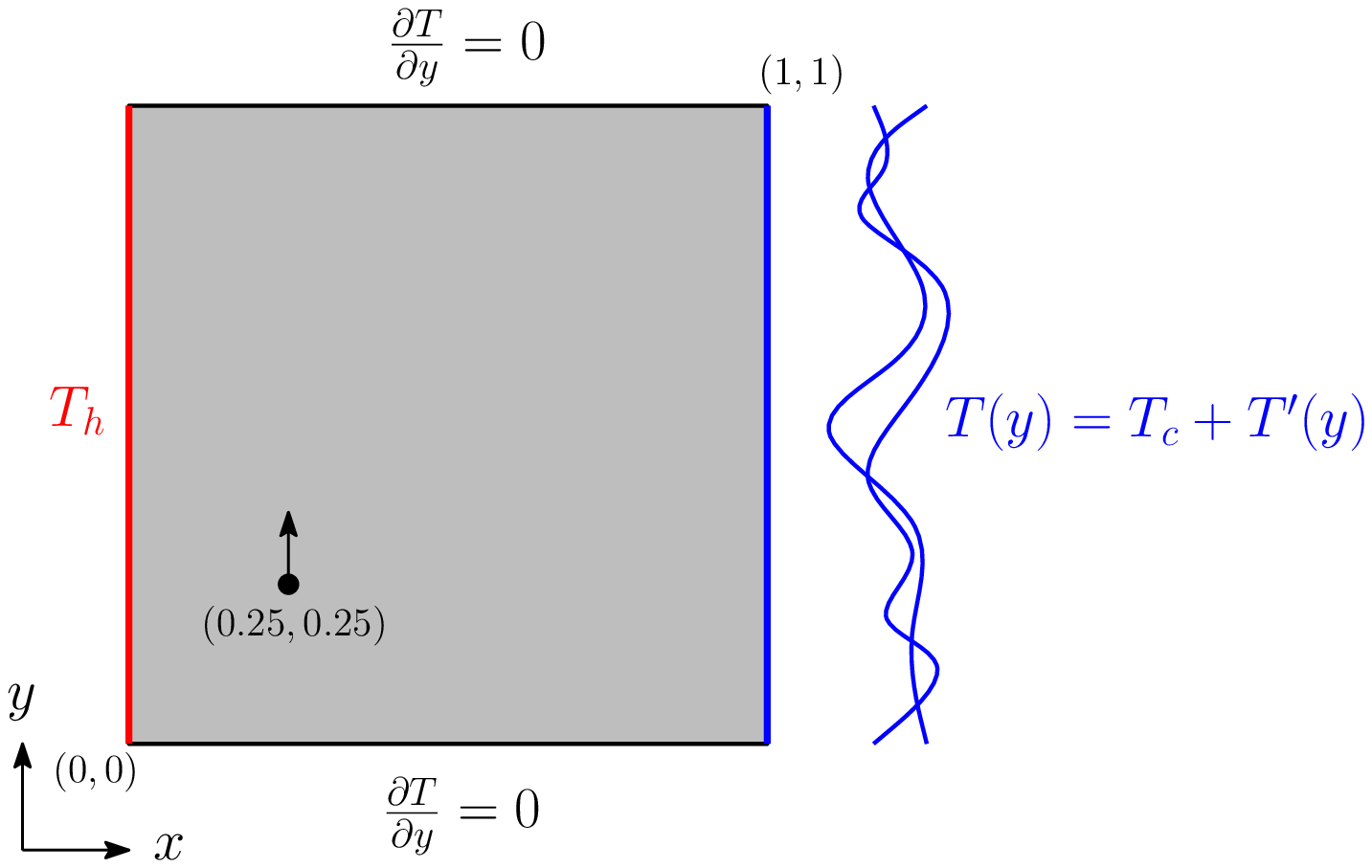}
}
\subfloat[An example of $T(x=1,y)$]{
\includegraphics[width=0.4\textwidth]{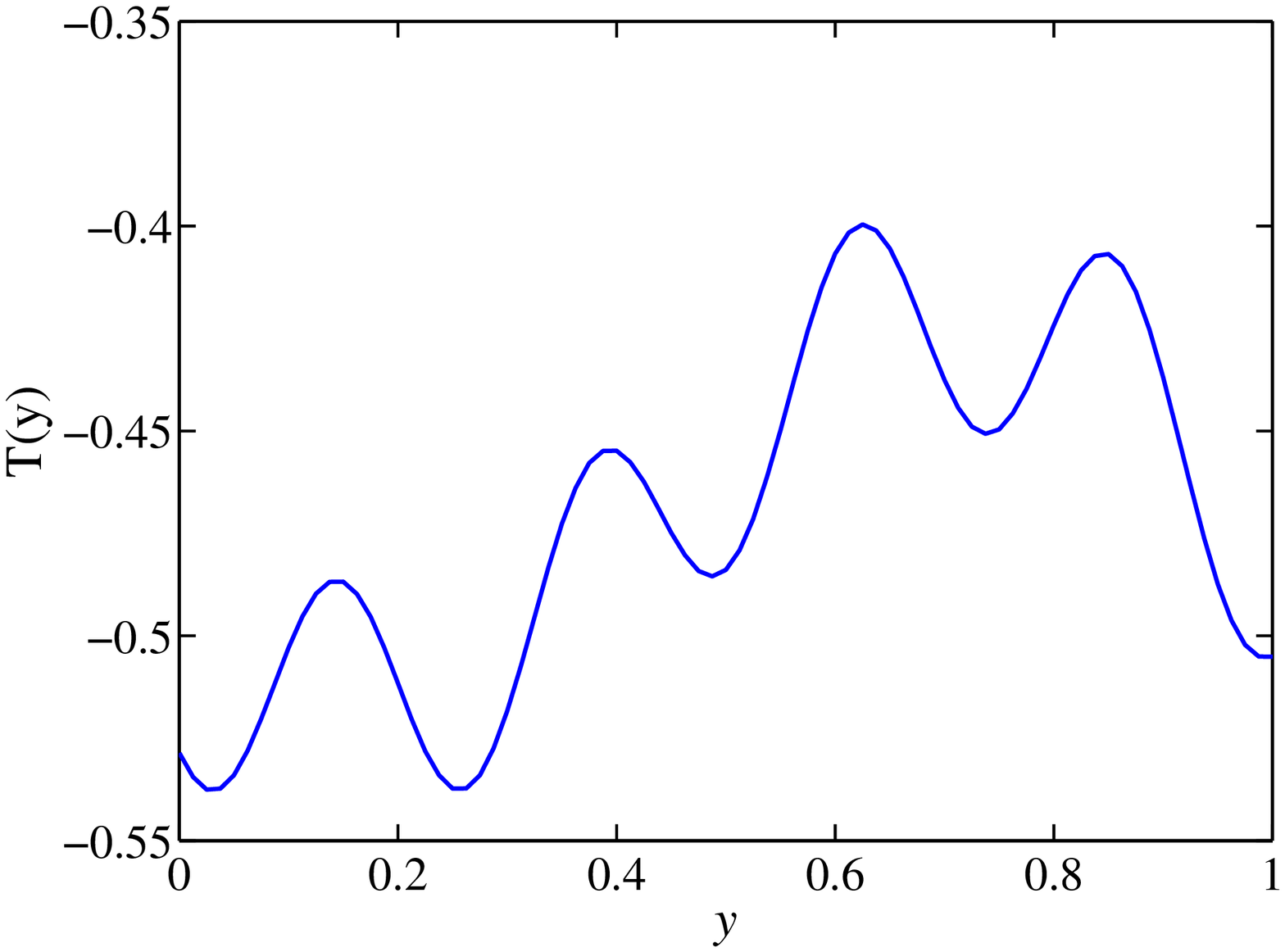}
\label{subfig:Tc}
}
\caption{Illustration of the problem, reproduced from Figure 6 of~\cite{Peng14}.}
\end{figure}

On the cold wall, a temperature distribution with stochastic fluctuations is applied,
\begin{equation}
T(x=1,y)=T_c+T'(y),
\label{eqn:coldwall}
\end{equation}
where $T_c = -0.5$ is a constant expected temperature, and $T_h = 0.5$ is the temperature on the hot wall. The fluctuation $T'(y)$ is given by the truncated Karhunen-Lo\`eve expansion
\begin{equation}
\label{eq:T'}
T'(y) = \sigma_T\sum_{i=1}^{d}\sqrt{\lambda_i}\varphi_i(y)\Xi_i,
\end{equation}
where $d=20$ and $\sigma_T=11/100$. Here, each $\Xi_i$ is assumed to be an independent and identically distributed uniform random variable on $[-1,1]$, with $\{\lambda_i\}_{i=1}^d$ and $\{\phi_i(y)\}_{i=1}^d$ the $d$ largest eigenvalues and the corresponding eigenfunctions of the exponential covariance kernel
\begin{equation}
C_{TT}(y_1,y_2) = \exp{\left(-\frac{\vert y_1-y_2\vert}{l_c}\right)},
\end{equation}
where $l_c=1/21$ is the correlation length. An example of cold boundary condition is shown in figure~\ref{subfig:Tc}. Our QoI is the vertical velocity component at $(0.25,0.25)$.  The QoI computations here do not solve this model directly, but instead use a surrogate solution computed using a basis of 2500 elements reduced from a total order 4 basis and a large number of samples.

\subsubsection{BASE-PC Iterations}
\label{subsubsec:sto_baspec}
\begin{figure}[ht]
\centering
\subfloat[RRMSE vs QoI evaluations]{\includegraphics[scale = 0.5]{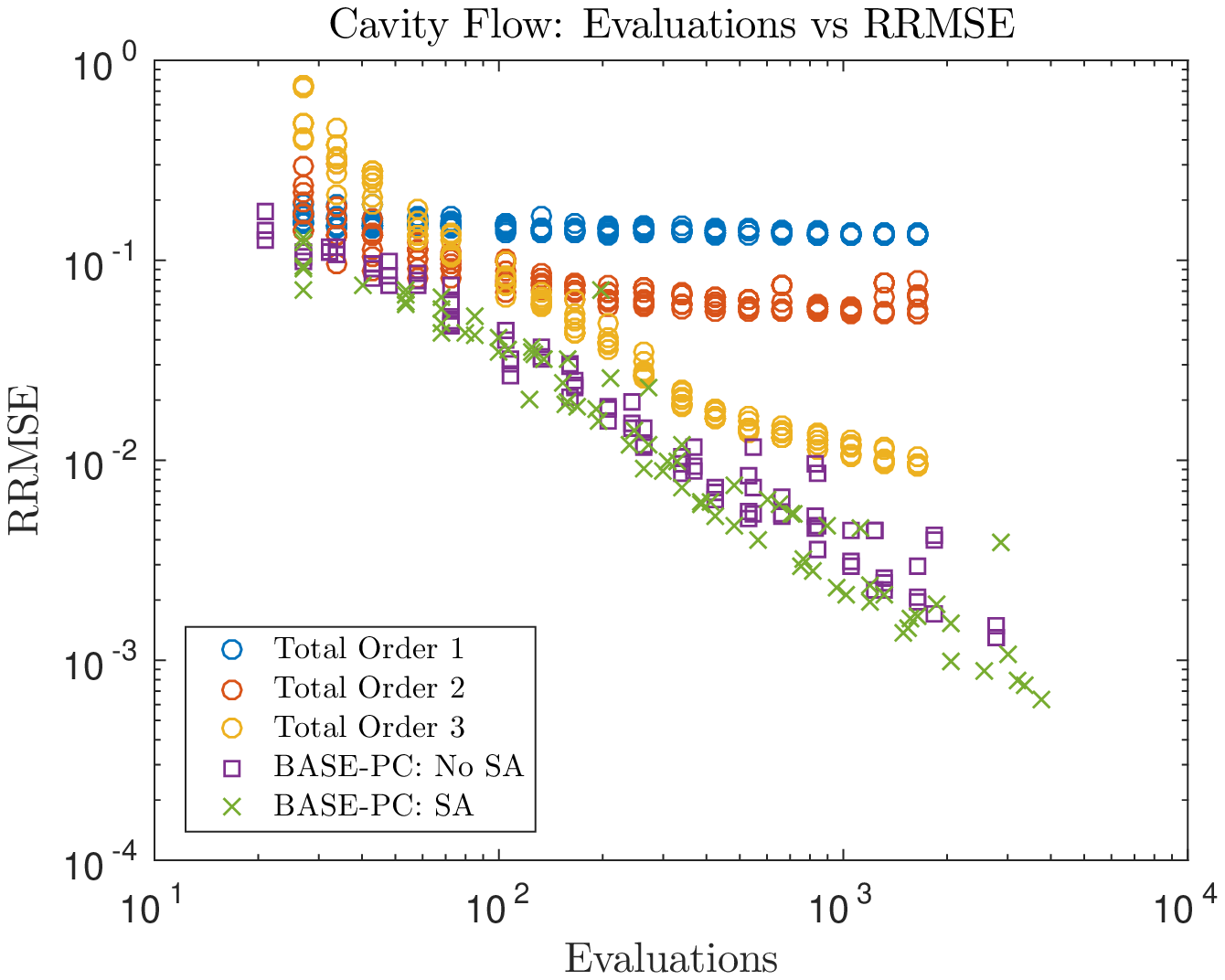}\label{fig:cavity_nevals}}
\subfloat[RRMSE vs number of basis elements]{\includegraphics[scale = 0.5]{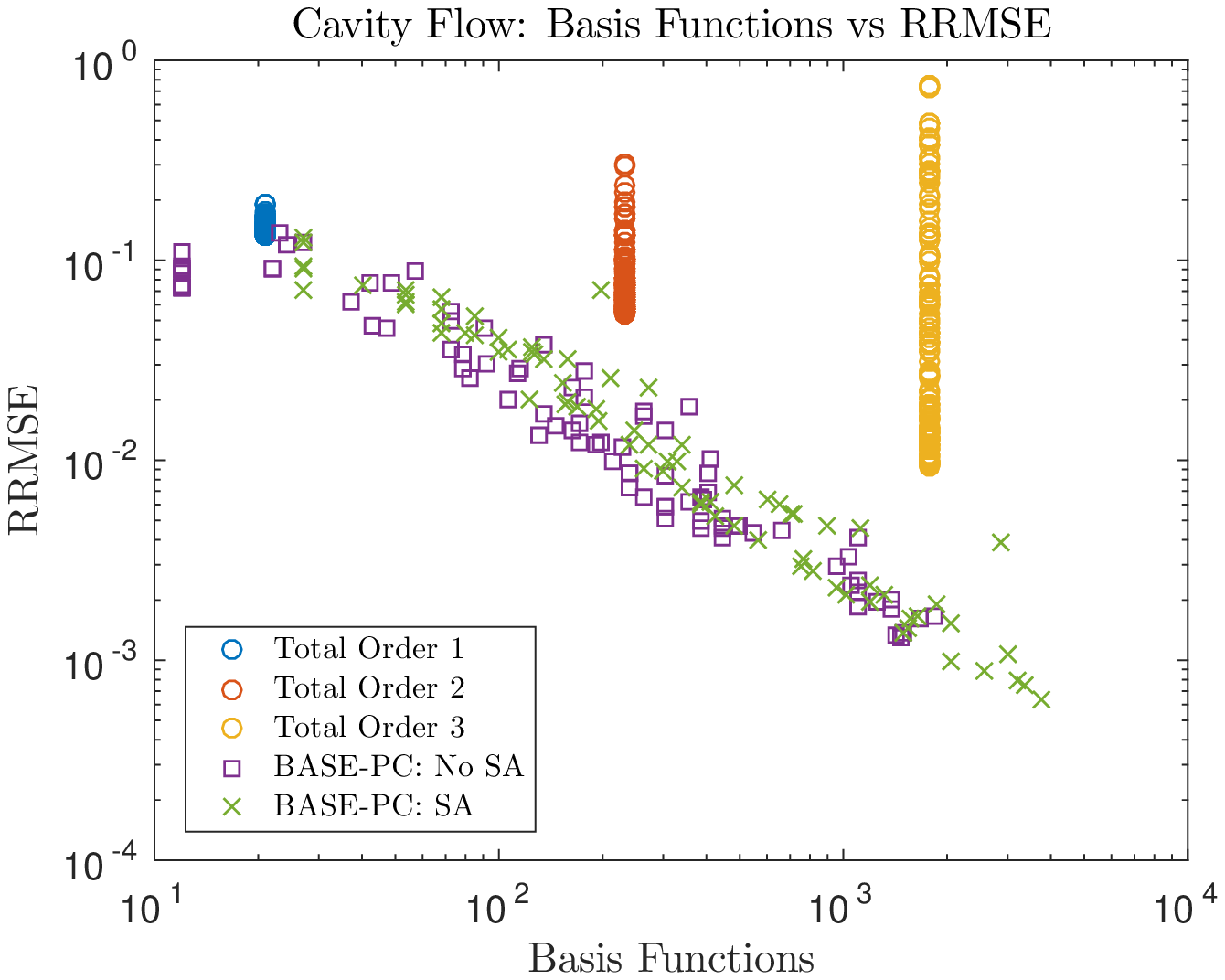}\label{fig:cavity_nelems}}

\subfloat[QoI evaluations vs number of basis elements]{\includegraphics[scale = 0.5]{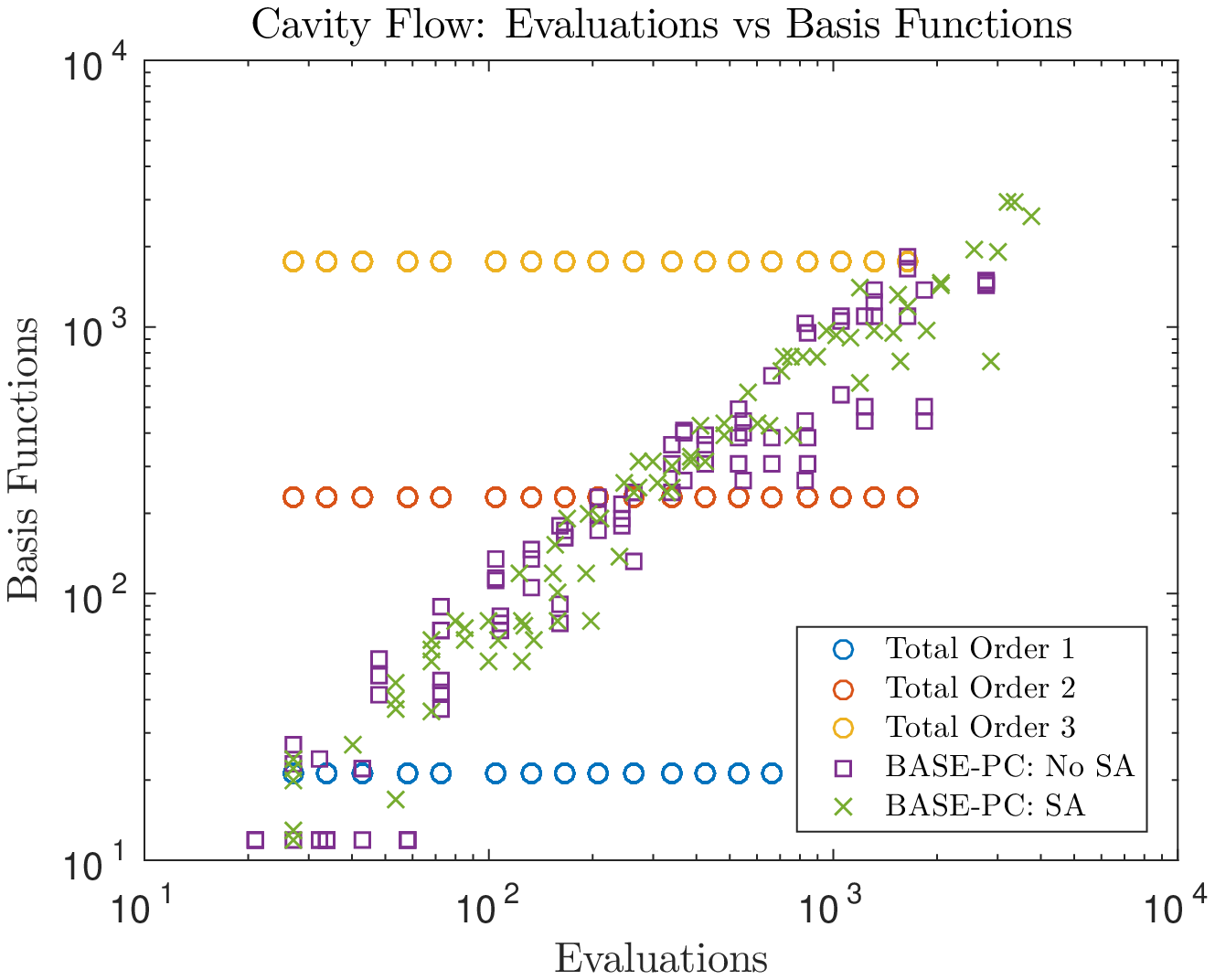}\label{fig:cavity_nevals_nelems}}
\subfloat[RRMSE vs Estimated RRMSE]{\includegraphics[scale = 0.5]{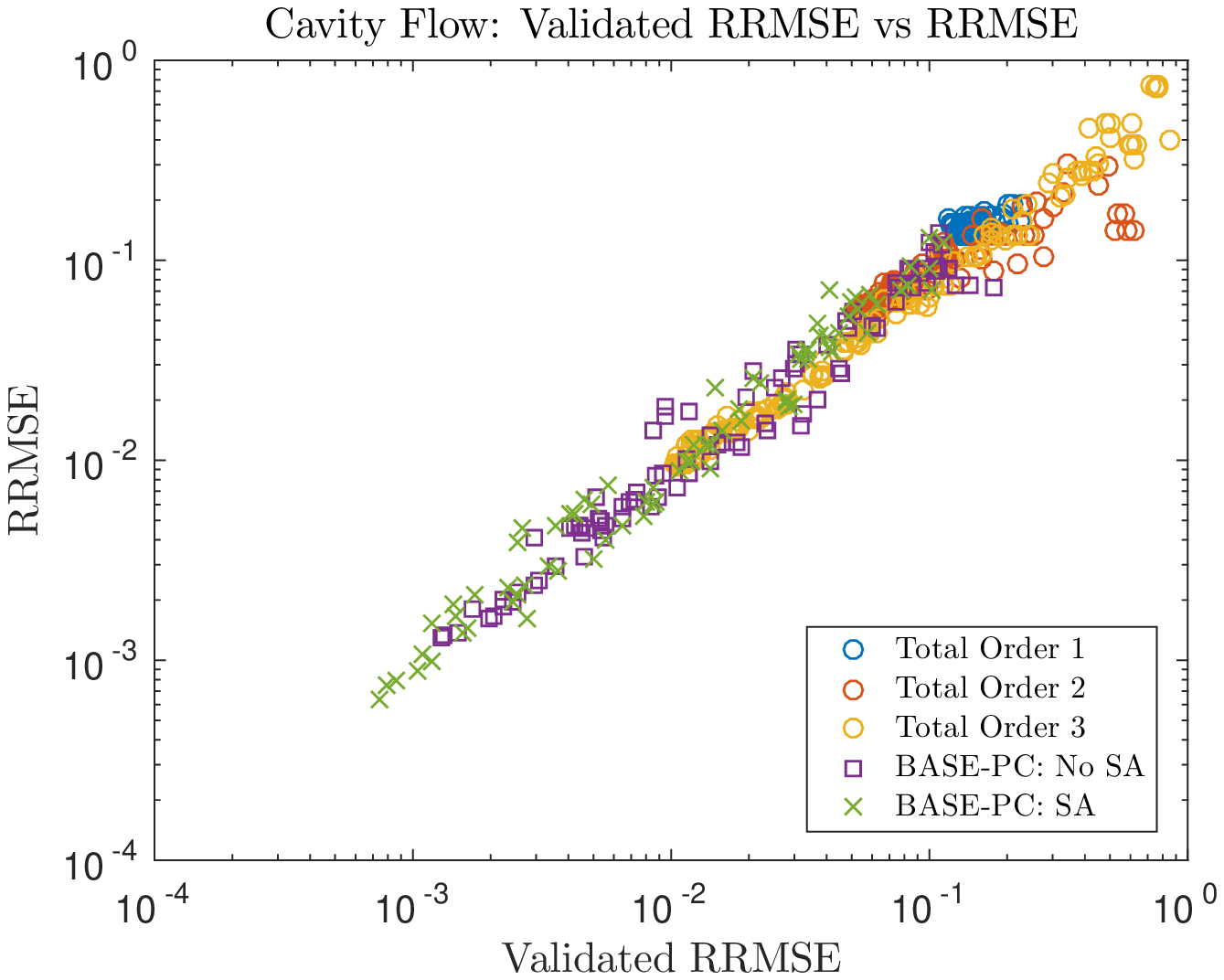}\label{fig:cavity_correlation}}
\caption{Comparisons of different methods for a cavity flow model with $d=20$.}
\label{fig:cavity_basis}
\end{figure}

The results of running the BASE-PC iterations are shown in Figure~\ref{fig:cavity_basis}, demonstrating dramatic improvement for adapted bases over the use of total order bases. This improvement is seen even when no sample adaptivity is done, i.e. when all samples are drawn from the orthogonality distribution. We note that the number of adapted basis elements is correlated strongly to the number of QoI evaluations, and that the correlation between the validated RRMSE and the actual RRMSE is also high for all methods. This problem is smooth in the input parameters, which facilitates an easy basis adaptation and leads to a smooth decay in RRMSE as the number of QoI evaluations increases for the basis adaptive methods. The non-adaptive total order bases are not tuned to the number of samples, nor distribute basis functions ideally between dimensions leading to a recovery with reduced effectiveness.
\subsection{Case III: 1000-Dimensional Manufactured Decay}
\label{subsec:sine_exponential_1000d}
As a demonstration of scaling for a high dimensional problem, consider
\begin{align}
\label{eqn:exp_sine_decay_def}
u(\bm{\Xi}) = \exp\left(2 - \mathop{\sum}\limits_{k=1}^d\frac{\sin(k)\Xi_k}{k}\right),
\end{align}
with $d=1000$. Here each $\Xi_k$ is independent and uniformly distributed on $[0,1]$. For computations at this dimensionality, order control is implemented for the basis expansion so that instead of increasing each $p_i$, a limited number of dimensions have increased $p_i$ at each iteration as dictated by Algorithm~\ref{alg:upper_bound}. Specifically, we set {\tt dim\_add} $= 20$.

\begin{figure}[ht]
\centering
\subfloat[RRMSE vs QoI evaluations]{\includegraphics[scale = 0.5]{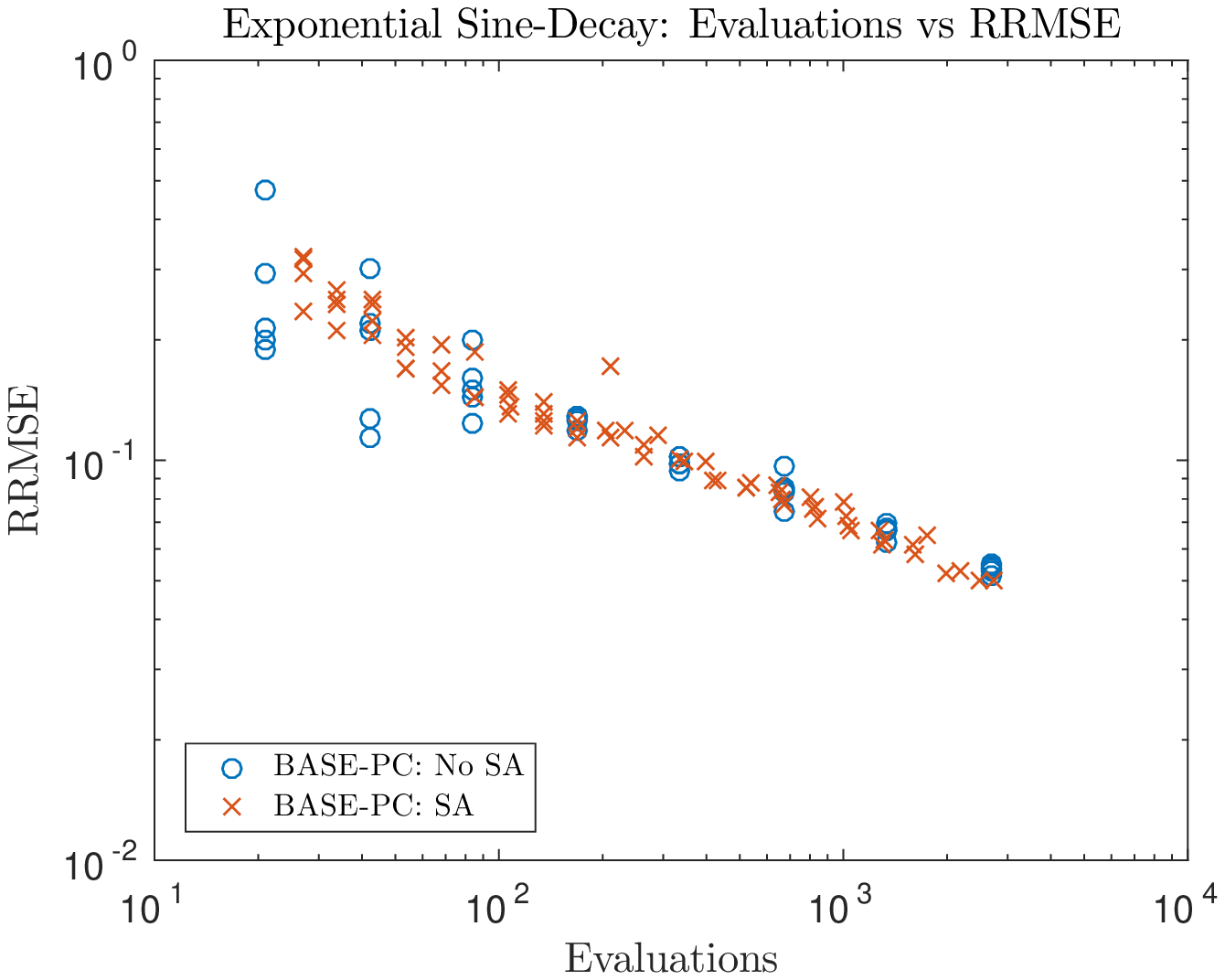}\label{fig:exp_nevals}}
\subfloat[RRMSE vs number of basis elements]{\includegraphics[scale = 0.5]{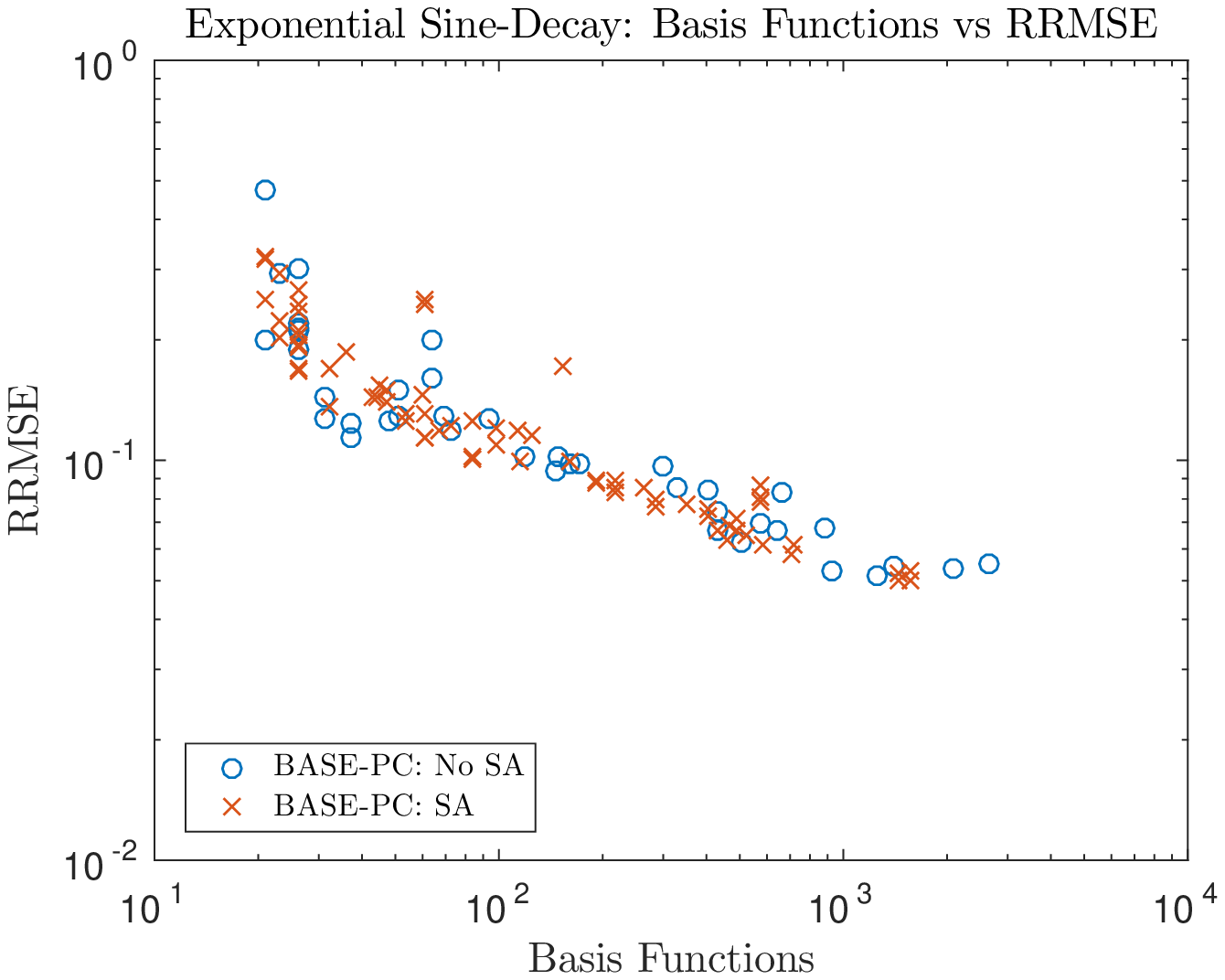}\label{fig:exp_nelems}}

\subfloat[QoI evaluations vs number of basis elements]{\includegraphics[scale = 0.5]{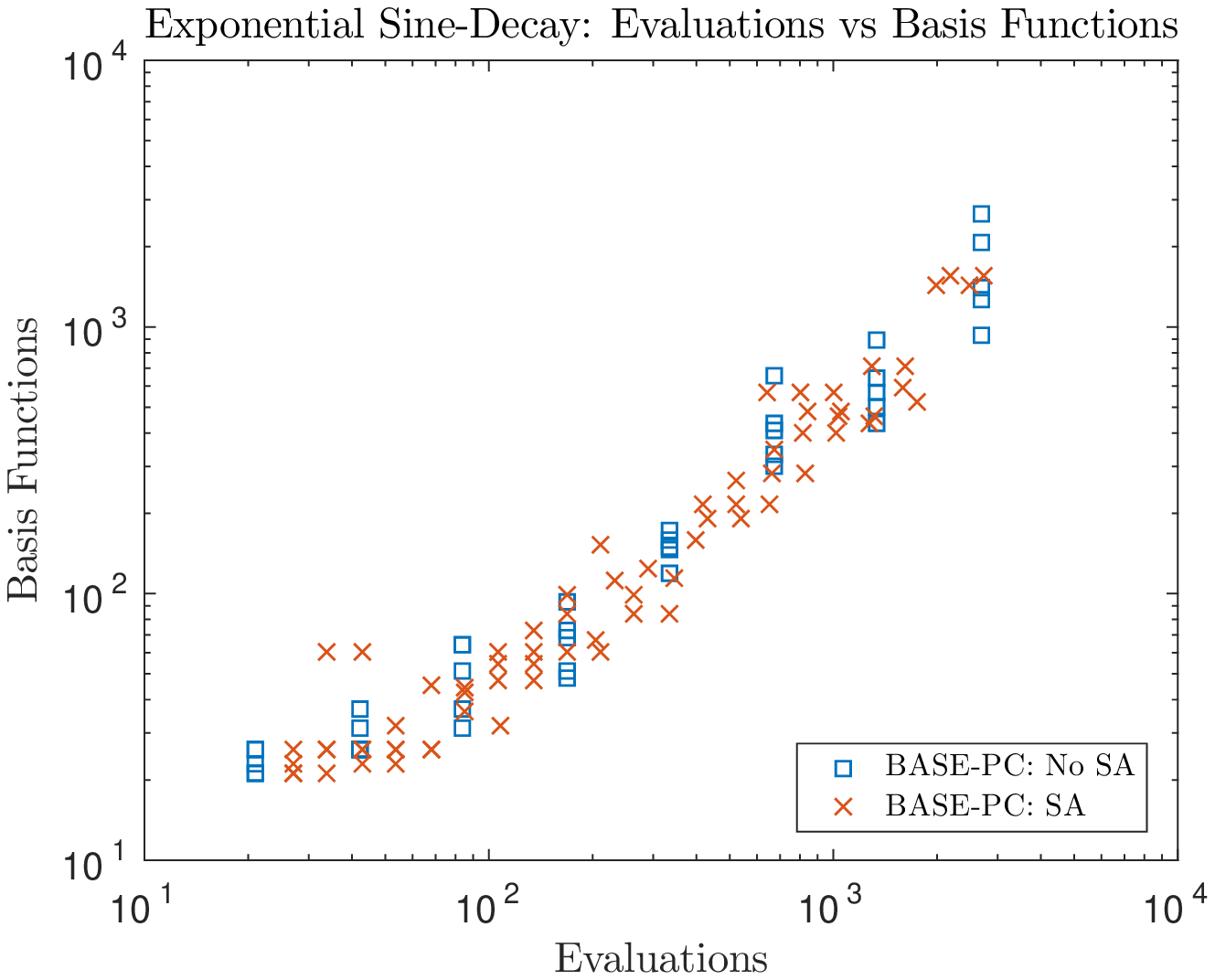}\label{fig:exp_nevals_nelems}}
\subfloat[RRMSE vs Estimated RRMSE]{\includegraphics[scale = 0.5]{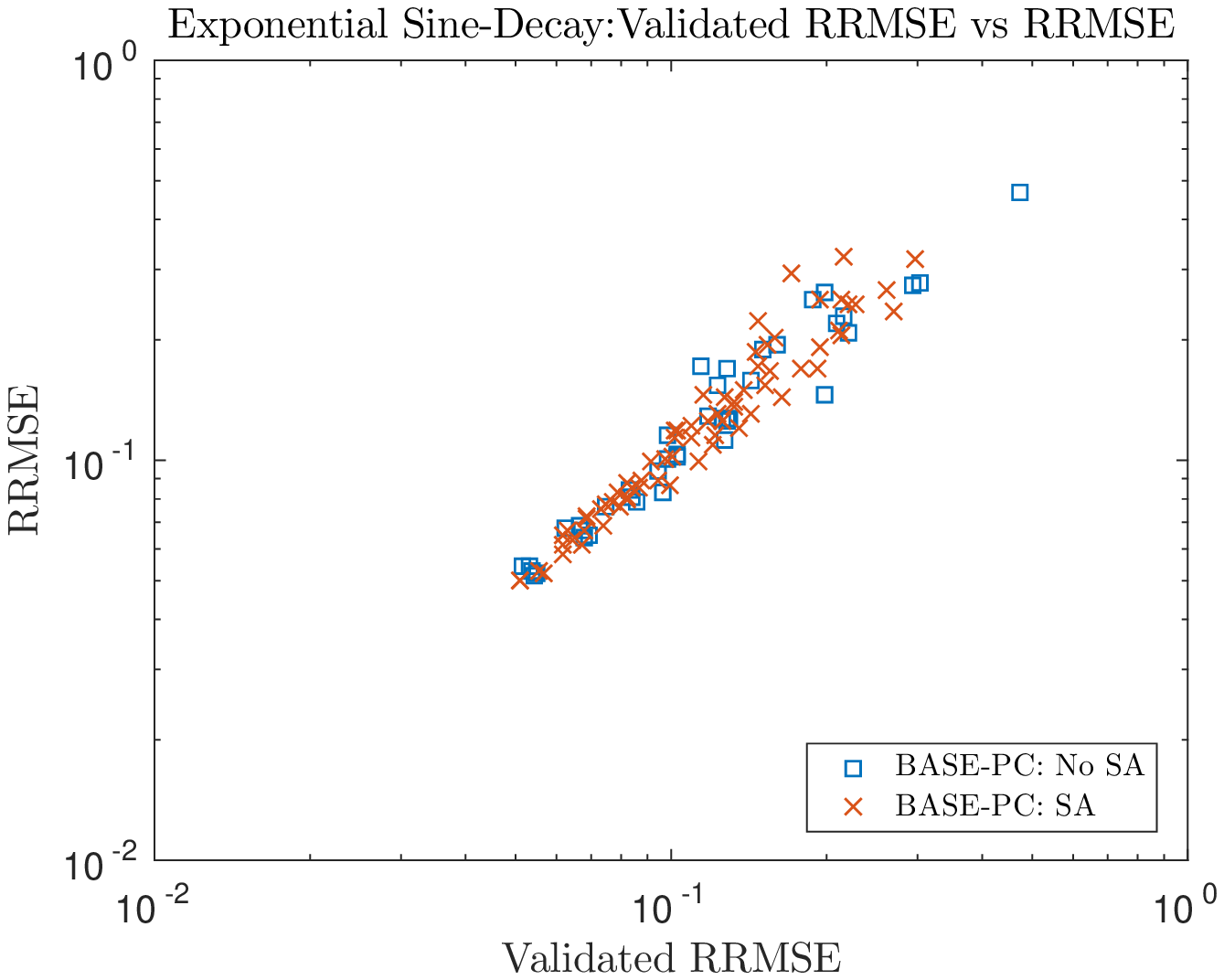}\label{fig:exp_correlation}}
\caption{Comparisons of different methods for (\ref{eqn:exp_sine_decay_def}) with $d=1000$.}
\label{fig:exp_basis}
\end{figure}

The results in Figure~\ref{fig:exp_basis} are a computation for a non-linear polynomial approximation in 1000 dimensions. We notice that the adaptive methods still exhibit a smooth reduction in RRMSE with regards to the number of QoI evaluations, although the rate of this reduction is not as large as that for the cavity flow problem in Section~\ref{subsubsec:sto_baspec}. This is coupled with a high correlation between the number of basis functions and QoI evaluations, as well as the estimated RRMSE and an accurate reference RRMSE.

\begin{figure}[ht]
\centering
\subfloat[Mean vs QoI evaluations]{\includegraphics[scale = 0.5]{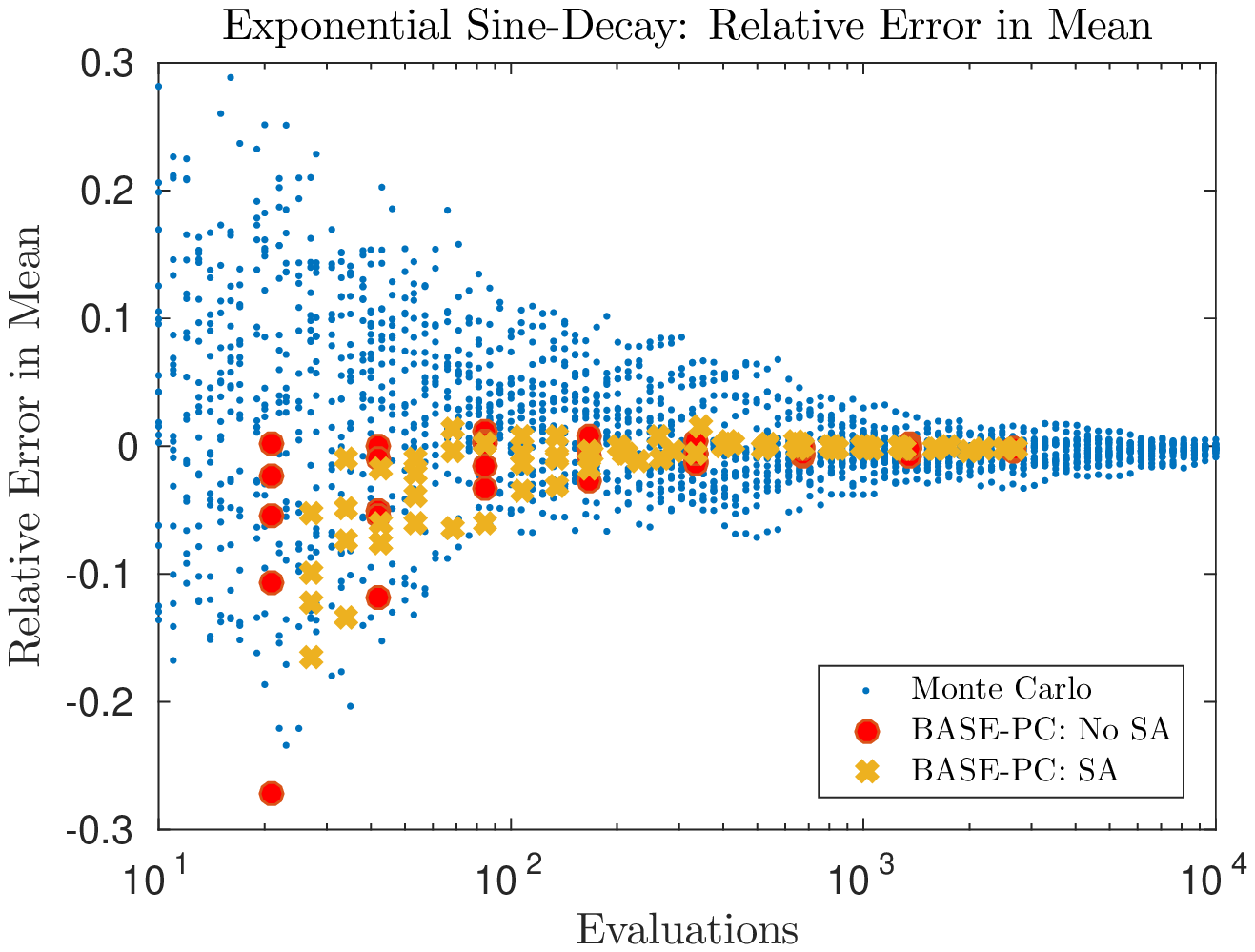}\label{fig:exp_sine_1000d_mean}}
\subfloat[Variance vs QoI evaluations]{\includegraphics[scale = 0.5]{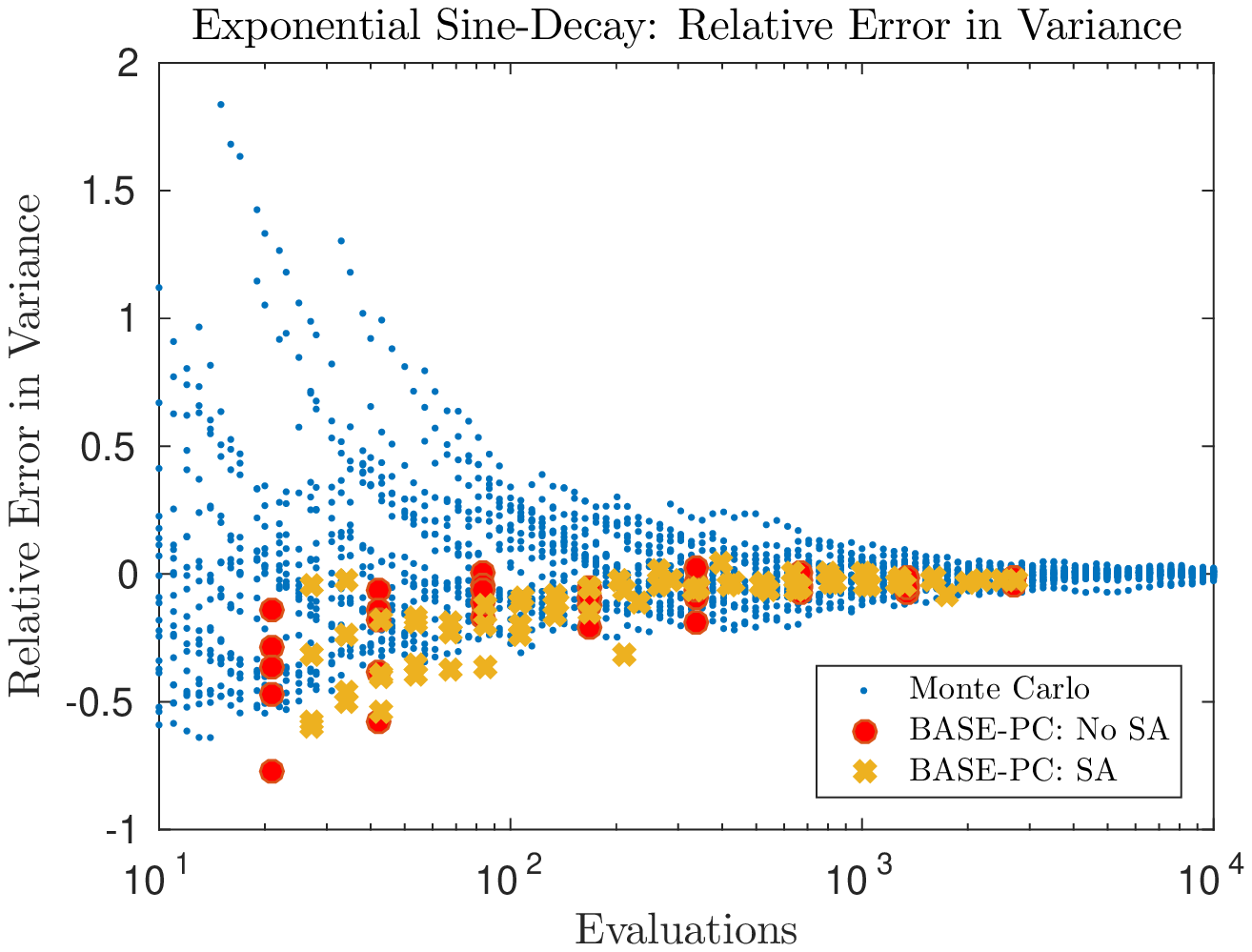}\label{fig:exp_sine_1000d_var}}
\caption{Estimates for mean and variance for (\ref{eqn:exp_sine_decay_def}) with $d=1000$.}
\label{fig:exp_sine_decay_mean_var_1000d}
\end{figure}

We may also consider how this method compares to Monte Carlo estimation of the first two moments of the distribution, given that this is a widely used approach for problems of this dimensionality. In Figure~\ref{fig:exp_sine_decay_mean_var_1000d} we see the comparison of errors in the mean and variance computations for this problem. We note that the BASE-PC iterations are generally more accurate in estimating the mean and variance than corresponding Monte Carlo computations, although the regression (\ref{eqn:ell1}) reduces $\|\bm{c}\|_1$ and does produce a bias to underestimate these quantities. This bias is negligible at larger sample sizes, but is significant at smaller sample sizes. Overall the BASE-PC moment estimates have lower error, and also have the benefit of producing a surrogate model that explains much of the variance in addition to estimating it.
\subsection{Case IV: Surface Adsorption}
\label{subsec:surface_adsorption}
While the previous examples are generally smooth and well approximated by low order polynomials, some QoI have stiff response to the input randomness and require high degree polynomials. One such problem is to quantify the uncertainty in the solution $\rho$ of the non-linear evolution equation  
\begin{align}
\label{eqn:reaction}
\left\{\begin{array}{l}\frac{d\rho}{dt} = \alpha(1-\rho) - \gamma\rho - \kappa(1-\rho)^2\rho, \\ 
\rho(t=0) = 0.9,\end{array}\right.
\end{align}
which models the surface coverage of certain chemical species, as examined in \cite{Makeev02,Lemaitre04b}. We consider uncertainty in the adsorption, $\alpha$, and desorption, $\gamma$, coefficients, and model them as shifted log-normal variables. Specifically, we assume
\begin{align*}
\alpha &= 0.1 + \exp(10\ \Xi_1),\\
\gamma &= 0.001 + 0.001\exp(10\ \Xi_2),
\end{align*}
where we consider $\Xi_1, \Xi_2$ as standard normal $\mathcal{N}(0,1)$ random variables; hence, the dimension of our random input is $d=2$. We note that this example differs from the corresponding example in~\cite{Hampton15}, which has $0.05\ \Xi_1$ and $0.05\ \Xi_2$ in place of $10\ \Xi_1$ and $10\ \Xi_2$ in the arguments of the exponentials. Also, the $0.001$ parameter multiplying $\gamma$ differs from $0.01$ in~\cite{Hampton15} which in this work somewhat reduces the relative variability with respect to $\gamma$ when compared to that of $\alpha$. In aggregate, the example here corresponds to significantly higher uncertainty in the input parameters. The reaction rate constant $\kappa$ in (\ref{eqn:reaction}) is assumed to be deterministic and is set to $\kappa = 10$. 

Our QoI is $\rho_c:=\rho(t=4,\Xi_1,\Xi_2)$, and to approximate this, we consider a Hermite PC expansion in the two variables, $\Xi_1$ and $\Xi_2$. This problem is interesting for its stiff transition and the need for high order polynomials in  $\Xi_1$, with a lower but still considerable order of polynomials in $\Xi_2$.

\begin{figure}[ht]
\centering
\subfloat[RRMSE vs QoI evaluations]{\includegraphics[scale = 0.5]{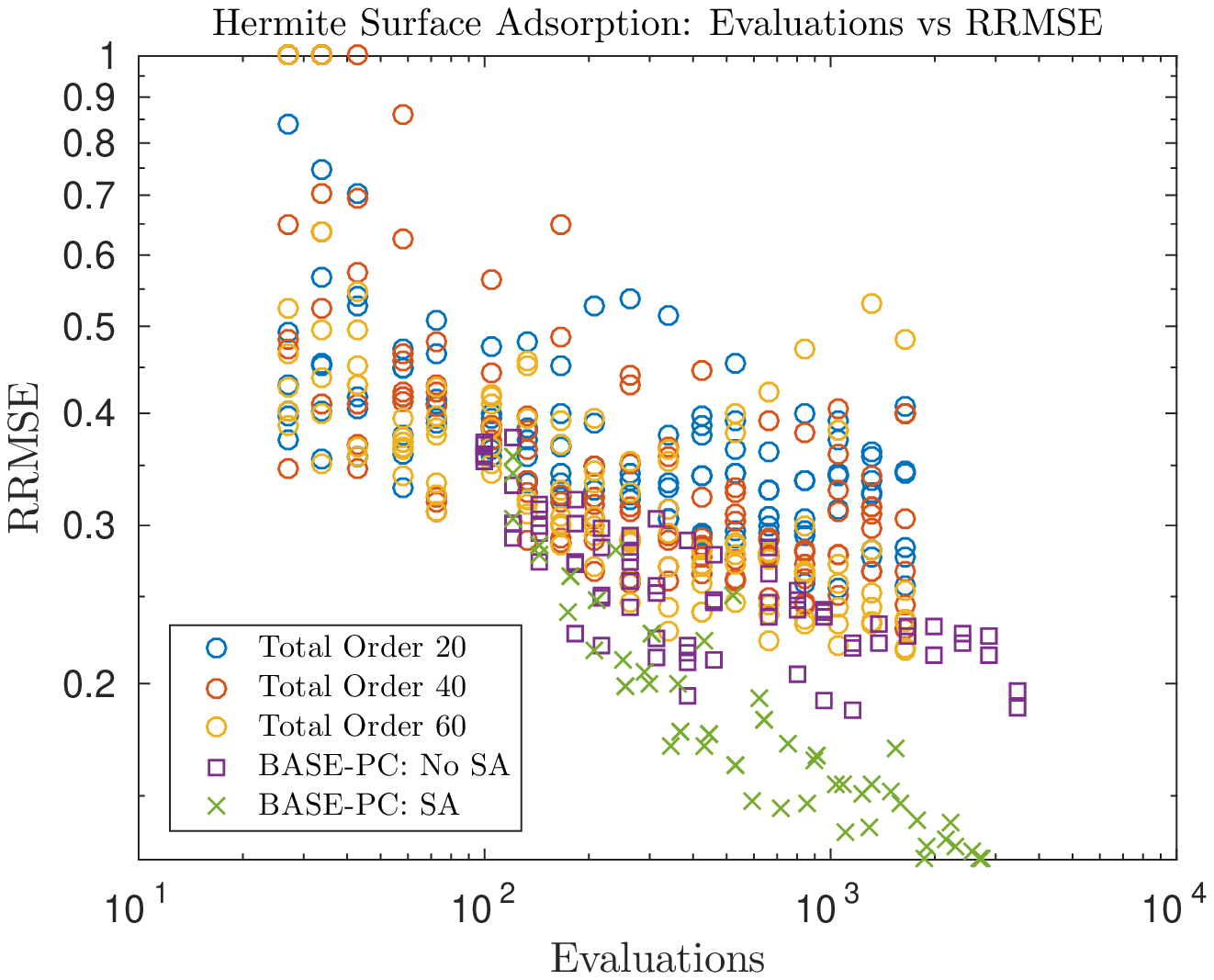}\label{fig:sa_herm_nevals}}
\subfloat[RRMSE vs number of basis elements]{\includegraphics[scale = 0.5]{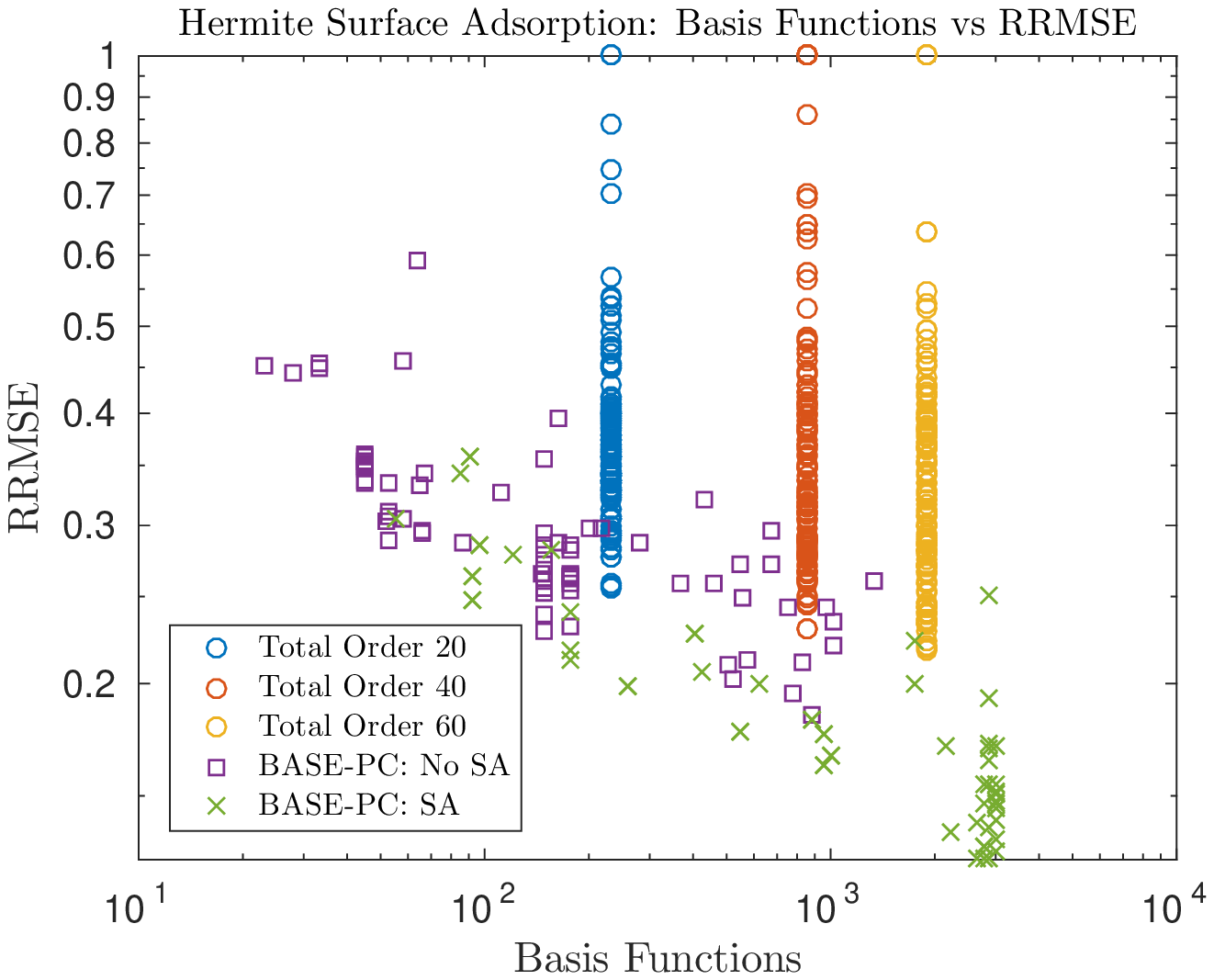}\label{fig:sa_herm_nelems}}

\subfloat[QoI evaluations vs number of basis elements]{\includegraphics[scale = 0.5]{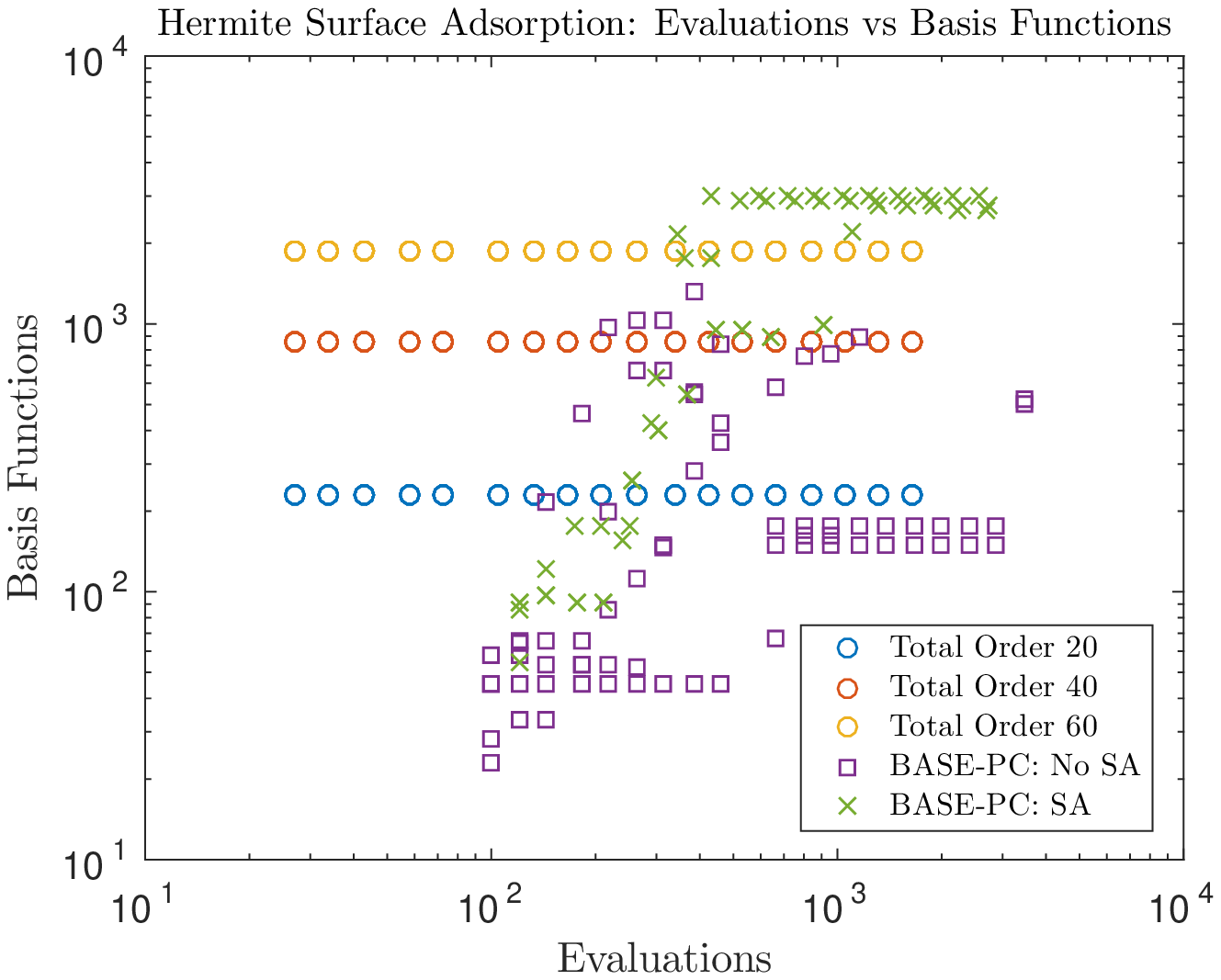}\label{fig:sa_herm_nevals_nelems}}
\subfloat[RRMSE vs Estimated RRMSE]{\includegraphics[scale = 0.5]{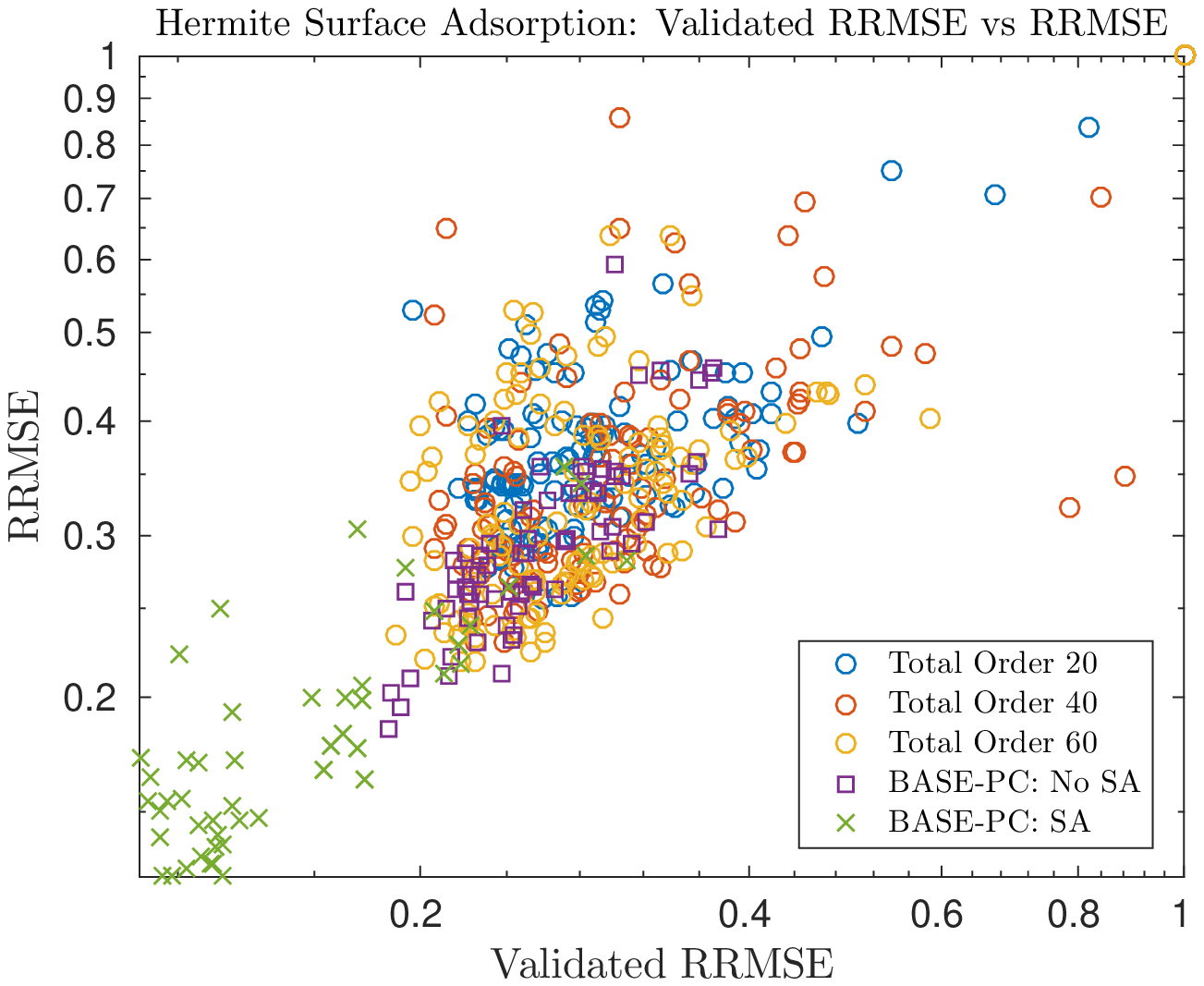}\label{fig:sa_herm_correlation}}
\caption{Comparisons of different methods for the surface adsoprtion model with Hermite polynomials.}
\label{fig:sa_herm_basis}
\end{figure}

For normally distributed input random variables, we utilize the associated Hermite polynomials for the approximation, which are known to be more difficult to use with polynomial approximation~\cite{Hampton14,Hampton15}; specifically, regression via (\ref{eqn:ell1}) using Hermite polynomials is quite sensitive for high order approximations. When high order Hermite polynomials are used, sampling from the orthogonality distribution is ineffective due to a dependence on exceptionally rare events for accurate approximations. The lack of smoothness in this problem is generally exacerbated by using Hermite polynomials, with results shown in Figure~\ref{fig:sa_herm_basis}. We note that while no method does particularly well for this problem, the BASE-PC method, specifically the sample adaptive version, significantly outperforms the total order computations.

\begin{figure}[ht]
\centering
\subfloat[QoI from surface adsorption model]{\includegraphics[scale = 0.5]{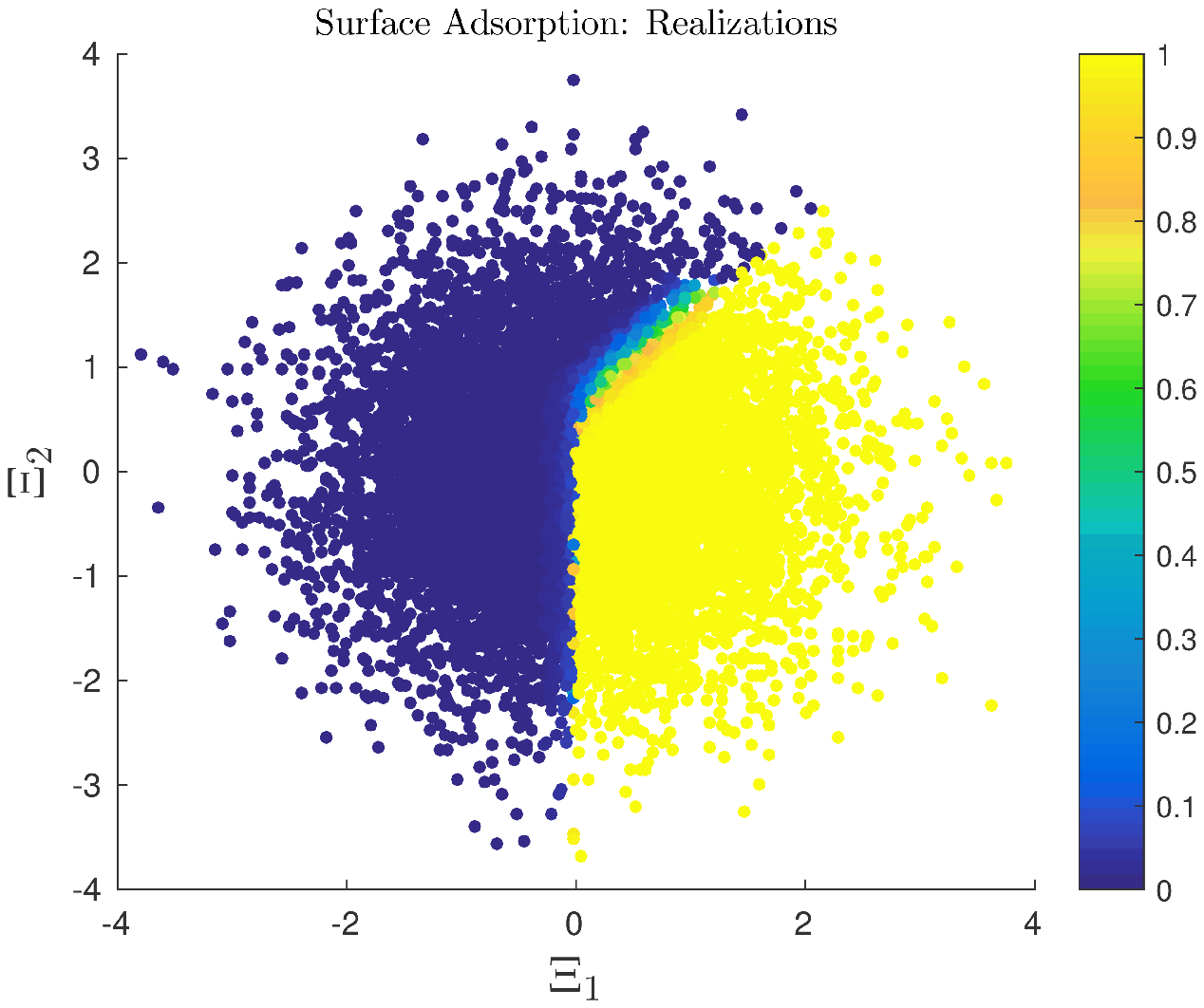}\label{fig:sa_realized_sa_true}}
\subfloat[Sample adaptive BASE-PC surrogate]{\includegraphics[scale = 0.5]{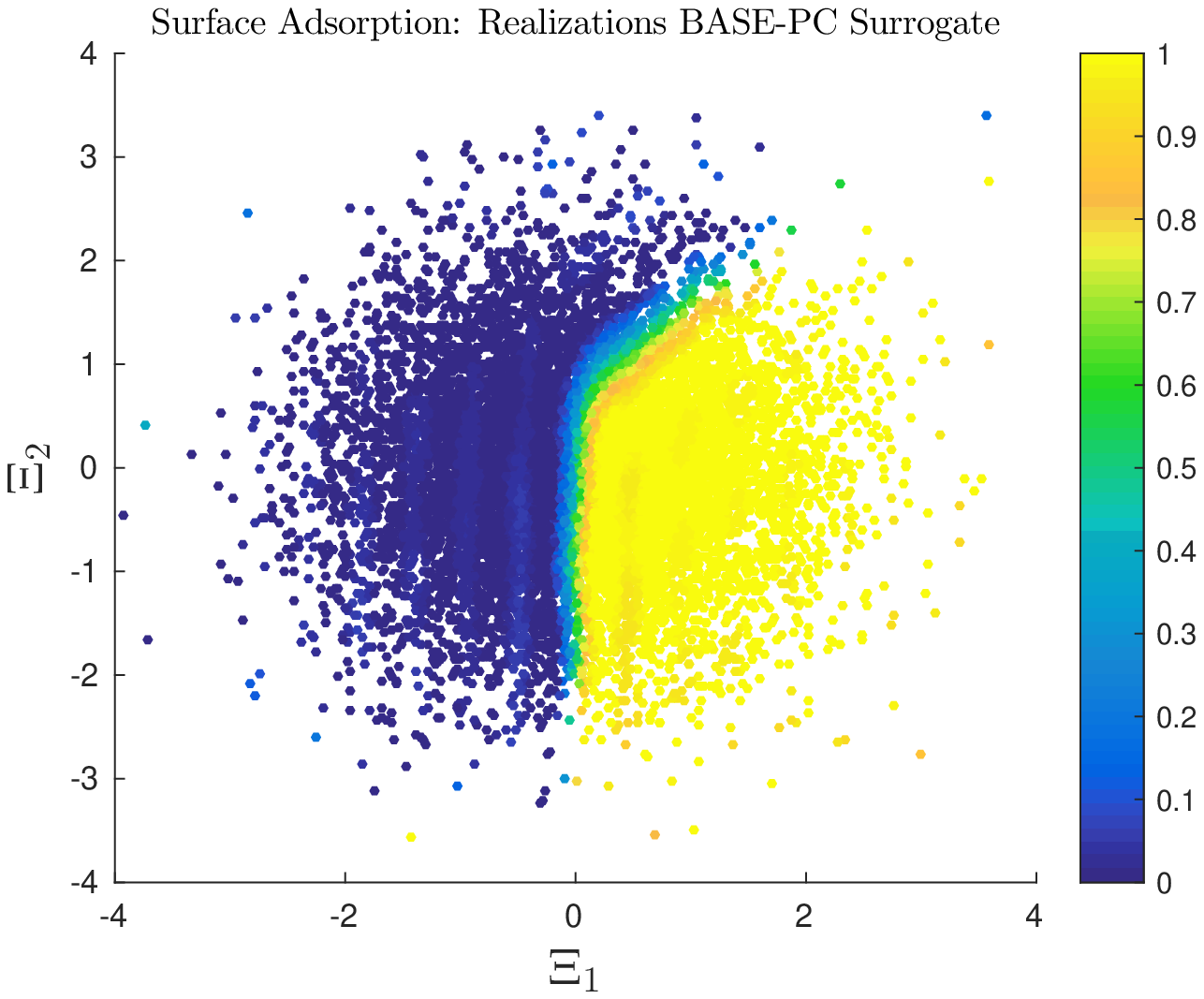}\label{fig:sa_realized_approximation}}
\caption{Comparison of QoI with sample adaptive BASE-PC surrogate}
\label{fig:sa_herm_realizations}
\end{figure}

Figure~\ref{fig:sa_herm_realizations} shows the realizations of the true model QoI and the realizations from a BASE-PC surrogate constructed using sample adaptation, where the realizations of the BASE-PC surrogate are set to $0$ if they are negative values, and set to $1$ if they are greater than $1$. This constraint is done as the physical model dictates values be in $[0,1]$, and this change makes the plots in Figure~\ref{fig:sa_herm_realizations} comparable as the BASE-PC surrogate takes values up to approximately $8$ and down to approximately $-1$. The figure shows that this QoI exhibits behavior that makes it very difficult to approximate by a polynomial, and indeed the utility of doing so for practical purposes is suspect. Here, we consider it as a contrast to the Franke function in Section~\ref{subsec:franke}. While the Franke function is smooth and well approximated by polynomials of modest order, this function has sharp transitions that approach discontinuity, and large areas of effectively no variation, which is a function that is not well approximated by polynomials; even here where polynomial orders in $\Xi_1$ reach the hundreds. The surrogate is noticeably deficient around the edge of the transition, and exhibits inaccuracy in approximating the constant regions, both of which are consistent with polynomial approximations to functions of this type. 

We also note that the use of the orthogonality distribution as a proposal distribution is lacking for an accurate coherence-optimal sampling here, tending to generate samples further towards the origin than an accurate coherence-optimal sampling. A more accurate sampling may improve solution recovery when utilizing the sample adaptive approach by better leveraging these rare events. Of some interest is noting that the reference RRMSE and validated RRMSE are significantly less correlated in this example, due to both estimates being significantly less accurate for this problem, though the BASE-PC method, particularly the sample adaptive version, is more accurate than the total order versions. We note that the use of sampling from the orthogonality distribution for proposal samples in the MCMC leads to the coherence-optimal sampling here being significantly less accurate than that of~\cite{Hampton15}.
\section{\texorpdfstring{Theoretical Exposition}{Theoretical Exposition}}
\label{sec:theory}
Here we present theoretical justification for the BASE-PC iteration, particularly with regards to the iterative basis adjustment and correction sampling.  We also address the recovery via $\ell_1$-minimization and its analysis that depends on sparsity, particularly as the goal of adapting a basis is to have contributions from as many basis functions as possible, i.e. to identify representations that are not sparse. However, it is still of great practical utility to be able to recover sparse solutions {\color{black}in an expanded basis}.

We present this analysis in the following sections. Section~\ref{subsec:theory_prelim} details some specifics of the coherence based approach we rely on here, as well as a notion of coupling that is key to the analysis of basis adaptation and some useful matrix bounds for the remaining sections. Section~\ref{subsec:theory_small_s} handles the recovery results for sparse solutions. Section~\ref{subsec:theory_large_s} details recovery results that are useful for recovering solutions that we consider non-sparse.  Section~\ref{subsec:theory_end} identifies a class of problems that under some assumptions may be recovered in a number of samples and basis functions that is independent of dimension.
\subsection{\texorpdfstring{Preliminaries}{Preliminaries}}
\label{subsec:theory_prelim}
Here we present some of the preliminaries used for our main results, including concepts, notation and a few results used in the remaining sections. We first note that all results here rely on a noise model that is at least with high probability uniformly bounded. {\color{black}Let $\bm{\psi}(\bm{\xi})$ denote the realized row vector that evaluates the basis functions in $\mathcal{B}_k$ at $\bm{\xi}$; let $\bm{w}(\bm{\xi})$ denote the weight associated with the coherence-optimal sampling associated with $\mathcal{B}_k$, and let $\epsilon_{\mathcal{B}_k}$ denote the truncation error associated with the basis $\mathcal{B}_k$ as from (\ref{eqn:pre_error}). For us, we then require that $\|\epsilon_{\mathcal{B}_k}(\bm{\xi})w(\bm{\xi})\bm{\psi}(\bm{\xi})\|_\infty \le \lambda$  
holds with high probability for some $\lambda$. We note that this is not a problem for most $u(\bm{\xi})$, and when using most practical distributions for $\bm{\xi}$ and the corresponding orthogonal polynomials and coherence-optimal weights, but could be an issue in more exotic cases.}
\subsubsection{\texorpdfstring{Sparse vs. Non-Sparse Recovery}{Sparse vs Non-Sparse Recovery}}
\label{subsubsec:sparse_v_nonsparse}
The BASE-PC method relies on the basis adaptation procedure to maintain a ratio of samples to basis functions that admits recovery, and we seek to identify a ratio of samples to basis functions that guarantees this stability. The basis adaptivity muddles the compressed sensing interpretation in that we actively seek to have the sparsity parameter be large relative to the number of basis functions. Because of this, we present our main results in both sparse and non-sparse cases. Here and throughout $s$ represents the sparsity parameter, and is broadly the number of non-zero coefficients needed to recover a QoI with a particular polynomial basis. Generally, $s$ increases as more accurate solutions are requested, and the relationship between $s$ and the accuracy of a solution is not addressed here. We do note that the tradeoff with $s$ can be partially interpreted in terms of truncation error as in (\ref{eqn:pre_error}), and this does show up in the presented results.

Heuristically, if basis adaptation is successful then we generally expect a large fraction of basis functions to be useful for recovering the QoI, and this falls into the non-sparse recovery framework. It is also useful to insure that when there is a relatively small number of useful basis functions in our basis, that we may still have {\color{black}a quality recovery}, and this recovery is referred to as sparse recovery. Another benefit of sparse recovery is with regards to how aggressively one may expand a basis at each iteration. An ability to accurately recover sparse reconstructions implies that we may add in a relatively large number of basis functions where few of them are expected to be useful in approximating the QoI. In terms of the BASE-PC method discussed here, this means that larger $\gamma$ and {\tt dim\_add} may be used in the \textit{basis\_expand} algorithm of Algorithm~\ref{alg:expand}. If the number of necessary basis functions is $s < 0.5|\mathcal{B}_k|$, i.e. $s$ is less than half the number of basis functions used at the $k$th iteration, then we consider this to be sparse recovery, and otherwise we consider $s$ to be non-sparse recovery. The surrogate identification may also be anticipated to be more robust with regards to the expansion and contraction parameters that determine the basis adaptation.

We note that sparse recovery has additional factors in $\log(s)$, that are unnecessary for the non-sparse recovery result, and if the non-sparse results are more favorable then those may be utilized freely{\color{black}, as may occur for $s$ near $0.5|\mathcal{B}_k|$.} We also note that there are some nuances that exist in the case of very sparse solutions; but that there is no issue with any of the results if we assume $s\ge 2$, and results can be defined for $s\ge 1$ with some changes to the presentation. We briefly remark on this later in Section~\ref{subsubsec:theory_coupling}. There is also some beneficial improvements to the number of samples when the sparsity is very high, such as when the adapted basis is of high quality and $s$ approaches $|\mathcal{B}_k|$, on which we also remark later in Section~\ref{subsubsec:theory_coupling}. The reason for both of these results arise from the probablistic approach to relevant bounds on the design matrix which we discuss over the course of the next several sections.
\subsubsection{\texorpdfstring{Coherence}{Coherence}}
\label{subsubsec:theory_coh_unif}
Let $\Omega$ be the domain of the random input being considered, $\mathcal{B}$ the current basis, and $\bm \psi(\bm{\xi})$ a $1\times|\mathcal{B}|$ vector whose entries are the evaluation of the basis functions $\psi_k$ at $\bm{\xi}$. Let $w(\cdot)$ denote the weight function associated with the importance sampling that determines how the {\color{black}$\bm\xi$} are drawn, so that $w(\bm{\xi})$ is the weight function evaluated at $\bm{\xi}$. Consider the definitions,
\begin{align}
\mu_{\infty} &:= \mathop{\max}\limits_{\bm{\xi}\in\Omega}\|w(\bm{\xi})\bm{\psi}(\bm{\xi})\|^2_{\infty};\\
\label{eqn:coherence_def}
\mu_2 &:= \mathop{\max}\limits_{\bm{\xi}\in\Omega}\|w(\bm{\xi})\bm{\psi}(\bm{\xi})\|^2_{2};\\
\label{eqn:coherence_def_2}
\mu_2(s) &:=  \mathop{\max}\limits_{\bm{\xi}\in\Omega}\mathop{\max}\limits_{|\mathcal{S}|\le s}\mathop{\sum}\limits_{k\in\mathcal{S}}|w(\bm{\xi})\bm{\psi}_k(\bm{\xi})|^2.
\end{align}
These represent the potential maximum of certain vector norms over potential rows in the matrix $\bm{D}$, where the coherence-optimal importance sampling is to minimize this maximum. We note that as in~\cite{CandesPlan,Hampton14}, the set $\Omega$ could be truncated if for example, these maximums are not bounded over the domain $\Omega$, as occurs with e.g. Hermite polynomials.  While $\mu_{\infty}$ has been used within the context of $\ell_1$-minimization~\cite{CandesPlan,Hampton14,RauhutWard}, and $\mu_2$ has been used within the context of $\ell_2$-minimization~\cite{CDL13,Hampton15,Narayan14}, it may be more appropriate to consider $\mu_2(s)$ in the case of $\ell_1$-minimization. We note that the importance sampling of~\cite{Hampton15} which is used here insures that $\mu_2 = |\mathcal{B}|$, the minimal possible value attainable by independent sampling. {\color{black} We note too that $\mu_2 = \mu_2(|\mathcal{B}|)$.} Further, straightforward bounds can be found relating these notions, as summarized by the following lemma.
\begin{lem}
\label{lem:coherence_relations}
With the coherence parameters defined as in (\ref{eqn:coherence_def}) it follows that,
\begin{align*}
&\max\left(\frac{s}{|\mathcal{B}|}\mu_2,\mu_{\infty}\right) &\le\mu_2(s) &\le& &\min\left(\mu_2,s\mu_{\infty}\right);\\ 
&\max\left(\frac{s}{|\mathcal{B}|}\mu_2,\mu_2(s)\right)&\le s\mu_{\infty} &\le& &\min\left(s\mu_2,s\mu_2(s)\right);\\
&\max\left(\mu_{\infty},\mu_2(s)\right) &\le \mu_2 & \le& &\min\left(|\mathcal{B}|\mu_{\infty},\frac{|\mathcal{B}|}{s}\mu_2(s)\right).
\end{align*}
\end{lem}
\proof
These results follow from standard inequalities of vector norms. $\blacksquare$

The quantities in the center of the inequality chain may each be used to bound $\ell_1$-recovery of a solution of sparsity $\lfloor s/2\rfloor$ with similar bounds, in a manner compatible with the analysis of~\cite{CandesPlan}. We denote any definition in the center of the above inequalities; one of $\mu_2(s)$, $s\mu_\infty$, or $\mu_2$; by $\mu$, and for simplicity of presentation we focus on {\color{black}$\mu=\mu_2(s)$} in what follows. The $\ell_2$-coherence-optimal sampling used in the examples here, is optimal with regards to minimizing $\mu_2$ over all independent random sampling distributions~\cite{Hampton15}. Note that $\mu_2(s)$ is the smallest of these three, but that $\mu_2(s)$ involves a maximum over a combinatorially large set $\{|\mathcal{S}|\le s\}$, that complicates the analysis. Interestingly, it is simple enough to perform a coherence-optimal sampling that minimizes $\mu_2(s)$, as for any realized candidate vector, $\bm{\psi}(\bm{\xi})$, the weight function, $w(\bm{\xi})$, and hence the MCMC sampling, involves only identifying the $s$ elements of the candidate row that have the largest absolute value. However, such a sampling is beyond the scope of this work, but could be useful in cases where a sparsity parameter $s$ is either assumed \textit{a priori} or estimated in some manner.

\subsubsection{\texorpdfstring{Matrix Bounds}{Matrix Bounds}}
\label{subsubsec:theory_essential}
Here we present matrix bounds that will be used to show our results in Section~\ref{subsec:theory_small_s} and Section~\ref{subsec:theory_large_s}. Before presenting a key matrix bound that will be used to justify our coherence-optimal sampling to minimize $\mu_2$, as well as our associated correction sampling, we discuss a normalization for the design matrix, denoted $\bm{D}$ that is made throughout. We utilize a bound in a probabilistic sense of the quantity $\|\bm{D}^T\bm{D}-\mathbb{E}(\bm{D}^T\bm{D})\|$, where in what remains, all unspecified matrix and vector norms are assumed to be $\ell_2$-norms. To consider the convergence of $\bm{D}^T\bm{D}$ to its mean in terms of the sample size, $N$, we normalize $\bm{D}^T\bm{D}$ so that,
\begin{align}
\label{eqn:measure_isometry}
 \mathbb{E}(\bm{D}^T\bm{D}) = \bm{I},
\end{align}
or {\color{black} at a minimum we require that this holds approximately.} 
We note that division of $\bm{D}$ by a constant is associated with a similar normalization on $\bm{Wu}$, and that there is no effect on the computed surrogate when this is accounted for. That is, this normalization is a theoretical convenience with no effect on the computed solution or its associated error. We also note that this normalization differs from that in (\ref{eqn:design_isotropy}), which would be inconvenient here.

For our purposes we let $\mathcal{S}$ denote a subset of the basis having size $|\mathcal{S}|$. We use the subscript of $\mathcal{S}$ to denote that the associated matrix is restricted to only those entries relevant for basis functions in $\mathcal{S}$.

The next probabilistic matrix bound is of a type that is useful for guaranteeing recovery of accurate, stable function approximations~\cite{Tropp12a,Vershynin2010,RauhutWard,CandesPlan,Hampton14,Hampton15}. Specifically, we cite results of Section 5.4 of~\cite{Vershynin2010}, with Theorem 5.44 of that work being directly applied here. We present that theorem here in a slightly different form, and as a lemma.
\begin{lem}~\cite{Vershynin2010}
\label{lem:vershynin544}
Let
\begin{align*}
\bm{E}_{\mathcal{S}} &:=\mathbb{E}\left(\bm{D}_{\mathcal{S}}^T\bm{D}_{\mathcal{S}}\right).
\end{align*}
There exists $\kappa>0$ depending only on $\|\bm{E}_{\mathcal{S}}\|^{-1/2}$, such that
\begin{align*}
  \mathbb{P}\Bigg(\|\bm{D}_{\mathcal{S}}^T\bm{D}_{\mathcal{S}}-\bm{E}_{\mathcal{S}}\| > t\Bigg)\le |\mathcal{S}|\exp\left(-\kappa tN\mu^{-1}\right).
\end{align*}
\end{lem}
\proof This is a rearrangement of Theorem 5.44 of~\cite{Vershynin2010} noting that in the context of that theorem, where $\bm{A}_i$ corresponds to the $i$th row of $\bm{D}_{\mathcal{S}}$, $\|\bm{A}_i\|_2 \le \sqrt{\mu}$ almost surely for all $i$. $\blacksquare$\\

We also show that samples generated from the correction sampling of Section~\ref{subsubsec:sample_id} will complement the old samples in a way such that (\ref{eqn:measure_isometry}) holds, and the realized $\bm{D}_{\mathcal{S}}^T\bm{D}_{\mathcal{S}}$ is near its mean. We first prove this for an individual iteration of sampling. Recall that the purpose of the correction sampling is not to alter moments, but to maintain low aggregate coherence in the samples. For what remains we assume that {\tt max\_sample\_ratio} from Section~\ref{subsubsec:sample_id} is set to infinity. This simplifies much of what follows, specifically we avoid technical complications that would arise from considering {\tt weight\_correction}.
\begin{lem}
\label{lem:iterative_adjustment}
Let $\bm{D}_{\mathcal{S},1}$ be the design matrix associated with an initial set of $N_1$ samples, and $\bm{D}_{\mathcal{S},2}$ that with a correction sampling as from Section~\ref{subsubsec:sample_id} using $N_2$ samples. Let $\bm{E}_{\mathcal{S},1}$ and $\bm{E}_{\mathcal{S},2}$ denote the expectations of $\bm{D}^T_{\mathcal{S},1}\bm{D}_{\mathcal{S},1}$ and $\bm{D}^T_{\mathcal{S},2}\bm{D}_{\mathcal{S},2}$, respectively. Let $\bm{D}_{\mathcal{S}}$ be the full design matrix, restricted to those entries relevant to $\mathcal{S}$ that are used for the computation of $\hat{u}$. For any fixed $t$,
\begin{align*}
\mathbb{P}\Bigg(\|\bm{D}_{\mathcal{S}}^T\bm{D}_{\mathcal{S}}-\bm{I}\| > t\Bigg)\le |\mathcal{S}|\mathop{\min}\limits_{\tau_1+\tau_2=t}\left(\exp\left(-\kappa_1 \tau_1N_1\mu_{1}^{-1}\right)+\exp\left(-\kappa_2 \tau_2N_2\mu_{2}^{-1}\right)\right),
\end{align*}
where $\kappa_i$ depends only on $\|\bm{E}_{\mathcal{S},i}\|^{-1/2}$ and $\mu_{i}$ is associated with samples from the corresponding distribution.
\end{lem}
\proof
Applying Lemma~\ref{lem:vershynin544} to $\bm{D}_{\mathcal{S},1}$ and $\bm{D}_{\mathcal{S},2}$, implies that
\begin{align*}
\mathbb{P}\Bigg(\|\bm{D}_{\mathcal{S},1}^T\bm{D}_{\mathcal{S},1}-\bm{E}_{\mathcal{S},1}\| > \tau_1\Bigg)\le |\mathcal{S}|\exp\left(-\kappa_1 \tau_1N_1\mu_{1}^{-1}\right);\\
\mathbb{P}\Bigg(\|\bm{D}_{\mathcal{S},2}^T\bm{D}_{\mathcal{S},2}-\bm{E}_{\mathcal{S},2}\| > \tau_2\Bigg)\le |\mathcal{S}|\exp\left(-\kappa_2 \tau_2N_2\mu_{2}^{-1}\right),
\end{align*}
where $\bm{E}_{\mathcal{S},1}$ and $\bm{E}_{\mathcal{S},1}$ are the associated expectations so that by the construction in Section~\ref{subsubsec:sample_id}, $\bm{E}_{\mathcal{S},1}+\bm{E}_{\mathcal{S},2}=\bm{I}$. Recall that $\mu_{1}$ and $\mu_{2}$ are the coherence parameters associated with the differing samples. Noting that
\begin{align*}
\bm{D}_{\mathcal{S}}^T\bm{D}_{\mathcal{S}} &= \bm{D}_{\mathcal{S},1}^T\bm{D}_{\mathcal{S},1} + \bm{D}_{\mathcal{S},2}^T\bm{D}_{\mathcal{S},2};\\
\bm{D}_{\mathcal{S}}^T\bm{D}_{\mathcal{S}} - \bm{I}  &= \bm{D}_{\mathcal{S},1}^T\bm{D}_{\mathcal{S},1} - \bm{E}_{\mathcal{S},1} + \bm{D}_{\mathcal{S},2}^T\bm{D}_{\mathcal{S},2} - \bm{E}_{\mathcal{S},2};\\
\|\bm{D}_{\mathcal{S}}^T\bm{D}_{\mathcal{S}} - \bm{I}\| &\le \|\bm{D}_{\mathcal{S},1}^T\bm{D}_{\mathcal{S},1}-\bm{E}_{\mathcal{S},1}\|_2 + \|\bm{D}_{\mathcal{S},2}^T\bm{D}_{\mathcal{S},2}-\bm{E}_{\mathcal{S},2}\|_2,
\end{align*}
completes the lemma. $\blacksquare$\\

Lemma~\ref{lem:iterative_adjustment} is somewhat unsatisfying in that there is no guarantee that $\mu_{1}$ and $\mu_{2}$ are well behaved, even though this is the key heuristic behind the correction sampling. Still, Lemma~\ref{lem:iterative_adjustment} communicates that a convergence of the gramian of the design matrix to identity is maintained by correction sampling. 

We note here that the use of multiple correction samplings leads to a sum of exponentials in Lemma~\ref{lem:iterative_adjustment}. Given the difficulty in addressing the individual sampling coherences, showing a convergence with this method as the number of iterations increases would be difficult. Instead of this, we argue differently, using the technology of coupling presented in Section~\ref{subsubsec:theory_coupling}. Before that however, we present one final result that we utilize when showing uniform recovery in the sparse case, as well as use for the non-sparse case. This also introduces key notation that is used in our approximation results, as well as demonstrating a key result for the recovery of solutions via $\ell_1$-minimization.

{\color{black} Let $\mathcal{B}$ denote the basis at a fixed iteration of BASE-PC.} Define $\tilde{u}$ to be the approximation in $\mathcal{B}$ that minimizes the RRMSE over all such approximations in that basis. Specifically, define $\mathcal{F}$ to be the space of possible approximations built from linear combinations of elements in $\mathcal{B}$, and {\color{black}then
\begin{align}
\label{eqn:u_tilde_def}
\tilde{u} &:= \mathop{\mbox{argmin}}\limits_{\hat{u}\in\mathcal{F}}\mbox{RRMSE}(\hat{u}).
\end{align}
%
Let $\hat{u}$} be the approximation computed at the same iteration of BASE-PC in the basis $\mathcal{B}$ using $N$ samples to form a design matrix $\bm{D}$. Using these definitions we may show a useful result that flows through the {\color{black}restricted isometry constant (RIC),} {\color{black} \cite{Candes06a, CS2},} which is denoted here by $\rho_s(\bm{D})$ and is defined to be the smallest number satisfying
\begin{align}
\label{eq:ric_def}
(1-\rho_s(\bm{D}))\|\bm{c}\|^2_2\le \|\bm{D} \bm{c}\|^2_2\le(1+\rho_s(\bm{D}))\|\bm{c}\|^2_2,
\end{align}
for all $\bm{c}$ having at most $s$ non-zero entries. Here, $\rho_s(\bm{D})$ yields a uniform bound on the spectral radius of the submatrices of $\bm{D}$ formed by selecting any $s$ columns. We occasionally shorten $\rho_s(\bm{D})$ to $\rho_s$, which should not be confusing in context. Related to an RIC is a restricted isometry property (RIP) that occurs when the RIC reaches a small enough threshold, and a RIP guarantees that $\ell_1$-minimization provides a stable approximation. An example of such a restricted isometry property is given in Theorem~\ref{thm:RauhutWard} from~\cite{RauhutWard}, restated here in our notation. This theorem shows that if $\rho_{2s}<3/(4+\sqrt{6})$, where $s$ is a sparsity parameter corresponding to how many basis functions are useful in building a surrogate approximation, then a stable recovery is assured.
\begin{thm}~\cite{RauhutWard}
\label{thm:RauhutWard}
Let $\tilde{\bm{c}}\in\mathbb{R}^{|\mathcal{B}|}$ represent the solution that produce $\tilde{u}$. Let $\bm{c}^{(s)}$ denote the best approximation to $\tilde{\bm{c}}$ in terms of minimizing $\|\tilde{\bm{c}}-\bm{c}^{(s)}\|_2$, where $\bm{c}^{(s)}$ has at most $s$ non-zero entries. Let $\hat{\bm{c}}$ be the solution to (\ref{eqn:ell1}), and let $\delta$ used to compute that solution, be chosen such that {\color{black} $\|\bm{W}(\bm{u}-\bm{\Psi}\tilde{\bm{c}})\|\le\delta\|\bm{Wu}\|_2$.} If
\begin{align*}
\rho_{2s}(\bm{D})&<\rho_{\star} := 3/(4+\sqrt{6})\approx 0.4652,
\end{align*}
then,
\begin{align*}
\|\tilde{\bm{c}}-\hat{\bm{c}}\|_2 &\le \frac{c_1}{\sqrt{s}}\|\bm{c}^{(s)}-\tilde{\bm{c}}\|_1 + c_2\textup{RMSE}(\tilde{u});\\
\|\tilde{\bm{c}}-\hat{\bm{c}}\|_1 &\le c_3\|\bm{c}^{(s)}-\tilde{\bm{c}}\|_1 + c_4\textup{RMSE}(\tilde{u})\sqrt{s},
\end{align*}
 where $c_1,c_2,c_3,$ and $c_4$ depend only on $\rho_{2s}$.
\end{thm}
We note that in the non-sparse case, we take $s\ge 0.5|\mathcal{B}|$ and the requirements on $\rho_{2s}$ are less stringent. For example by Theorem 1 of~\cite{Foucart10} we could take $\rho_{2s} \le 4/(6+\sqrt{6})\approx 0.4734$. We also note that in this case $2s\ge |\mathcal{B}|$, and so the condition here translates to requiring $\rho_{|\mathcal{B}|} < \rho_{\star}$, which is an isometry condition with no ``restriction'' to vectors of a particular sparsity. Utilizing a RIC with $s=|\mathcal{B}|$ is useful here where we do not want to assume sparsity and still want to guarantee a stable solution to (\ref{eqn:ell1}). We note that such a condition is also useful for guaranteeing solutions computed via least-squares regression~\cite{CDL13, Hampton15}, although we do not consider such solutions here. {\color{black} We conclude this section by noting that the condition on $\delta$ is not generally an issue, as cross-validation is chosen so as to minimize $\textup{RRMSE}(\hat{u})$, and when cross-validation has accurate validation, this  loosely corresponds to minimizing $\|\tilde{\bm{c}}-\hat{\bm{c}}\|_2$, so that even if $\delta$ does not satisfy the condition, the bound on $\|\tilde{\bm{c}}-\hat{\bm{c}}\|_2$ will still be satisfied regardless of which $\delta$ is chosen.}
\subsubsection{\texorpdfstring{Coupling}{Coupling}}
\label{subsubsec:theory_coupling}
We assume that the aggregate samples at each iteration of correction sampling closely resemble an independent sample. Heuristically, this is justified as the introduced dependence is given in terms of (\ref{eqn:sample_split_dist}), which is mild. Rigorously, we assume the existence of at least one of several couplings~\cite{lindvall2002coupling} between samples, one corresponding to that of the BASE-PC iterative correction samplings, and the other a set of independent samples drawn from a distribution, that considering (\ref{eqn:sample_split_dist}) should closely coincide with the coherence-optimal distribution for the particular working basis at that iteration. Unfortunately, comparing dependent distributions and coupled independent distributions is difficult to interpret and analyze, and a method for constructing a coupling is currently unavailable. As a result, we assume that a desired coupling exists with a few parameters, leaving as an open problem the verification of the existence of such couplings, as well as any construction of such a coupling. We operate under the heuristic that our coupling is such that the coupled independent distribution is near the coherence-optimal distribution, which is validated by the correction sampling implied by  (\ref{eqn:sample_split_dist}).

Specifically, a coupling here refers to a joint distribution from which random variables are drawn, so that they are dependent in a way that is favorable. Here, we want random variables drawn via the correction sampling distributions to behave similarly to random variables drawn independently from a particular distribution, which for the moment we denote $g_\star$. As the coupled samples are drawn independently, we can deploy powerful existing analysis. As we can bound the error for solutions computed using the samples drawn from $g_\star$, we can in turn bound convergence for those drawn via the correction sampling. We note that coupling may be done between individual realizations of $\bm{\xi}^{(i)}$, or by coupling the entire pool of realized samples $\{\bm{\xi}^{(i)}\}_{i=1}^{N_k}$, as long as the {\color{black} coupled samples} respect that they are drawn independently from some $g_\star$. This provides significant {\color{black}freedom} in how couplings may be identified or constructed.

We now present the couplings that we consider here. Let $\bm{D}$ and $\bm{D}_\star$ be design matrices associated with a common basis $\mathcal{B}$. Let $\bm{D}_\star$ be generated from independent, identically distributed, random sampling, and $\mu_\star(s)$ be the coherence associated with this distribution and basis, as by (\ref{eqn:coherence_def_2}).   If there exists a $\beta$ and $\kappa_t > 0$ {\color{black}such that}
\begin{align}
\label{def:couple_non_sparse}
\mathbb{P}\Bigg(\|\bm{D}^T\bm{D}-\bm{D}_{\star}^T\bm{D}_{\star}\| > t\Bigg)\le \beta \exp\left(-\kappa_t N_k\mu^{-1}_{\star}(|\mathcal{B}|)\log^{-1}(|\mathcal{B}|)\right),
\end{align}
then we say that $\bm{D}$ is \textit{non-sparse-coupled} to $\bm{D}_\star$ with coupling constants $\beta$ and $\kappa_t$. This coupling is so named as it is most useful when considering non-sparse recovery. {\color{black} Let subscript $\mathcal{S}$ denote taking the submatrix associated with columns in $\mathcal{S}$.} Another form of coupling is given by,
\begin{align}
\label{def:couple_strong}
\mathop{\sup}\limits_{|\mathcal{S}|\le s}\mathbb{P}\Bigg(\|\bm{D}_{\mathcal{S}}^T\bm{D}_{\mathcal{S}}-\bm{D}_{\mathcal{S},\star}^T\bm{D}_{\mathcal{S},\star}\| > t\Bigg)\le \beta \exp\left(-\kappa_t N_k\mu^{-1}_{\star}(s)\log^{-1}(|\mathcal{B}|)\log^{-3}(s)\right),
\end{align}
and if this holds, then we say that $\bm{D}$ is \textit{$s$-coupled} to $\bm{D}_\star$.  This form of coupling is useful for considering recovery uniformly over coefficient supports in the case of sparse recovery. We also consider another form of coupling that is weaker than $s$-coupling, in that it requires the supremum of (\ref{def:couple_strong}) to hold over a smaller set. Specifically, fix a set $\mathcal{S}_{0}$, corresponding to a fixed support set that is good for building an approximation to the QoI. Define $\mathcal{S}_r$ to be the set of $\mathcal{S} := \mathcal{S}_{0}\cup\mathcal{R}$, where $|\mathcal{R}| \le r$. Let $\bm{D}$, $\bm{D}_{\star}$, and $\mu_\star$ be as before. If there exists a $\beta$ such that for some $\kappa_t > 0$,
\begin{align}
\label{def:couple_sr}
\mathop{\sup}\limits_{\mathcal{S}\in\mathcal{S}_r}\mathbb{P}\Bigg(\|\bm{D}_{\mathcal{S}}^T\bm{D}_{\mathcal{S}}-\bm{D}_{\mathcal{S},\star}^T\bm{D}_{\mathcal{S},\star}\| > t\Bigg)\le \beta\exp\left(-\kappa_t N_k\mu^{-1}_{\star}(s+r)\log^{-1}(|\mathcal{B}|)\right),
\end{align}
then we say that $\bm{D}$ is \textit{$(s,r)$-coupled} to $\bm{D}_\star$. This recovery is useful for the non-uniform version of sparse recovery, that is, when we consider the recovery of a single QoI. As the set $\mathcal{S}_r$ has comparatively fewer sets over which to take the supremum; the $(s,r)$-coupling is generally weaker than the $(s+r)$-coupling. 

We remark again that the authors are unaware of how to identify such couplings or in how to bound the relevant $\beta$ and $\kappa$ parameters associated with them. Intuitively, we expect the proposed sampling to behave similarly to independent sampling, and this framework can make the concept of similarity to independence explicit. Here, the difference between the iteratively adjusted sample and independent samples is by the relationship in (\ref{eqn:sample_split_dist}), and so we expect the samples to behave similarly to independent samples, which is seen experimentally, where the two sets are indistinguishable in appearance.

The following theorem utilizes each of the above couplings to achieve a corresponding conclusion. Specifically, it links the non-independent random sampling that we use with the independent sampling that is a common assumption in most recovery theorems. This performs the heavy lifting for showing the recovery results in Sections~\ref{subsec:theory_small_s} and~\ref{subsec:theory_large_s}. {\color{black} We note that while $\mu_{\star}$, $\beta_{\star}$ and $\kappa^{\prime}_t$ do depend on the couplings, and hence on $k$, we suppress this dependence for notational brevity.}

\begin{thm}
\label{thm:sample_efficacy}
For the $k$th iteration of sample expansion and solution computation, let $\mathcal{B}_k$ denote the basis; $N_k$ denote the total number of samples; and $\bm{D}_k$ denote the design matrix. Fix $t > 0$, and assume that at least one of the three couplings (\ref{def:couple_non_sparse}), (\ref{def:couple_strong}) or (\ref{def:couple_sr}) exists, with the corresponding coupling constants for $t\ge \epsilon_t$ for some unspecified $\epsilon_t$ that is bounded away from zero. Let $s, |\mathcal{B}| > 1$. There exists $\kappa^{\prime}_{t}>0$ depending {\color{black}on} $t$, $\beta$, and $\kappa_{t}$; but independent of the other variables such that if non-sparse-coupling holds,
\begin{align}
\label{eqn:thm_efficacy_1}
\mathbb{P}\Bigg(\|\bm{D}_{k}^T\bm{D}_{k}-\bm{I}\|> t\Bigg)\le 2\beta_\star\exp\left(-\kappa^{\prime}_{t} N_k\mu^{-1}_{\star}(|\mathcal{B}_k|)\log^{-1}(|\mathcal{B}_k|)\right),
\end{align}
where $\beta_\star = \max(\beta, 1)$. Let $\bm{D}_{k,\mathcal{S}}$ corresponds to the columns of $\bm{D}_k$ corresponding to basis functions in $\mathcal{S}$. If the $s$-coupling of (\ref{def:couple_strong}) holds then with $\kappa^{\prime}_t$ having the same dependency as before,
\begin{align}
\label{eqn:thm_efficacy_2}
\mathop{\sup}\limits_{|\mathcal{S}|\le s}\mathbb{P}\Bigg(\|\bm{D}_{k,\mathcal{S}}^T\bm{D}_{k,\mathcal{S}}- \bm{I}\| > t\Bigg)\le 2\beta_\star \exp\left(-\kappa^{\prime}_{t} N_k\mu^{-1}_{\star}(s) \log^{-1}(|\mathcal{B}_k|)\log^{-3}(s)\right), 
\end{align}
where $\beta_\star = \max(\beta, C_t)$ for some unspecified universal $C_t$. If the $(s,r)$-coupling of (\ref{def:couple_sr}) holds with $r = C s$, where $C$ is near unity but has a mild dependence on $(s, |\mathcal{B}_k|, N_k, \mu_{\star}(s))$, then with $\kappa^{\prime}_t$ having the same dependency as before,
\begin{align}
\label{eqn:thm_efficacy_3}
\mathop{\sup}\limits_{\mathcal{S}\in\mathcal{S}_r}\mathbb{P}\Bigg(\|\bm{D}_{k,\mathcal{S}}^T\bm{D}_{k,\mathcal{S}}- \bm{I}\| > t\Bigg)\le 2\beta_{\star}\exp\left(-\kappa^{\prime}_{t}N_k\mu^{-1}_{\star}(s)\log^{-1}(|\mathcal{B}_k|)\right),
\end{align}
where $\beta_\star = \max(\beta, C_t)$, for some unspecified universal $C_t$ having a minor dependence on $(s, |\mathcal{B}_k|, N_k, t)$.
\end{thm}
\proof
We first define $\bm{D}_{k^{\star}}$ to be a design matrix made from $N_k$ samples drawn independently from the coupled distribution for samples at the $k$th iteration, denoted $g_{k^{\star}}$, using the basis $\mathcal{B}_k$.  We consider first the non-sparse-coupling. We apply Lemma~\ref{lem:vershynin544} to this matrix to get that there exists $\kappa^{\prime\prime}>0$, which in this case is a modest universal constant, such that
\begin{align*}
\mathbb{P}\Bigg(\|\bm{D}_{k^{\star}}^T\bm{D}_{k^{\star}}-\bm{I}\| > t\Bigg)\le |\mathcal{B}_k|\exp\left(-\kappa^{\prime\prime} tN_k\mu^{-1}_{\star}(|\mathcal{B}_k|)\right), 
\end{align*}
and that this holds for all $t\ge t_\epsilon$ for some unspecified $t_\epsilon>0$. We now consider the coupling between the original matrix $\bm{D}_k$ and its coupled, independently sampled matrix, $\bm{D}_{k^{\star}}$. This depends on the type of coupling considered, and we consider first the non-sparse-coupling. For a fixed $t^\prime$, there exists a $\kappa_{t^\prime}$, such that
\begin{align*}
\mathbb{P}(\|\bm{D}_{k}^T\bm{D}_{k}-\bm{D}_{k^{\star}}^T\bm{D}_{k^{\star}}\| > t^{\prime}) &\le \beta |\mathcal{B}_k|\exp(-\kappa_{t^{\prime}}N_k\mu^{-1}_{\star}(|\mathcal{B}_k|)).
\end{align*}
Now,
\begin{align*}
\mathbb{P}\Bigg(\|\bm{D}_{k}^T\bm{D}_{k}-\bm{I}\|> t \Bigg)&\le \mathbb{P}\Bigg(\|\bm{D}_{k}^T\bm{D}_{k}-\bm{D}_{k^{\star}}^T\bm{D}_{k^{\star}}\| + \|\bm{D}_{k^{\star}}^T\bm{D}_{k^{\star}}-\bm{I}\| > t\Bigg),\\
&\le \mathop{\min}\limits_{\stackrel{t_1+t_2 = t}{t_1\ge\epsilon_t}}|\mathcal{B}_k|\left(\beta\exp\left(-\kappa_{t_1}N_k\mu^{-1}_{\star}(|\mathcal{B}_k|)\right) + \exp\left(-\kappa^{\prime\prime} t_2 N_k\mu^{-1}_{\star}(|\mathcal{B}_k|)\right) \right).
\end{align*}
Recall that $\kappa^{\prime\prime}$ is a universal constant. For some $\kappa^{\prime}_t>0$, dependent on $t$, and $\kappa_{t_1}$ as a function of $t_1$ for $t_1\in[\epsilon_t,t]$,
\begin{align*}
 \mathbb{P}\Bigg(\|\bm{D}_{k}^T\bm{D}_{k}-\bm{I}\|> t \Bigg) &\le  2\beta_{\star}|\mathcal{B}_k|\exp\left(-\kappa^{\prime}_t N_k\mu^{-1}_{\star}(|\mathcal{B}_k|)\right),
\end{align*}
where $\beta_{\star} = \mathop{\max}(\beta,1)$. We consider the transfer of $|\mathcal{B}_k|$ into the exponential as $\log(|\mathcal{B}_k|)$ and note that for $|\mathcal{B}_k|>1$ this can be handled by changing the constant $\kappa^{\prime}_t$. For the case $|\mathcal{B}_k|=1$, there is no need to move $|\mathcal{B}_k|$ into the exponential. As a result, this shows (\ref{eqn:thm_efficacy_1}), giving for a newly defined $\kappa^{\prime}_t$ and $|\mathcal{B}_k|>1$
\begin{align*}
 \mathbb{P}\Bigg(\|\bm{D}_{k}^T\bm{D}_{k}-\bm{I}\|> t \Bigg) &\le  2\beta_{\star}\exp\left(-\kappa^{\prime}_t N_k\mu^{-1}_{\star}(|\mathcal{B}_k|)\log^{-1}(|\mathcal{B}_k|)\right).
\end{align*}
To show (\ref{eqn:thm_efficacy_2}), (\ref{eqn:thm_efficacy_3}), we assume either appropriate coupling and let $\bm{D}_{k^{\star},\mathcal{S}}$ denote the submatrix of $\bm{D}_{k^{\star}}$ corresponding to restricting to the columns associated with basis functions in $\mathcal{S}$. We must address the bound as a supremum over choices of $\mathcal{S}$. In the case of (\ref{eqn:thm_efficacy_2}) a similar argument as above leads to,
\begin{align*}
\mathop{\sup}\limits_{|\mathcal{S}|\le s}\mathbb{P}\Bigg(\|\bm{D}_{k,\mathcal{S}}^T\bm{D}_{k,\mathcal{S}}- \bm{I}\| > t\Bigg)&\le \mathop{\min}\limits_{\stackrel{t_1+t_2 = t}{t_1\ge\epsilon_t}}\Bigg\{\mathop{\sup}\limits_{|\mathcal{S}|\le s}\mathbb{P}\Bigg(\|\bm{D}_{k,\mathcal{S}}^T\bm{D}_{k,\mathcal{S}}-\bm{D}_{k^{\star},\mathcal{S}}^T\bm{D}_{k^{\star},\mathcal{S}}<t_1\Bigg)\cdots\\
&+ \mathop{\sup}\limits_{|\mathcal{S}|\le s}\mathbb{P}\Bigg(\|\bm{D}_{k^{\star},\mathcal{S}}^T\bm{D}_{k^{\star},\mathcal{S}}-\bm{I}\|<t_2\Bigg)\Bigg\},
\end{align*}
where the difference between showing (\ref{eqn:thm_efficacy_2}) and (\ref{eqn:thm_efficacy_3}) is taking a supremum over $\mathcal{S}$ belonging to different sets. 

The first term on the right hand side is already accounted for by the definition of $s$-coupling, but the second term is a subtle term to bound. The analogy between the first and second terms of the right hand side also occurs with regards to (\ref{eqn:thm_efficacy_3}) and $(s,r)$-coupling. The couplings are defined in such a way that bounds for the first and second term are compatible so that the bounds in (\ref{eqn:thm_efficacy_2}) and (\ref{eqn:thm_efficacy_3}) are closely connected with bounds associated with the independent samples from $g_{\star}$, with corrections to the constant $\beta_{\star}$ and $\kappa_t^{\prime}$ that account for the coupling. Specifically for $s$-coupling we claim that for some  $\kappa^{\prime\prime}_t$ and $C_t$ depending only on $t$ that
\begin{align*}
\mathop{\sup}\limits_{|\mathcal{S}|\le s}\mathbb{P}\Bigg(\|\bm{D}_{k^{\star},\mathcal{S}}^T\bm{D}_{k^{\star},\mathcal{S}}-\bm{I}\|<t\Bigg) &\le C_t \exp\left(-\kappa^{\prime\prime}_t N_k\mu^{-1}_{\star}(s) \log^{-1}(|\mathcal{B}_k|)\log^{-3}(s)\right).
\end{align*}
And that the $s$-coupling insures then that with the potential changes in constants that (\ref{eqn:thm_efficacy_2}) holds. A similar bound holds for $(s,r)$-coupling and (\ref{eqn:thm_efficacy_3}) with the optimization over a different set. We now argue that both such bounds hold for the independently generated rows, showing the theorem.

Recall that the coupling is defined such that the matrix $\bm{D}_{k^{\star}}$ has independent rows. This allows tighter bounds on this quantity than the na\"{i}ve union bound over all sets satisfying $|\mathcal{S}|\le s$, which would introduce a pessimistic order in the bound. The specifics of these tighter bounds are detailed and not presented here, but we point the interested reader to Talagrand's majorizing measures~\cite{talagrand96, talagrand01} as well as works of Rudelson and Vershynin~\cite{rudelson99_2, rudelson99, rudelson07, Vershynin2010}. We also point to Section 8.6 of~\cite{Rauhut10} for results that more directly translate to our use. We note that these results are typically presented in terms of the coherence parameter in (\ref{eqn:coherence_def}), but that the proofs translate to those of (\ref{eqn:coherence_def_2}). We also note that these results often go further and show results in terms of the sample size needed for an effective recovery, so that the result posted here is an intermediate result. For the most directly applicable results with respect to (\ref{eqn:thm_efficacy_2}) we point to Theorem 8.4 of~\cite{Rauhut10} and the closely connected Proposition 7.1 of~\cite{RauhutWard}. For results directly applicable to (\ref{eqn:thm_efficacy_3}) we reference Section 2.3 of~\cite{CandesPlan} and the associated proofs. $\blacksquare$

We add some remarks regarding Theorem~\ref{thm:sample_efficacy}. We note here that the relationship of $\beta_\star$ in terms of the maximum of $\beta$ and another parameter, and similarly the potential decrease of $\kappa_t$ to $\kappa^\prime_{t}$ is because our analysis compares to independent samples. In the case that $C_t \ge \beta$, and {\color{black}a similar relationship on the associated coefficients of exponential decay}, then the recovery requires a similar number of samples to that of independent sampling. It is not clear in practice how small the corresponding $\beta$ and $\kappa_t$ values can be, but we note that if the sampling is itself independent, {\color{black}then} the trivial coupling of samples to themselves yields a value of $\beta = 0$. In such a case, or in the relaxed case that $\beta$ is below some threshold while $\kappa_t$ is above {\color{black}some threshold, the recovery} bound may be reworked to be identical to independent sampling, modulo a constant factor of $2$ by the proof technique here. As a result, if the non-independent sample used in {\color{black}BASE-PC} resembles an independent sampling, then it may be expected that the $\beta$ and $\kappa_t$ values are small enough that the independent sampling result dominates the recovery. This appears to hold for the examples in Section~\ref{sec:examples}, and it is suspected to hold in some generality.

We {\color{black}additionally note} that this theorem is written in terms of $\mu_{\star}(s)$, corresponding to the definition of coherence in (\ref{eqn:coherence_def_2}), this may be bounded in terms of the other coherence definitions via Lemma~\ref{lem:coherence_relations}. We also note that these bounds are problematic if $s$ or $|\mathcal{B}| = 1$, but that this is an artifact of bounds within the proof, and could be removed by replacing the corresponding $\log(1)$ terms with $1$.

Theorem~\ref{thm:sample_efficacy} is sufficient to bound errors for each iteration, and we are equipped to show results for recovery in both the sparse and non-sparse cases. 
\subsection{\texorpdfstring{Sparse Recovery}{Sparse Recovery}}
\label{subsec:theory_small_s}
Here we consider the recovery of solutions when the sparsity parameter $s$ {\color{black}satisfies $s< 0.5|\mathcal{B}_k|$. We consider} the case of uniform and non-uniform recovery, where uniform recovery refers to the ability of the matrix $\bm{D}_k$ to recover any signal of sparsity $s$. This first result corresponds to uniform recovery. We recall from Lemma~\ref{lem:coherence_relations} that $\mu_{\star}(s) \le s\mu_{\infty}$ and $\mu_{\star}(s) \le \mu_2$. The coherence-optimal relationship in practice leads to $\mu_{\star}(2s)$ being proportional or nearly proportional to $2s$, and the $\ell_2$-coherence optimal sampling used here insures that $\mu_{\star}(2s) \le |\mathcal{B}_k|$ for all $s$.
\begin{cor}{\bf Uniform Sparse Recovery:}
\label{cor:sample_efficacy_sparse_uniform}
Let $t<3/(4+\sqrt{6})$, and assume that a $2s$-coupling holds with regards to Theorem~\ref{thm:sample_efficacy}. For some $C$, let $N_k$ be such that,
\begin{align}
\label{eqn:nk_for_recovery}
N_k\ge (C+\log(2\beta_\star))(\kappa^{\prime}_t)^{-1}\mu_{\star}(2s)\textup{log}^3(2s)\textup{log}(|\mathcal{B}_k|).
\end{align}
Then for the $k$th iteration of {\color{black}BASE-PC, it follows that, with probability
\begin{align}
\label{eqn:bound_prob}
p_k \ge 1 - \exp(-C),
\end{align}
the computed surrogate $\hat{u}_k$ satisfies}
\begin{align}
\label{eqn:rrmse_bound}
\mbox{RRMSE}(\hat{u}_k) \le D_1\mbox{RRMSE}(\tilde{u}_k) + D_2\frac{\|\bm{c}_k^{(s)}-\tilde{\bm{c}}_k\|_1}{\sqrt{s}\sqrt{\mathbb{E}(u^2(\bm{\xi}))}},
\end{align}
where $D_1$, $D_2$ are constants that depend only on $t$; $\hat{u}_k$ is the approximation computed via BASE-PC; and $\tilde{u}_k$ is an optimal approximation as in (\ref{eqn:u_tilde_def}). This result holds uniformly over any $\tilde{\bm{c}}_k$. 
\end{cor}
\begin{remark}
We note that $\|\bm{c}_k^{(s)}-\tilde{\bm{c}}_k\|_1/\sqrt{\mathbb{E}(u^2(\bm{\xi}))}$ converges to zero as $s\rightarrow |\mathcal{B}_k|$, even without the $\sqrt{s}$ term. Though we do not write it that way, this result may be used with a minimization over a range of $s$ such that (\ref{eqn:nk_for_recovery}) is satisfied.
\end{remark}
\proof
As the assumptions of Theorem~\ref{thm:sample_efficacy} is satisfied, we may rewrite (\ref{eqn:thm_efficacy_2}) as
\begin{align*}
\log\left(\mathop{\sup}\limits_{|\mathcal{S}|\le{\color{black}2s}}\mathbb{P}\Bigg(\|\bm{D}_{k,\mathcal{S}}^T\bm{D}_{k,\mathcal{S}}- \bm{I}\| > t\Bigg)\right) &\le \log(2\beta_\star) -\kappa^{\prime}_{t} N_k\mu^{-1}_{\star}({\color{black}2s}) \log^{-1}(|\mathcal{B}|)\log^{-3}({\color{black}2s}),\\
&\le  \log(2\beta_\star) - (C+\log(2\beta_\star)),\\
&= -C.
\end{align*}
where the second inequality follows from (\ref{eqn:nk_for_recovery}). From this it follows that
\begin{align*}
\mathop{\sup}\limits_{|\mathcal{S}|\le {\color{black}2s}}\mathbb{P}\Bigg(\|\bm{D}_{k,\mathcal{S}}^T\bm{D}_{k,\mathcal{S}}- \bm{I}\| > t\Bigg) & \le  \exp(-C),\\
\mathop{\sup}\limits_{|\mathcal{S}|\le {\color{black}2s}}\mathbb{P}\Bigg(\|\bm{D}_{k,\mathcal{S}}^T\bm{D}_{k,\mathcal{S}}- \bm{I}\| \le t\Bigg) & \ge  1 - \exp(-C).
\end{align*}
We note that if $\mathop{\sup}\limits_{|\mathcal{S}|\le {\color{black}2s}}\|\bm{D}^T\bm{D} - \bm{I}\| \le t$ then it follows that $\rho_{{\color{black}2s}} \le t$. We may then apply Theorem~\ref{thm:RauhutWard} to get that,
\begin{align*}
\|\hat{\bm{c}}-\tilde{\bm{c}}\| & \le \frac{c_3}{\sqrt{s}}\|\bm{c}_k^{(s)}-\tilde{\bm{c}}_k\|_1 + c_2\mbox{RMSE}(\tilde{u}_k);\\
\sqrt{\mathbb{E}(\hat{u}(\bm{\Xi})-\tilde{u}(\bm{\Xi}))^2} &\le \sigma_{\min}(\bm{D})^{-1}\|\hat{\bm{c}}-\tilde{\bm{c}}\|,\\
&\le (1-t)^{-1}\left(\frac{c_3}{\sqrt{s}}\|\bm{c}_k^{(s)}-\tilde{\bm{c}}_k\|_1 + c_2\mbox{RMSE}(\tilde{u}_k)\right);\\
&\le \frac{C_3}{\sqrt{s}}\|\bm{c}_k^{(s)}-\tilde{\bm{c}}_k\|_1+ C_2\mbox{RMSE}(\tilde{u}_k),
\end{align*}
where the precise value of $C_2$ depends on $t$, and $c_2$, and similarly $C_3$ depends on $c_3$ and $t$. As
\begin{align*}
\sqrt{\mathbb{E}(\hat{u}(\bm{\Xi})-u(\bm{\Xi}))^2} &\le \sqrt{\mathbb{E}(\hat{u}(\bm{\Xi})-\tilde{u}(\bm{\Xi}))^2} + \sqrt{\mathbb{E}(\tilde{u}(\bm{\Xi})-u(\bm{\Xi}))^2},\\ 
&\le (C_2+1)\mbox{RMSE}(\tilde{u}_k) + \frac{C_3}{\sqrt{s}}\|\bm{c}_k^{(s)}-\tilde{\bm{c}}_k\|_1 .
\end{align*}
Dividing both sides by $\sqrt{\mathbb{E}(u^2(\bm{\Xi}))}$ and setting $D_1 = (C_2+1)$, $D_2 = C_3$, shows (\ref{eqn:rrmse_bound}). $\blacksquare$

We may also address non-uniform recovery which theoretically requires fewer samples, and is especially useful within the context of UQ where design matrices are rarely used to recover large numbers of vastly differing QoIs. We show this scaling in the next result, which shows that several log terms may be removed from $N_k$.
\begin{cor}{\bf Non-Uniform Sparse Recovery:}
\label{cor:sample_efficacy_sparse_non_uniform}
Assume that the $(s,r)$-coupling holds with regards to Theorem~\ref{thm:sample_efficacy}. Let $t=1/4$, for some $C$ let $N_k$ be such that,
\begin{align}
\label{eqn:nk_for_recovery_nonunif}
N_k\ge (C+\log(2\beta_\star))(\kappa^{\prime}_t)^{-1}\mu_{\star}({\color{black}s+r})\textup{log}(|\mathcal{B}_k|).
\end{align}
Using the same notation as Theorem~\ref{thm:RauhutWard} and Corollary~\ref{cor:sample_efficacy_sparse_uniform}, we have for the $k$th iteration of BASE-PC, with probability
\begin{align}
\label{eqn:bound_prob_non_unif}
p_k \ge 1 - \exp(-C),
\end{align}
the computed surrogate $\hat{u}_k$ satisfies
\begin{align}
\label{eqn:rrmse_bound_2}
\mbox{RRMSE}(\hat{u}_k) \le D_1\mbox{RRMSE}(\tilde{u}_k) + D_2\frac{\|\bm{c}_k^{(s)}-\tilde{\bm{c}}_k\|_1}{\sqrt{s}\sqrt{\mathbb{E}(u^2(\bm{\xi}))}},
\end{align}
where $D_1$, $D_2$ are constants that depend on $t$ and have a mild dependence on $(s, |\mathcal{B}_k|, N_k)$; {\color{black}$\hat{u}_k$} is the approximation computed via BASE-PC; and $\tilde{u}_k$ is an optimal approximation as in (\ref{eqn:u_tilde_def}).
\end{cor}
\begin{remark}
We note that the use of $t=1/4$ is for compatibility with the theory presented in~\cite{CandesPlan}, and that this value could be taken larger. We also note that the dependency of $D_1$ and $D_2$ on $(s, r, N, |\mathcal{B}_k|)$ is {\color{black} never larger than $D_3\log^2(|\mathcal{B}_k|)$ for some unspecified $D_3$}. Finally, as in Corollary~\ref{cor:sample_efficacy_sparse_uniform}, a similar optimization over $s$ may be performed.
\end{remark}
\proof
This proof is similar to that of Corollary~\ref{cor:sample_efficacy_sparse_uniform}, but utilizing the fact that we are restricting our coefficient support. Technical details are omitted, as the proof relies on different optimizations and estimates that are generally more favorable with regards to the constants. We point the interested reader to~\cite{CandesPlan} and the proofs leading up to Theorem 1.3 there, as those are sufficient. We note how that paper is presented mostly in terms of the LASSO estimator, which is a dual form of the $\ell_1$-minimization problem in (\ref{eqn:ell1}), but the results translate without issue. We bring special attention to the weak RIP of Section 2.3 of that paper which is the motivation for the $(s,r)$-coupling and its use in Theorem~\ref{thm:sample_efficacy}. $\blacksquare$

Corollaries~\ref{cor:sample_efficacy_sparse_uniform} and~\ref{cor:sample_efficacy_sparse_non_uniform} suggest that a number of samples that scales nearly linearly with the sparsity of the problem is sufficient to guarantee recovery, and demonstrates requirements for stability with respect to the correction sampling described in Section~\ref{subsubsec:sample_id}.  As seen in Section~\ref{sec:examples}, the BASE-PC iteration often selects bases with a number of elements that scale nearly linearly with the number of samples, suggesting that in those cases the desired sparsity parameter is a fraction of the total number of basis functions.
\subsection{\texorpdfstring{Non-Sparse Recovery}{Non-Sparse Recovery}}
\label{subsec:theory_large_s}
Here we consider the recovery of solutions when the sparsity parameter $s$ satisfies {\color{black}$s\ge 0.5|\mathcal{B}|$. We} note that the results in this section use the non-sparse-coupling. We note here that the indepedent coherence-optimal sampling gives $\mu_2(|\mathcal{B}_k|)=|\mathcal{B}_k|$, which is the theoretical minimum. The correction sampling aims to admit a coupling so that $\mu_{\star}(|\mathcal{B}_k|)$ remains near $|\mathcal{B}_k|$.
\begin{cor}{\bf Non-Sparse Recovery:}
\label{cor:sample_efficacy_non_sparse}
Let $t=3/(4+\sqrt{6})$, and assume that non-sparse-coupling holds and that the assumptions of Theorem~\ref{thm:sample_efficacy} and Theorem~\ref{thm:RauhutWard}. For some $C$, let $N_k$ be such that,
\begin{align}
\label{eqn:nk_for_recovery_non_sparse}
N_k\ge (C+\log(2\beta_\star))(\kappa^{\prime}_t)^{-1}\mu_{\star}(|\mathcal{B}_k|)\textup{log}(|\mathcal{B}_k|).
\end{align}
Then for the $k$th iteration of BASE-PC it follows that, with probability
\begin{align}
\label{eqn:bound_prob_non_sparse}
p_k \ge 1 - \exp(-C),
\end{align}
the computed surrogate $\hat{u}_k$ satisfies
\begin{align}
\label{eqn:rrmse_bound_non_sparse}
\mbox{RRMSE}(\hat{u}_k) \le D_1 \mbox{RRMSE}(\tilde{u}_k),
\end{align}
where $D_1$, depends only on $t$; $\hat{u}_k$ is the approximation computed via BASE-PC; and $\tilde{u}_k$ is an optimal approximation as in (\ref{eqn:u_tilde_def}). We note that this result holds uniformly over all $\tilde{u}_k$.
\end{cor}
\proof
There are no significant differences between this proof and that of Corollary~\ref{cor:sample_efficacy_sparse_uniform}. The primary difference is that the proof of Corollary~\ref{cor:sample_efficacy_sparse_uniform} requires a supremum over certain support sets, while this proof assumes the largest possible support, which leads to a bound that is more favorable than setting $s=|\mathcal{B}|$ in Corollary~\ref{cor:sample_efficacy_sparse_uniform}. We note that in the context of that Corollary, $\|\bm{c}_k^{(s)}-\tilde{\bm{c}}_k\|_1 = 0$, which explains the disappearance of the corresponding $D_2$ term. $\blacksquare$
\subsection{\texorpdfstring{Dimension Independent Scaling}{Dimension Independent Scaling}}
\label{subsec:theory_end}
We conclude this section with a theoretically satisfying guarantee for sufficiently smooth $u$, specifically that recovery may be achieved with a number of samples that does not depend on the dimension of the problem. If an exponentially decaying bound can be guaranteed for the coefficients, then to reach a particular RRMSE, $|\mathcal{B}|$ scales independently of dimensionality, and so too does the necessary $N$ to achieve a particular RRMSE, assuming the basis is approximately identified{\color{black}, and that a quality non-sparse-coupling exists.}
\begin{lem}
\label{lem:curse_dimensionality_assumption}
Let $\bm{i}$ be the $d\times 1$ vector that indexes the order of the basis function in each of $d$ dimensions, and let $c_{\bm{i}} = \mathbb{E}(u(\bm{\Xi})\psi_{\bm{i}}(\bm{\Xi}))$ denote the corresponding coefficient for the most accurate reconstruction of the surrogate, $\hat{u}$. If there exists $B>0$ and $\alpha >0$ such that
\begin{align}
\label{eqn:curse_assumption}
|c_{\bm{i}}|\le B\ \textup{exp}\left(-\alpha\mathop{\sum}\limits_{k=1}^d k^2 i_k\right),
\end{align}
then for any $\epsilon\in (0,0.9)$, there exists an anisotropic order basis, $\mathcal{B}_\epsilon$, such that
\begin{align}
\label{eqn:basis_size_condition}
|\mathcal{B}_\epsilon| \le \epsilon^{-\nu},
\end{align}
where $\nu\ge0$ depends only on $D$, $\alpha$ and $\mathbb{E}(u^2(\bm{\Xi}))$. With $\tilde{u}_{\mathcal{B}}$ as in (\ref{eqn:u_tilde_def}), it follows that
\begin{align}
\label{eqn:u_tilde_cond}
\mbox{RRMSE}(\tilde{u}_{\mathcal{B}}) \le \epsilon.
\end{align}
\end{lem}
\begin{remark}
 We note that requiring $\epsilon\in(0,0.9)$ is due to an estimate involving $\log(\epsilon)$ that may produce issues for $\epsilon$ near $1$. Specifically, in (\ref{eqn:basis_size_condition}), for $\epsilon$ near $1$ we would expect a basis having $1$ basis function to suffice. While this could be guaranteed with an arbitrary basis, this is difficult to guarantee with the anisotropic order basis, as the only available such basis is the basis with $\bm{p}= \bm{0}$, which is a basis consisting only of a constant term. In some cases, this basis function may not contribute to an accurate approximation, that is $\tilde{u}$ built in this basis may still have $\mbox{RRMSE}(\tilde{u}) = 1$. Bounding $\epsilon$ away from $1$ removes this issue, which is not of much practical interest when compared to the case of $\epsilon$ approaching $0$. Another fix to the issue would be a bound such as $C\epsilon^{-\nu}$, but we avoid this approach due to an already large number of constants being used in this analysis.
\end{remark}

\proof Consider using exact projection coefficients, i.e $c_k = \mathbb{E}(u(\bm{\Xi})\psi_k(\bm{\Xi}))$, 
 and  including the basis functions associated with largest magnitude coefficients until the RRMSE is less than $\epsilon$.  We bound the size of such a desired basis by considering how many terms of the sum, $ki_k$, return values less than any given threshold $M$, that is we consider all basis functions associated with $\bm{i}$ satisfying
\begin{align}
\label{eqn:m_def_bound}
\mathop{\sum}\limits_{k=1}^dk^2 i_k \le M.
\end{align}
We note that this set of basis functions corresponds to an anisotropic order basis with each $p_k = M/k^2$. We may bound the number of such functions, denoted by $B_M$, independently of dimension. Specifically, the set of all terms that have non-zero order in exactly one dimension is bounded by,
\begin{align*}
M \mathop{\sum}\limits_{k=1}^{\infty}k^{-2} = \frac{M\pi^2}{6},
\end{align*}
which follows from (\ref{eqn:m_def_bound}) by considering how many $i_k$ satisfy the relationship in each dimension, with
\begin{align*}
\mathop{\sum}\limits_{k=1}^{\infty} k^{-2} = \pi^2/6,
\end{align*}
being a classical result. Combinatorially, we may then bound the set of terms with non-zero order in exactly {\color{black}$l$ dimensions that satisfy (\ref{eqn:m_def_bound}) by $(M\pi/6)^l/l!$, that is for $\bm{i}_l$ having at most $l$ non-zero entries,
\begin{align}
\label{eqn:a_neat_bound}
\left|\left\{\bm{i}_l : \mathop{\sum}\limits_{k=1}^dk^2 i_k = M\right\}\right|&\le \left(\frac{M\pi^2}{6}\right)^l (l!)^{-1}.
\end{align}
}Note that the constant term in the basis, $\bm{i} = \bm{0}$, corresponds to including indices with non-zero order in zero dimensions. Summing over basis functions that include elements in any of $l$ dimensions for $l\ge 0$ gives that,
\begin{align}
\label{eqn:bm_bound}
B_M &\le  \mathop{\sum}\limits_{l=0}^{\infty}\left(\frac{M\pi^2}{6}\right)^l (l!)^{-1} = \mbox{exp}\left(\frac{M\pi^2}{6}\right),
\end{align}
which we note does not depend on $d$. 

We consider now how large $M$ should be to insure that $\mbox{RRMSE}(\tilde{u}_{\mathcal{B}})$ is below $\epsilon$. From (\ref{eqn:a_neat_bound}) we can define $\tilde{u}_M$ to be the function approximation that uses all coefficients satisfying (\ref{eqn:m_def_bound}), and note that with (\ref{eqn:curse_assumption}),
\begin{align*}
\mbox{MSE}(\tilde{u}_M) &\le B^2\mathop{\sum}\limits_{l > M} \left(\frac{\pi^2e^{-\alpha}}{6}\right)^{2l} (l!)^{-2}.
\end{align*}
The convergence here is spectral as $M\rightarrow\infty$. Thus there exists a $\nu>0$, depending on $\alpha$ and $B$, such that for any $\epsilon\in (0,0.9)$, and for all $M\ge -\nu\log(\epsilon)$,
\begin{align*}
\mbox{MSE}(\tilde{u}_M) &\le \epsilon.
\end{align*}
We note that to strengthen this to the $\mbox{RRMSE}$ in (\ref{eqn:u_tilde_cond}), we may still take $M\ge -\nu\log(\epsilon)$, and need only potentially increase $\nu$ while adding a dependence on  $\mathbb{E}(u^2(\bm{\Xi}))$.

These two results bound $M$ sufficiently for (\ref{eqn:u_tilde_cond}), and the number of basis functions that satisfy (\ref{eqn:m_def_bound}). Hence we have that $\mathcal{B}_\epsilon$ satisfying $\mbox{RRMSE}(\tilde{u}_{\mathcal{B}})$ is an anisotropic order basis defined by taking each 
\begin{align*}
p_k = M/k^2 = -\nu\log(\epsilon)/k^2.
\end{align*}
Hence, using $M=-\nu\log(\epsilon)$ in  (\ref{eqn:bm_bound}) it follows that for any $\epsilon \in (0,0.9)$,  that $|\mathcal{B}_{\epsilon}| \le \epsilon^{-\nu}$ where $\nu$ depends on $\alpha$, $B$, and $\mathbb{E}(u^2(\bm{\Xi}))$. We remark that these $p_k$ are not necessarily integers, and that requiring $p_k$ to be integers would in turn require a modest increase to $\nu$.  $\blacksquare$\\

The following corollary then shows that the influence of dimensionality has the potential to be significantly reduced when considering basis adaptation. Under the assumptions of Lemma~\ref{lem:curse_dimensionality_assumption}, an anisotropic order basis exists for an accurate approximation with a number of basis functions independent of dimension, which implies that a number of samples to guarantee an accurate computation is also independent of dimension. Note that computations in Section~\ref{sec:sampling} scale favorably in dimension due to the $d$ parameters to define the anisotropic total order basis, so that the computations in the BASE-PC iteration scale well with dimension.

This identifies a class of problems where BASE-PC may achieve accurate results with a benign scaling in dimension. The issue that prevents a stronger statement to this effect is that there is no guarantee provided that such a basis can be identified by the BASE-PC iteration. However, the search of BASE-PC that continually minimizes the estimate of $\mbox{RRMSE}(\hat{u})$ is reasonable, and in practice it has consistently found quality bases.
\begin{cor}
\label{cor:curse_break}
Let the assumptions of Lemma~\ref{lem:curse_dimensionality_assumption} and Theorem~\ref{thm:sample_efficacy} be satisfied {\color{black} such that there exists a non-sparse-coupling at each iteration of BASE-PC. Let} $\epsilon\in(0,0.9)$. Let $t<3/(4+\sqrt{6})$.  There exists a $\nu >0$, and an anisotropic order basis $\mathcal{B}_\epsilon$ satisfying (\ref{eqn:basis_size_condition}). {\color{black}Fix $\epsilon\in(0,0.9)$. If at the $k$th iteration of BASE-PC it follows that $\mathcal{B}_\epsilon \subset \mathcal{B}_k$, and $\hat{u}$ is an approximation in $\mathcal{B}_k$ computed using }
\begin{align}
\label{eqn:nk_bound_cb}
N_k\ge -\nu\log(\epsilon)(C+\log(2\beta_\star))(\kappa^{\prime}_t)^{-1}\mu_{\star}(|\mathcal{B}_k|).
\end{align}
{\color{black}samples then} it follows that, with probability
\begin{align}
\label{eqn:bound_prob_curse_break}
p_k \ge 1 - \exp(-C),
\end{align}
the computed surrogate, $\hat{u}_k$ satisfies
\begin{align}
\label{eqn:rrmse_cond_cb}
\mbox{RRMSE}(\hat{u}) \le \epsilon.
\end{align}
\end{cor}
\begin{remark}
We note that in the event a coupling exists such that $\beta_{\star}$,  $\kappa^{\prime}_t$ and $\mu_{\star}(|\mathcal{B}_k|)$ are independent of $d$ the dimension of the $\bm{\Xi}$, then no statement of this corollary depends on $d$. That is, if the problem exhibits a certain decay in the importance of dimension and order, made explicity in Lemma~\ref{lem:curse_dimensionality_assumption}; and {\color{black} quality couplings exist} to independent samples, as from Theorem~\ref{thm:sample_efficacy}; then there exists an anisotropic order basis such that it is possible to guarantee recovery for problems of arbitrarily high dimensions with a finite number of basis functions and samples.
\end{remark}

\proof From Lemma~\ref{lem:curse_dimensionality_assumption}, we have a bound on the size of the desired basis as it scales with $\epsilon$ for an anisotropic order basis, $\mathcal{B}_\epsilon$ given as in Lemma~\ref{lem:curse_dimensionality_assumption}, and satisfying (\ref{eqn:basis_size_condition}), so that
\begin{align*}
|\mathcal{B}| &\le \epsilon^{-\nu}; &\log(|\mathcal{B}|) &\le -\nu\log(\epsilon).
\end{align*}

To insure (\ref{eqn:rrmse_cond_cb}) holds, take a larger basis {\color{black} for $\mathcal{B}_\epsilon$} whose optimal approximation has an error of $\epsilon/D_1$, where {\color{black}$D_1$} is as from Corollary~\ref{cor:sample_efficacy_non_sparse}. This effect guarantees that (\ref{eqn:rrmse_cond_cb}) holds while requiring a further increase in $\nu$, due to the need for a larger basis. Note that $\kappa_{t^{\prime}}$ is as from Theorem~\ref{thm:sample_efficacy}{\color{black}, when using the non-sparse-coupling}. Then with this basis, and a number of samples satisfying (\ref{eqn:nk_bound_cb}), the computed approximation satisfies (\ref{eqn:rrmse_cond_cb}) with probability at least as large as in (\ref{eqn:bound_prob_curse_break}). $\blacksquare$
\section{\texorpdfstring{Conclusions}{Conclusions}}
\label{sec:conc}
A definition for anisotropic order~\cite{AnisotropicTotalOrderCite} basis is presented as being compatible with accurate PC expansions and having a number of parameters that scales as the problem dimension, allowing a tunable basis which limits the number of unnecessary basis functions in our active basis while still admitting accurate approximations. Using this basis, an adaptive-sampling is identified so that at each iteration all samples taken up to that point are effectively used in the computation of the surrogate solution. If the basis adaptation is successful, then we have {\color{black} performed a theoretical analysis for both sparse and non-sparse recovery. Further, under some assumptions, when recovering a solution from a class of smooth functions, both the size of a necessary basis and the overall number of samples necessary to compute surrogates to desired accuracy does not depend on the dimension of the problem, representing a significant result with respect to the so-called curse of dimensionality. 

Also,} as the design matrix has fewer basis functions and samples than standard PCE approaches, the computation of the coefficients needed to construct the surrogate scales relatively well with the dimension of the problem. Although no guarantee is provided that a successful basis adaptation can be identified in any given number of basis adaptation iterations, the deployed heuristic of greedily searching to minimize an estimate of $\mbox{RRMSE}(\hat{u})$ is numerically seen to perform well for the examples considered. The scaling is significantly more favorable than the exponential growth in basis functions when considering total order expansions.

Numerically we see that a smoothness of the problem in terms of its polynomial coefficients is more informative of the success of this method than the dimensionality of the problem, and that for problems which utilize high-order basis functions that the proposed correction sampling is of significant assistance for recovering a quality approximation when compared to sampling from the orthogonality distribution.
\section*{Acknowledgements}
The work of JH was supported by the DARPA EQuiPS project.

This material is based upon work supported by the U.S. Department of Energy Office of Science, Office of Advances Scientific Computing Research, under Award Number DE-SC0006402, and NSF grant CMMI-145460.
\label{sec:acknow}

\bibliographystyle{elsarticle-num}
\bibliography{citations}
\end{document}